\documentclass[12pt]{report}

\renewcommand{\arraystretch}{1.5}
\usepackage{epstopdf}

\usepackage{graphicx}
\usepackage{epsfig,color}
\usepackage{amsmath}

\newcommand{\redt}{\textcolor{red}}
\newcommand{\bluet}{\textcolor{blue}}

\newcommand{\half}{{\textstyle{\frac{1}{2}}}}
 \newcommand{\thalf}{{\frac{3}{2}}}
 \newcommand{\fhalf}{{\frac{5}{2}}}

\newcommand{\beq}{\begin{equation}}
\newcommand{\enq}{\end{equation}}
\newcommand{\beqa}{\begin{eqnarray}}
\newcommand{\beqast}{\begin{eqnarray*}}
\newcommand{\enqa}{\end{eqnarray}}
\newcommand{\enqast}{\end{eqnarray*}}
\newcommand{\nn}{\nonumber}
\newcommand{\req}[1]{(\ref{#1})}
\newcommand{\lb}{\label}

\newcommand{\pa}{\partial}

\newcommand{\bec}{\begin{center}}
\newcommand{\enc}{\end{center}}
\newcommand{\beqo}{\begin{quote}}
\newcommand{\enqo}{\end{quote}}

\newcommand{\mbf}[1]{\mathbf{#1}}

\newcommand{\cL}{{\cal L}}

\newcommand{\cD}{{\cal D}}

\newcommand{\cN}{{\cal N}}

\newcommand{\cS}{{\cal S}}
\newcommand{\cT}{{\cal T}}

\newcommand{\al}{\alpha}
\newcommand{\be}{\beta}
\newcommand{\ga}{\gamma}
\newcommand{\de}{\delta}
\newcommand{\ep}{\epsilon}
\newcommand{\vep}{\varepsilon}
\newcommand{\ze}{\zeta}
\newcommand{\et}{\eta}

\newcommand{\ka}{\kappa}
\newcommand{\la}{\lambda}
\newcommand{\rh}{\rho}
\newcommand{\si}{\sigma}

\newcommand{\ta}{\tau}

\newcommand{\ph}{\phi}
\newcommand{\vp}{\varphi}
\newcommand{\vph}{\varphi}
\newcommand{\ch}{\chi}
\newcommand{\ps}{\psi}

\newcommand{\Ga}{\Gamma}
\newcommand{\De}{\Delta}

\newcommand{\La}{\Lambda}

\newcommand{\Si}{\Sigma}

\newcommand{\Ph}{\Phi}

\newcommand{\Om}{\Omega}

\begin{document}

\begin{titlepage}
\begin{center}
{\Huge A very Practical Guide to Light Front Holographic QCD }

\bigskip

{\Large
Liping Zou and H.G. Dosch~\footnote{permanent address: Institut f. Theoretische Physik der Universit\"at Heidelberg, Germany.}}

Institute of Modern Physics, Chinese Academy of Sciences\\
Lanzhou

\end{center}

\vspace{4cm}

\begin{quote}
The aim of these lectures is to  convey a working
knowledge of Light Front Holographic QCD and Supersymmetric  Light
Front Holographic QCD. We first give an overview of holographic QCD in general and then concentrate on the application of the holographic methods on QCD quantized in the light front form. We show  how the implementation of the supersymmetric algebra fixes the interaction and how one can  obtain hadron mass spectra with the minimal number of parameters. We also treat  propagators and compare the holographic approach with other non-perturbative methods. In the last chapter we describe the application of Light Front Holographic QCD to  electromagnetic form factors. 
\end{quote}

\end{titlepage}

\tableofcontents

 \sloppy

\chapter{Introduction}

\section{Preliminary  Remarks}
Light Front Holograph QCD~\cite{Brodsky:2003px,Brodsky:2014yha} is
a model theory, which tries to explain non-perturbative features
of the quantum field theory for strong interactions, QCD.  Like in
all realistic quantum field theories, also in QCD perturbation
theory is the only analytical method to obtain rigorous numerical
results. Unfortunately the most interesting questions in particle
physics, like the calculation of hadron masses, cannot be solved
by perturbation theory. The only rigorous method to do that are
very elaborate numerical calculations with supercomputers. These calculations are performed in
Euclidean
space-time and the continuum is approximated by a lattice,  a set  of discrete
points and links between.

In order to get some insight into the structure of the most
interesting  phenomena, one has to make specific models and
approximations. An especially important approach is the
semiclassical approximation of a quantum field theory. Here the
complicated structure of the interaction, which notably involves
virtual particle creation  and annihilation (loops), is
approximated by a potential in a Schr\"odinger-like quantum
mechanical equation. All the results on the structure of atoms
and molecules,  which follow in principle from quantum
electrodynamics (QED),  are not obtained by calculating
complicated Feynman diagrams, but by solving the Schr\"odinger or
Dirac equation with the electromagnetic potentials.  This does not
mean that quantum field theory is obsolete, since firstly it is
used to derive the potentials in the Schr\"odinger equation (in
the simplest case by one photon exchange), secondly important
constraints on the solutions, like those of the Pauli principle,
can only be derived from quantum field theory and finally, quantum
field theory is used to improve the semiclassical results, as is
done for instance by the calculation of the Lamb shift in
QED.

Light front holographic QCD allows to obtain a semiclassical
approximation to QCD. Since the quarks which constitute ordinary
matter are very light, their mass is only a few MeV, the
kinematics is ultra-relativistic. In that case  the so called
Light Front Quantization  is the  easiest way to obtain a
semiclassical approximation. In it this form of quantization
the commutators of the quantum fields are not defined at equal (ordinary)
time, but at equal ``light front time'', which is the sum of the
ordinary time and one of the space coordinates.

The basis of light front holographic QCD is the "holographic
principle". It states that certain aspects of a quantum field
theory in four space-time dimensions  can be obtained as limiting
values of a five dimensional theory~\footnote{The name is derived
from "hologram" which  is a two dimensional picture which contains
the information of a three dimensional object}.  In our case the
basis is the Maldacena conjecture~\cite{Maldacena:1997re}, which
states the equivalence of a {\bf five dimensional classical}
theory with a {\bf four dimensional quantum field} theory. The
five dimensional {classical} theory has a non-Euclidean geometry
(the so called Anti-de-Sitter metric), the four dimensional {
quantum field} theory is a quantum gauge theory, like QCD, but it
has not $N_c=3$ colours, but $N_c \to \infty$, it has conformal
symmetry (that is it has no scale) and furthermore it is
supersymmetric, that is to each fermion field there exist also
bosonic fields with properties governed by a ``supersymmetry''.
Unfortunately this "superconformal" quantum gauge
theory\footnote{The name AdS/CFT correspondence, frequently used
for this holographic approach, comes from the {\bf A}nti-{\bf
d}e-{\bf S}itter metric and the {\bf C}onformal {\bf F}ield {\bf
T}heory.} with infinitely many colours is rather remote from QCD.
Therefore in Light Front Holographic QCD (LFHQCD) one chooses a
``bottom-up'' approach, that is one modifies the five dimensional
classical theory in such a way as to obtain from this modified
theory and the holographic principle realistic features of hadron
physics. This reduces the power to explain structural features of
hadron physics, since just this observed structures
are used as input to determine the modifications of the classical
5-dimensional theory. This shortcoming is removed in
supersymmetric light front holographic QCD (SuSyLFHQCD), which
forms the the main subject of these lectures. Here the
implementation~\cite{Dosch:2015nwa} of superconformal symmetry on
the semiclassical theory fixes the necessary modifications
completely. In this SuSyLFHQCD the number of parameters is just
the one dictated by QCD itself (like in lattice QCD). In the limit
of massless quarks one has the universal scale (fixed for instance
by one hadron mass), and for massive quarks one has also the quark
masses as parameters.  It
should be noted that the underlying supersymmetry is a symmetry
between wave functions of observed mesons and observed baryons
and not a supersymmetry of fields. Therefore no new particles like
``squarks'' or ``gluinos'' have to be introduced.

The derivation of semiclassical equations for hadron physics is
certainly a big achievement of the holographic principle, but not
the only one. The correspondence allows in principle to determine
all matrix elements of the quantum field theory by the classical
solution of the five-dimensional theory. Therefore  one can also
calculate form factors in
LFHQCD~\cite{Brodsky:2007hb,Sufian:2016hwn}.

A limitation on the accuracy of the numerical results is the limit
of infinitely many colours. This limit is well studied in the
framework of conventional QCD~\cite{tHooft:1973alw} and leads
typically to errors of the order of 10\% of the hadronic scale or
around 100 MeV.

Since the aim of this notes is  to convey a practical working knowledge as fast as
possible, this  necessitates many omissions of more subtle points.
Also the quoted literature is mostly confined to subjects directly
related to the material, which is  explicitly treated in these notes,  but the quoted literature allows easily
to find more sources and to expand the knowledge.
%%%%%%%%%%%%%%%%%%%%%%%%%%%%%%%%%%%%%

%

\section{Old string theory in strong interactions}
Before QCD emerged as a consistent theory based on quark and gluon
fields in the early 1970ies, there was  another approach to strong interaction physics, which
did not search for elementary particles at all. The basis of this
approach was \textbf{duality}.
\begin{figure}
\begin{center}
\includegraphics*[width=5cm]{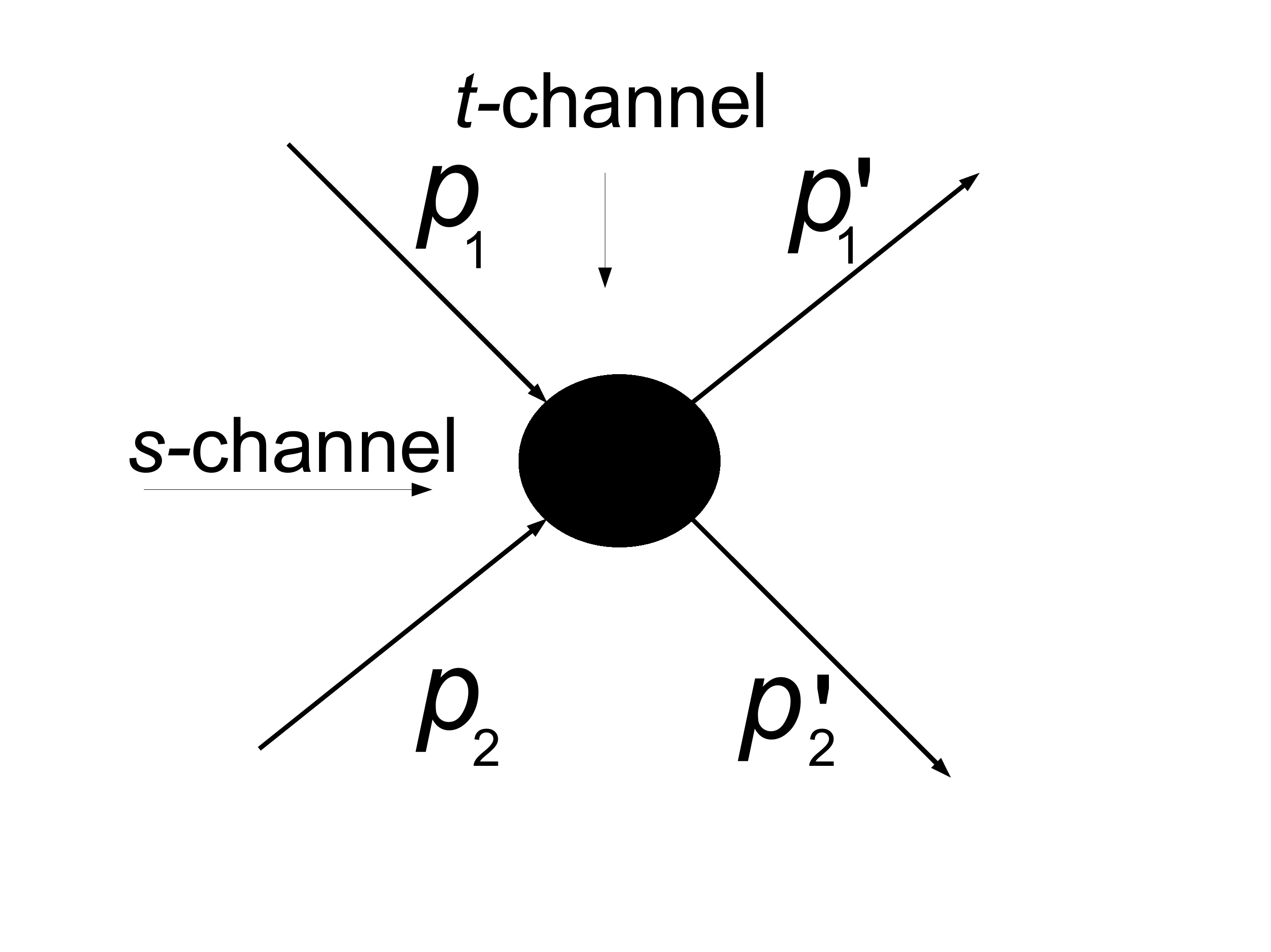}
\end{center}
\caption{Elastic scattering amplitude. \lb{scat}}
\end{figure}
For an elastic  scattering amplitude $T(s,t)$, Fig. \ref{scat}, which depends on the total energy $s=(p_1+p_2)^2=(p'_1+p'_2)^2$ and the momentum transfer $t=(p'_1-p_1)^2 =(p'_2-p_2)^2$, we have two salient features:\\
1) For low values of $s$ we observe resonances, for instance in
$N-\pi$ scattering the $\Delta$ and higher resonances. This means
that there are poles in the variable $s$ at the resonance masses. The
amplitude $T(s,t)$  behaves near the resonance as
$$T(s,t) \sim  \frac{A}{s-m^2_R}$$
For unstable resonances $m_R$ has an imaginary part.

2) For high values of $s$ we have Regge behaviour, that is in that
limit the amplitude  behaves like
$$T(s,t) \sim s^{\al(t)}, $$
the function $\al(t)$ is called a Regge trajectory.

This gives a good description of high energy scattering, that is
for large values of $s$ and for negative values of $t$. For positive values of
$t$, which can occur in annihilation, a resonance pole with total angular momentum
$J$ occurs at those values of $t$, where $\al(t)$ is a nonnegative
integer $J$. It turned out that linear trajectories, that is \beq
\al(t) = \al_0 + \al' \cdot t  \lb{lintr} \enq give a good
description of the data. $\al_0$ is called the intercept and $\al
'$ the slope of the trajectory. The concept of duality was
developed  as an  attempt to unify  these two seemingly very
different features.

\begin{figure}
\setlength{\unitlength}{1cm}
\begin{center}
\begin{picture}(10,5)(0,0)
 \put(0,0){\vector(1,0){7}}
 \put(0,-1){\vector(0,1){5}}
  \put(1.5,1){\circle*{0.2}}
\put(1.5,2){\circle*{0.2}} \put(1.5,3){\circle*{0.2}}
\put(1.5,4){\circle*{0.2}}  \put(2.5,2){\circle*{0.2}}
\put(2.5,3){\circle*{0.2}}
  \put(2.5,4){\circle*{0.2}}
\put(3.5,3){\circle*{0.2}} \put(3.5,4){\circle*{0.2}}
 \put(4.5,4){\circle*{0.2}}
 %\put(6.5,2){\circle*{0.2}}
%\put(3.5,3){\circle*{0.2}} \put(4.5,3){\circle*{0.2}}
%\put(5.5,3){\circle*{0.2}} \put(6.5,3){\circle*{0.2}}
% \put(4.5,4){\circle*{0.2}}
 % \put(5.5,4){\circle*{0.2}} \put(6.5,4){\circle*{0.2}}
\put(7.2,0){\Large $t$}
 \put(0,4.2){\Large $J$}
 \put(0,-0.5){\line(1,1){5}}
  \put(0,0.5){\line(1,1){4}}
  \put(0,1.5){\line(1,1){3}}  \put(0,2.5){\line(1,1){2}}
   %\put(5,-0.5){\line(1,1){2}}
 % \put(3,4.6){Regge traj.} \put(5.5,4.6){daughter traj.}
\end{picture}
\end{center}
\caption{Trajectories in the Veneziano model.\label{vene}}
\end{figure}
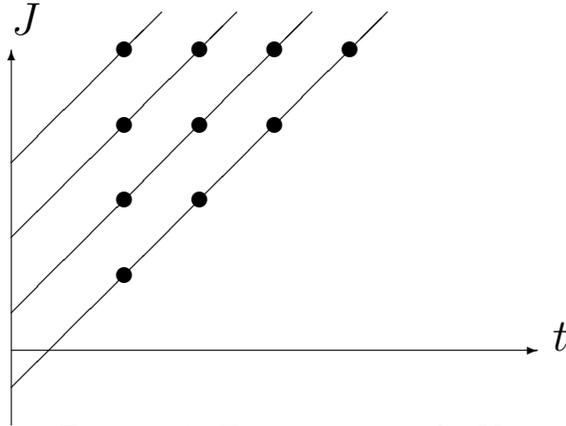

An important  model for scattering amplitudes which shows this
dual behaviour is the \textbf{Veneziano model }$V(s,t)$
\cite{Veneziano:1968yb}. It consists of a sum of expressions like

\beq T(s,t) = \frac{\Gamma(1-\al(s)) \Gamma(1-\al(t))}
{\Gamma(2-\al(s) - \al(t))} = \frac{\Gamma(1-\al_0-\al' s))
\Gamma(1-\al_0-\al' t)} {\Gamma(2-\al_0-\al' s-\al_0-\al'
t)}\label{veneziano} \enq
 with the linear trajectory $ \al(x) = \al_0+\al ' x$.
 $\Gamma(z)$ is the Euler Gamma function which for integer values is
 the factorial,
 $\Gamma(z+1)=z!$. From the properties of the $\Gamma$ function
 follows:
for large values of $s$ and negative values of $t$  the amplitude
$T(s,t)$  shows Regge behaviour, and
 it has   resonance poles for for integer values of  $\al(s)$ or $\al(t)$.
 These poles lie on straight lines, the lowest one is
 called the Regge trajectory, the  ones above it are called daughter
 trajectories, see Fig. \ref{vene}.
%%%%%%%%%%%%%%%%%%%%%%%%%%%%%%%%%%%%%%
%%%%%%%%%%%%%%%%%%%%%%%%%%%%%%%%%%%%%\%%%%%%%%%%
%%% up to here

It was soon realized, that the Veneziano model corresponds to a
string theory, where the rotation of the string gives the
resonances along the Regge trajectories and the vibrational modes
yield the daughter trajectories, see figure \ref{rotvib}.

\begin{figure}
\begin{center}
\includegraphics*[width=5cm]{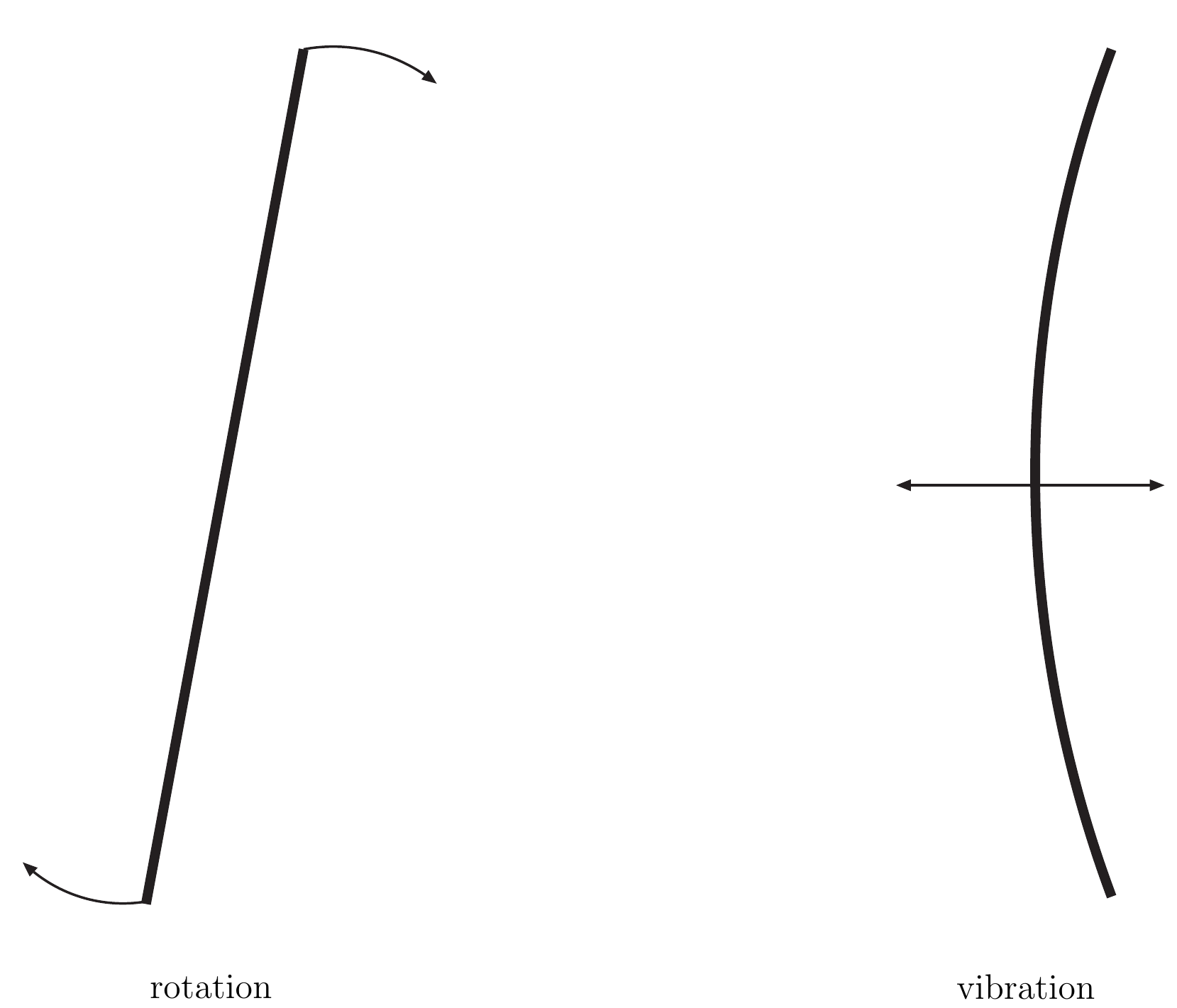}
\end{center}
\caption{Rotational and vibrational modes  of a string.
\lb{rotvib}}
\end{figure}

In this approach the hadrons are not point-like objects nor %are they
 composed of point-like objects (elementary quantum fields),
but they are inherently extended objects: strings.

\begin{figure}[h]

\begin{center}
\includegraphics*[width=4cm]{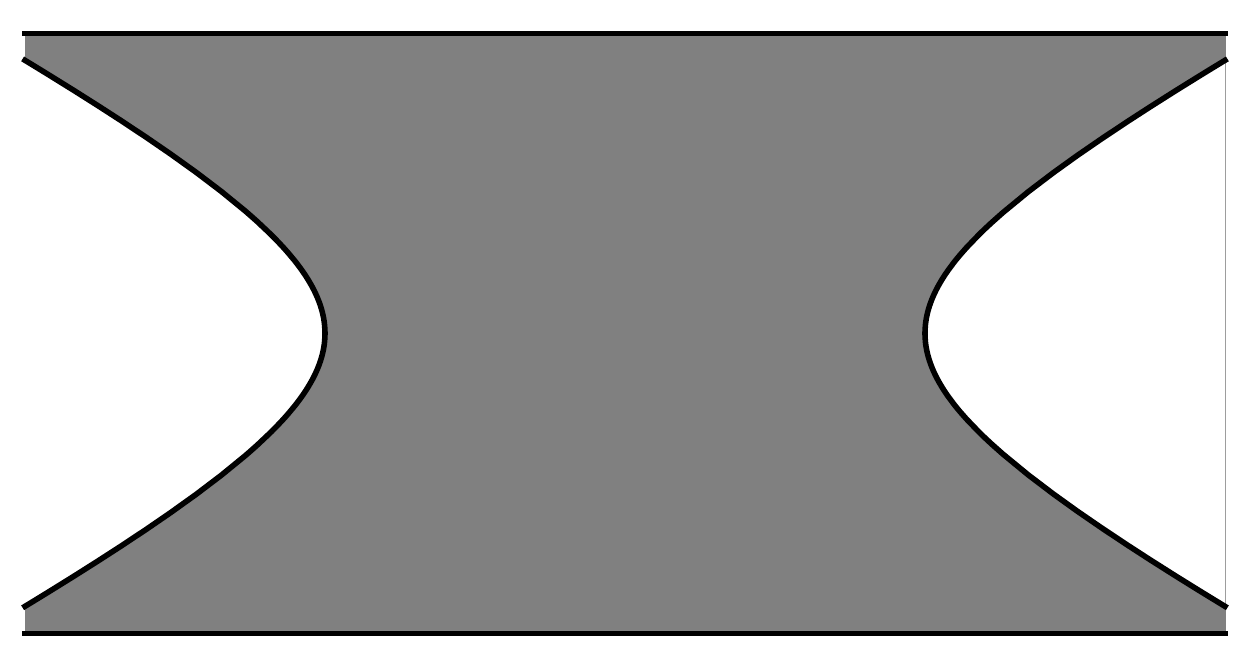}~~~~~
\includegraphics*[width=4cm]{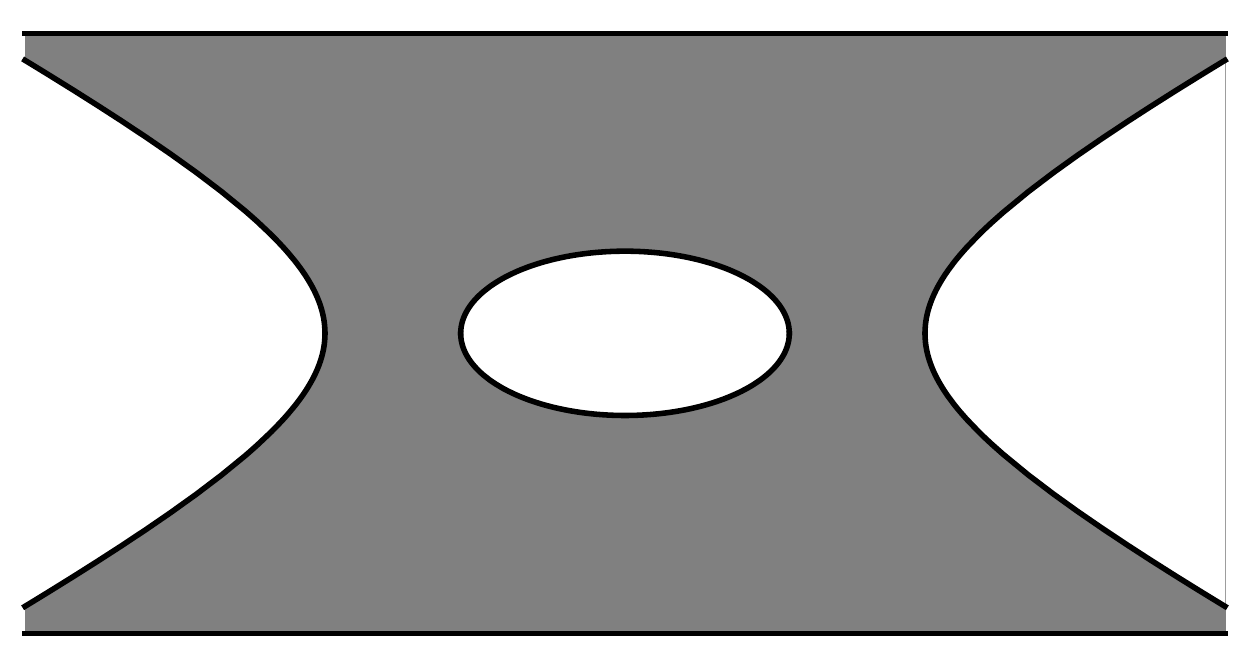}
\end{center}
\caption{Veneziano model(left) and quantum corrections(right) in string theory.
\label{venezian}}
\end{figure}

One imprtant  result of the classical relativistic string is that the angular
momentum is proportional to the squared mass of the string, $J
\sim m^2$ ; this  is just the Regge behaviour. The Veneziano model
corresponds to a classical string theory, quantum corrections to
it are shown in figure \ref{venezian}.

\begin{figure}[h]

\begin{center}
\includegraphics*[width=7cm]{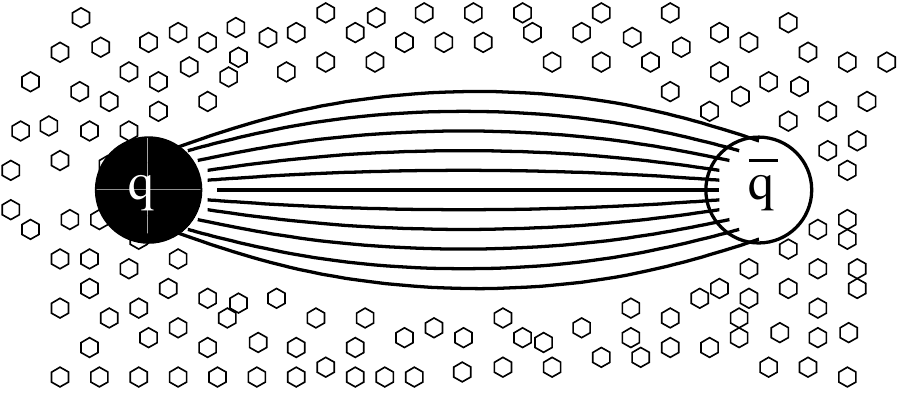}
\end{center}
\caption{Formation of a colour electric flux tube \label{flux}.}
\end{figure}

Big hopes were put in the Veneziano model and its development, but
soon it turned out that it was not the most adequate theory for
strong interactions. Beyond internal difficulties one reason was
that Quantum Chromodynamics (\textbf{QCD}) came out as a strong
competitor and now this field theory is generally considered as
the correct theory of strong interactions. String theory however
developed in a completely different direction and it is nowadays
considered as the best candidate for a quantum theory of
everything (TOE), that is of all interactions, including gravity.
 But string theory in strong interaction physics was never
completely dead. The reason is that many aspects of
non-perturbative QCD seem to indicate that hadrons have indeed
stringlike features. The most popular model for confinement, the
t'Hooft-Mandelstam model (see figure \ref{flux}) is based on the
assumption that that the colour-electric force lines are
compressed (by monopole condensation) into a flux tube which
behaves in some respect indeed like a string.

Also the particular role of  quarks as confined particles shows
some analogy with a string picture. If you split a hadron, you do
not obtain quarks, but again  hadrons. In a similar way, if you
cut a string you do not obtain two ends, but two strings again. We
shall see in the next subsection, that string theory plays, at
least indirectly again a role in strong interaction physics
through the holographic approach.

\section{AdS/CFT \label{adsgen}}
 String theory became very esoteric. Firstly for consistency reasons the basic theory
 had to be supersymmetric, and secondly the theory had to be formulated in a space-time
 with much more dimensions than 4. The only reason that it was pursued further,
 apart from the purely mathematical interest, was that restricted to 4 dimensions it
 yielded a gauge quantum field theory, that is a quantum field theory like QCD.
\begin{quote}
{\small {\bf Supersymmetry} is a symmetry which relates particles
with different spin. There exists  a theorem of Coleman and
Mandula which says that such a symmetry is impossible. The only
way out is to extend the concept of symmetry, which is generated
by an algebra of commuting generators, to a supersymmtry which is
generated by commuting and anticommuting operators. To each
particle with integer spin there must be also particles with half
integer spin. Unfortunately the fields of the observed particles
with different spin cannot be related by supersymmetry (susy). A
big hope of LHC was to find supersymmetric partners of existing
particles, but it was not realized up to now. In our approach supersymmetry
plays an important role, but not as a symmetry  of quantum
fields, but of wave functions.

In the case of {\bf higher dimensions} all the dimensions except
those of space-time are supposed to be ``rolled up'' that they
cannot be observed with present day technology, and most probably
with the  technology of the next centuries. Some years ago there
was  hope  that some of the dimensions, only to be perceived   by
gravity, might be macroscopic (for instance 10$^{-6}$ m). But this
hope did not realize. }
\end{quote}

The present renewed interest of phenomenologically oriented
physicists in this seemingly esoteric field came through
another esoteric principle, the holographic principle: One can
sometimes obtain results of a theory in a space of $d$ dimensions
easier, if one considers it as a limit of a problem in a space of
higher dimension. This principle was first applied to the
thermodynamics of black holes.
 The application to strong interactions goes back to a conjecture made by Maldacena, later elaborated by
Gubser Klebanov and Polyakov, and Witten
1998\cite{Maldacena:1997re,Gubser:1998bc,Witten:1998qj}
\footnote{A more recent short review is \cite{Ramallo:2013bua}, a
very complete description can be found in the book of Ammon and
Erdmenger \cite{AE15}, for non-specialists see e.g.
\cite{Gubser:2009md} and the very short article
\cite{Klebanov:2009zz}.}
\footnote{The seminal paper by Maldacena
received 13233 citations until end of 2017, that is the record for
a theoretical paper.}. It states that a certain string theory is
equivalent to a certain Yang-Mills theory. Many people tried to
bring this mathematically
 high-brow theory down to earth and try to learn from string
 theory some aspects of nonperturbative QCD.

The basis for the application of the holographic principle to
solve quantum field theories is the following. There are good
reasons to believe, that a certain superstring (Type II B) theory
in ten dimensions is dual to a highly supersymmetric (N=4) gauge
theory (Maldacena conjecture). Duality here means, that the
classical solutions of the 5-dimensional
 gravitational theory\footnote{ A gravitational theory is a theory where the interaction
  is due to the (non-Euclidean) metric, like the gravitation in our 4-dimensional world can be
  derived from the metric.}
determine the properties of the confined objects in the
4-dimensional field theory. This sounds very promising. The five
dimensional gravitational theory is rather simple, it is based on
the  metric of a 5-dimensional space, the so called Anti-de-Sitter space,
AdS$_5$. The dual quantum field theory is very far from QCD.
It is a gauge quantum field theory, but it is a conformal
theory that contains no mass scale and therefore cannot
give rise to hadrons with finite masses.
 Furthermore it is supersymmetric and has an infinite number of colours.

\begin{figure}[h]

\begin{center}
\includegraphics*[width=15cm]{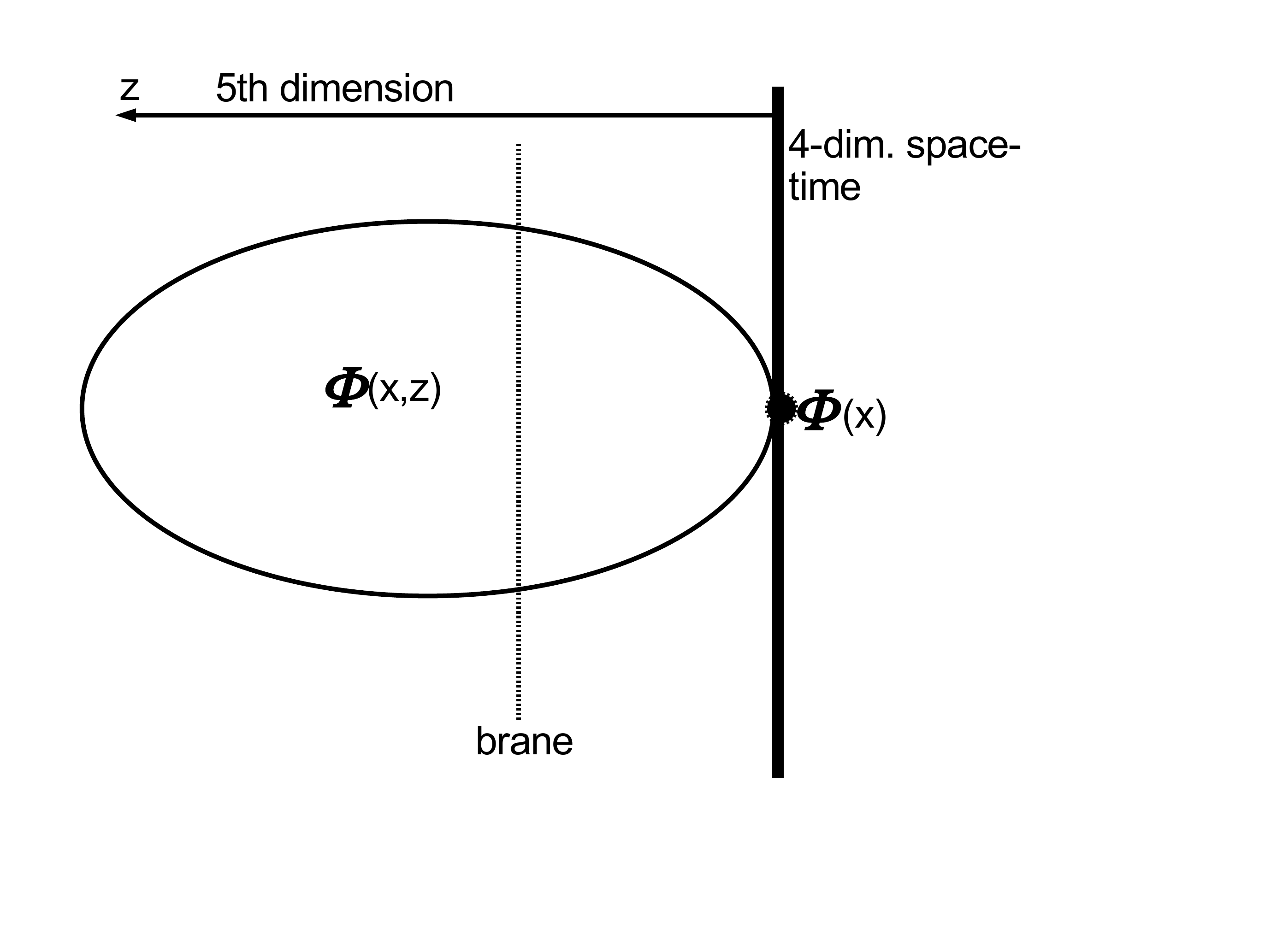}
\end{center}
\caption{A local QCD operator extended in the AdS$_5$ (bulk).
\label{bulk}}
\end{figure}

The relation between the two very different theories comes over
the so called  D-branes. A D-brane is a hyper-surface on which
open strings end. Since energy and momentum flows from the string
to the D-branes they are also dynamical objects. The D stands by
the way for Dirichlet, since the Dirichlet boundary conditions on
the D-brane are essential for string dynamics. In the mentioned
case the D3-branes have 3 space and one time direction and they
are boundaries of a 5 dimensional space with maximal symmetry
 (we shall come to this back in detail). In
figure \ref{bulk} a D1-branes (1 space, 1 time dimension) are
shown, at which an open string ends (picture at a fixed time).

There are two very different approaches to apply  the holographic
 principle to a more realistic situation:

\begin{itemize}
    \item The top-down approach:  One looks for a superstring theory which has as limit on
    a D3 brane realistic QCD or at least a similar theory. This
    approach is very difficult and has to our knowledge not yet led
    to phenomenologically very useful results.
    \item The bottom-up approach: One starts with QCD, or at least a theory near QCD,  and tries to
    construct at least an approximate string theory which one can
    solve and obtain nonperturbative results for QCD.
\end{itemize}

Needless to say that we follow here the bottom-up approach.

The procedure we adopt will be the following: We construct
operators in AdS$_5$ which correspond to \textbf{local} QCD
operators, e.g. a vector field $ \bar \psi(x) \gamma_M \psi(x)$, and
study the behaviour of this operator in the 5 dimensional space
(the so called bulk), and hope to get information on the
properties of confined objects.

%%%%%%%%%%%%%%%%%%%%%%%%%%%%%%%%%%%%%%
\chapter{Some mathematical preparations}

\section{The general claim}
The AdS-CFT correspondence claims, that in a certain limit the
essential results  of a quantum field theory, like propagators,
bound state poles etc, can be obtained  from the classical
solutions of the higher dimensional gravitational theory. This can
be very concisely formulated in terms of the generating
functionals. We shall come back to that in more detail in Chapt.
\ref{c6} and give here only a short overview.

A generating functional $Z[j]$ contains all information on the
observables of  quantum field theory. For a free theory we
have $Z[j]= e^{\int dx\,dy \, j(y) D(x-y) j(x)}$, where $D(x-y)$
is the free propagator.

The correspondence statement claims: \beq Z_{FT}[j] = i
S_{AdS}[\Ph_{cl}] _{\Big/
 \Phi_{cl} \stackrel{z \to  0 }{ \longrightarrow} j} \enq
where $\Phi_{cl}$ is the solution of the classical equations of
motion derived from the action in the AdS$_5$.

We shall exploit that relation in chapter \ref{c6}  in order to
calculate propagators. But here we take a very practical
attitude. We construct in AdS$_5$ the action for a field with the
same quantum numbers as the one we want to investigate in the
4-dimensional quantum field theory. The classical equations of
motion, the solutions of which minimize the action, are the bound
state wave equations for the hadrons. But before we come to that, we have to make some
preparatory steps.

\section{Metric in 5-dimensional Anti-de-Sitter space.}

\subsection{Euclidean Metric}
The line element in Minkowski metric, that is our usual
relativistic space time continuum, is given by: \beq ds^2 =
-(dx^1)^2 -(dx^2)^2 -(dx^3)^2 +(dx^4)^2 = \sum_{\mu\nu=1}^4
\et_{\mu\nu}\, dx^\mu dx^\nu \label{line_euclid} \enq $g_{\mu\nu}$
is called the metric tensor. In Minkowski space the metric tensor in Cartesian coordinates 
is particularly simple and given by:
 \renewcommand{\arraystretch}{0.8}
 \beq \{\et_{\mu\nu}\}= \left\{\begin{array}{c c c c} -1 & 0 & 0 & 0\\
0 & -1 & 0 & 0\\0 & 0 & -1 & 0\\0 & 0 & 0 & 1\\
\end{array} \right\}
\label{eta}  \enq

Since $\{\et^{\mu\nu}\}$ is not positive definite, it is called a
 pseudo-Euclidean metric tensor and geometry in Minkowski space is
 called pseudo-Euclidean.

The metric tensor in a Minkowski space with 3 space, 1 time
variable and an additional spacelike 5th coordinate  is
\beq  \lb{e5} \{\et_{MN}\}= \left\{\begin{array}{c c c cc} -1 & 0 & 0 & 0& 0\\
0 & -1 & 0 & 0& 0\\0 & 0 & -1 & 0& 0\\0 & 0 & 0 & 1& 0\\0 & 0 & 0 & 0&-1\\
\end{array} \right\}
 \enq

\subsection{Non-Euclidean metric: Anti-de-Sitter space \lb{adsmath}}

In non-Euclidean geometry, the elements of the metric are no
longer constants, but may differ from point to point. The line
element of (\ref{line_euclid}) therefore becomes:

\beq (ds)^2 =\sum_{M N} g_{M N}(x)\, dx^M\ dx^N \label{line} \enq

where $g_{MN}$ is a symmetric matrix function, $g_{M N}(x)= g_{N
M}(x)$.

The metric tensor in AdS$_5$ is~\footnote{The specific form of the metric depends naturally on the choice of coordinates. In AdS$_5$ we always choose the so called Poincar\'e coordinates} : \beq  g_{MN} =
\frac{R^2}{z^2}\left\{\begin{array}{c c c cc} -1 & 0 & 0 & 0& 0\\
0 & -1 & 0 & 0& 0\\0 & 0 & -1 & 0& 0\\0 & 0 & 0 & 1& 0\\0 & 0 & 0 & 0&-1\\
\end{array} \right\} = \frac{R^2}{z^2}  \, \eta_{MN}\, \label{ads3} \enq

Here $z=x_5$ is the fifth variable, normally called the hologtraphic variable,  $R$ is a measure for the curvature of the space.

The modulus of the determinant of the metric tensor is
accordingly: \beq |g|= \left(\frac{R^2}{z^2}\right)^{5} \lb{det}
\enq

%{ Caution}

The inverse metric tensor is given by upper indices: \beq
\{g_{MN}\}^{-1} \equiv g^{MN} \enq

that is $\sum_{N=1}^5  g_{MN} \, g^{NA} = \de_M^A$, in Euclidean
metric one has $\et_{MN}=\et^{MN}$.

In the future we will use the Einstein convention: over upper and
lower indices with equal name will be summed, that is e.g. \beq
g_{MN}\,g^{NA}\equiv \sum_{N=1}^5  g_{MN} \, g^{NA}. \enq

One calls lower indices covariant indices and upper indices
contravariant indices. With the metric tensor and its inverse one
can transform a covariant into a contravariant index and vice
versa: $a^M= g^{MN}a_N, \; a_M= g_{MN}a^N$.

With the help of the metric tensor we can construct easily
invariants from covariant and contravariant quantities. If $a_M$
and $b_M$ are covariant vectors in AdS, then the product
$g^{MN}\,a_M\,b_N $ is an invariant, like the 4-product of two
Lorentz vectors  in Minkowski space, $\et^{\mu \nu} a_\mu \,
b_\nu$, is an invariant.

 The invariant volume element
in AdS$_5$ is the Euclidean volume element $d^4x dz$ multiplied by
the square root of the modulus of the determinant of $g_{MN}$. For
the metric tensor \req{ads3} the determinant is the product of the
diagonal elements and hence we obtain as invariant volume element
of AdS$_5$: \beq dV = d^4x dz \sqrt{|g|} = d^4x\,
dz\left(\frac{R}{z}\right)^5 \lb{volume} \enq where we have
inserted  the modulus of the determinant of $g^{MN}$ with the
relation \req{det}.

Since in the following we shall always jump between the
4-dimensional Minkowski space and the 5-dimensional space AdS$_5$
we will introduce the following conventions:

The Greek indices, $\mu,\nu, \al \dots$ run from 1 to 4, where
$x^4 = c\,t$ is the timelike variable. In the 5-dimensional space,
we shall use capital Latin letter, $M,N, A \dots$ which run from 1
to 5, and $x^5=z$.

\section{Relation between AdS and CFT parameters \lb{ads-cft}}

Before coming to the relation, we have to introduce the Planck
units. They are named, since Plank was the first to look for
natural units, which are independent of human standards (like
meter, second etc) and he realized that with his constant $\hbar$
he could achieve it.

In conventional units we have three dimensionful quantities: mass
$[m]$, time $[t]$, and length $[l]$. We have as fundamental
constants the velocity of light $c$,  Planck's constant $\hbar$,
and Newton's constant of gravity $G_N$. The natural unit for the
velocity is certainly the velocity of light $c$ ,% that is $[t] = [l]/c$,  
 the natural unit of  Energy is  $[m]\,c^2$, for the action
$[E][t]$ the natural unit is $\hbar$. Therefore we can reduce the
three dimensions to only one, e.g. the length. We have $[t]=[l]/c;
[m]= \frac{\hbar}{c [l]}$. The gravitational constant is defined
in Newton's law: $F= \frac{m_1\,m_2}{r^2}\, G_N$. With that we
can obtain as natural unit for the remaining dimension, the
length, the Planck length: \beq \l_P= \sqrt{\frac{\hbar }{c^3}
\,G_n\; }  \approx 1.6\times 10^{-33} cm \enq The natural unit for
the mass and the time are the Planck mass $m_P$ and the Planck
time $t_P$: \beqa
m_P&=& \sqrt{\frac{\hbar c}{G_N} }\approx 1.2 \times 10^{14} GeV/c^2 \approx 2.2 \times 10^{-5} g\\
t_P &=&\frac{l_P}{c}  \approx 5.4 \times 10^{-44} s \enqa

From string theory one deduces the following relation between rhe AdS$_5$ and CFT quantities:

\beq \left\{\frac{l_P}{R}\right\}_{AdS} =
\left\{\frac{\sqrt{\pi}}{2^{1/8} N_c^{1/4}}\approx
\frac{1.6}{N_c^{1/4}}\right\}_{CFT} \enq

In a gravitational theory quantum effects can be neglected if
$\l_P \ll R$. In our universe, where the curvature radius is
infinity or very large and the Planck length is tiny, quantum
effects are completely negligible. But in the very early universe,
say one Planck time $t_P$ after the big bang, quantum gravity must have played
a crucial role, therefore we cannot say anything about the big
bang properly.

But back to AdS, in order to treat gravity in AdS  as a classical
theory, the number of colours must be huge ($> 10^4$), so the application in our
world, where the gauge theory QCD has three colours, $N_C=3$,
seems to be hopeless. But fortunately gouge  theories with $N_C
\to \infty$ are well studied and might give results not so far
from $N=3$, typical deviations may be $1/N_c^2 \approx 0.11$. We
come to this important point in the next subsection.

\section{Gauge Theory in the limit $N_c \to \infty$}
\begin{figure}
\begin{center}
\includegraphics*[width=9cm]{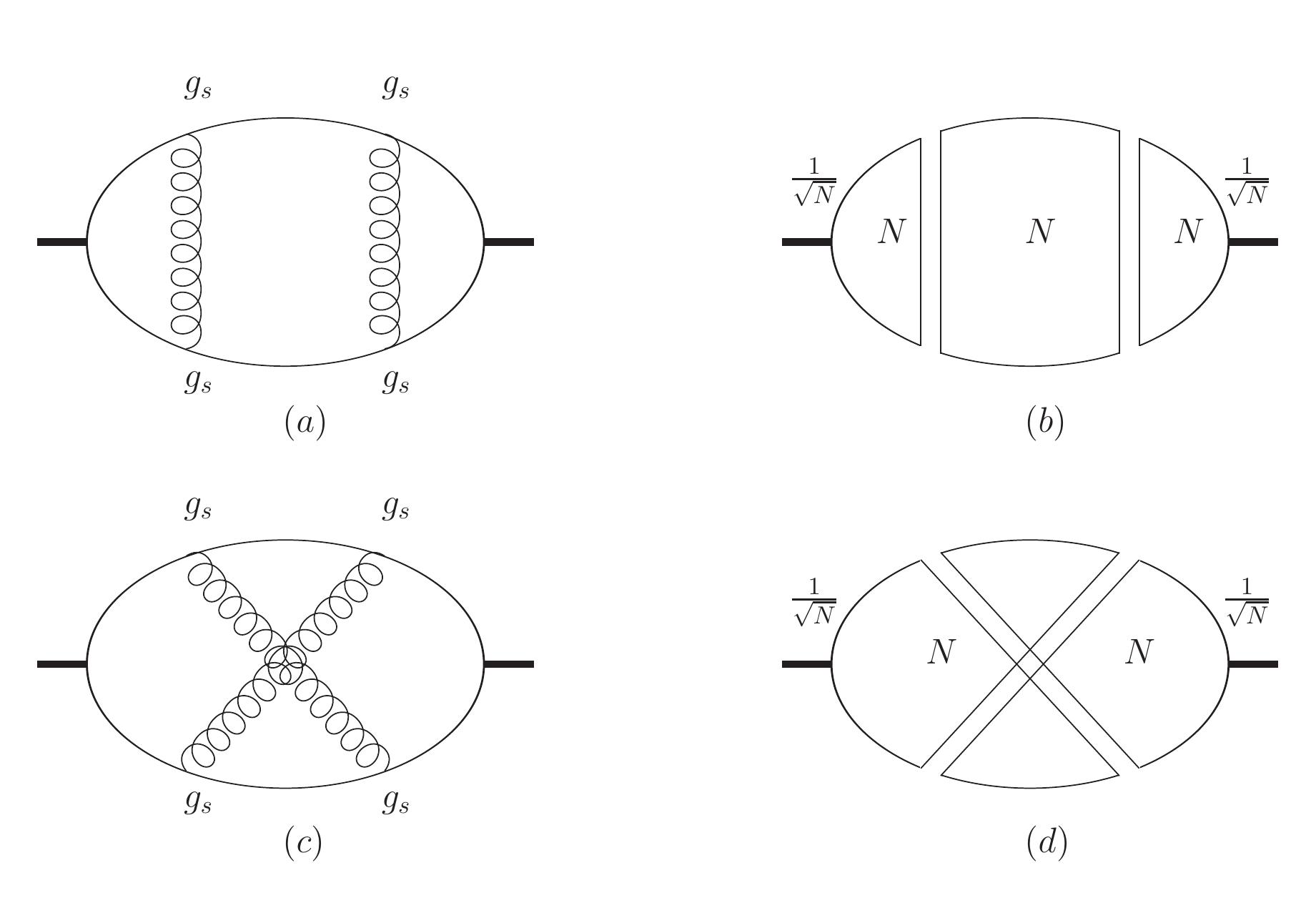}
\includegraphics*[width=9cm]{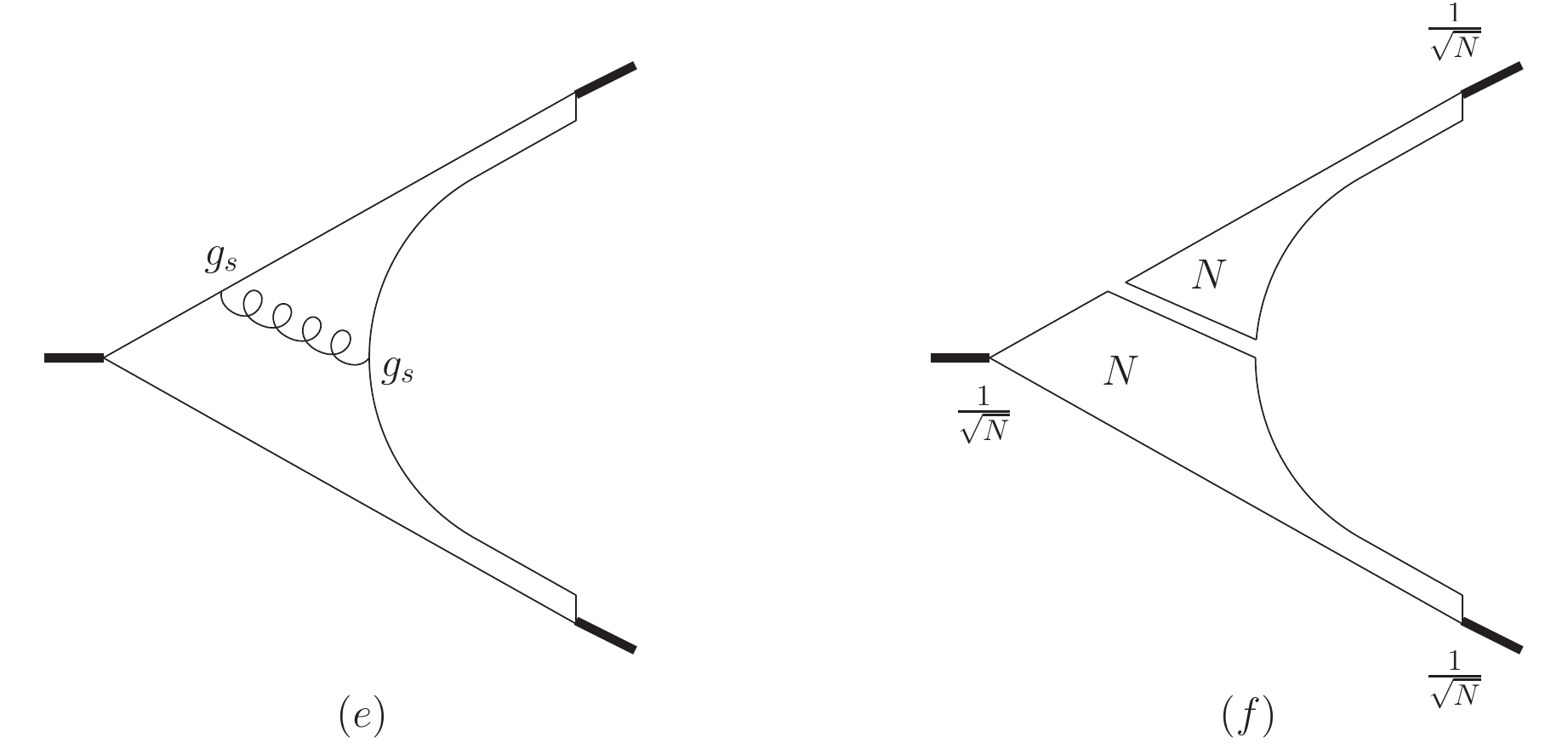}
\caption{Evaluation of the colour factor for several diagrams. In
the diagrams (a) and (c)  we have a factor $g_s^4$ from gluon
exchange. According to (b), where gluon lines have been
replaced by double lines of quarks, we have 3 loops and hence a
factor $N_c^3$. Together with the normalization factors this
yields the overall colour expression  $\frac{1}{N_C} N_C^3 \,g_s^4
$ which is finite in the t'Hooft limit \req{thooft}. For the
aplanar diagram (c) we have only two loops, see (d) and therefore
the contribution vanishes in the $N_c\to \infty$ limit. In the
decay diagram (c) we have $g_s^2$, two loops and three vertices.
This yields the overall factor $\frac{1}{N_c^{3/2} }\, N_c^2
\,g_s^2$ which vanishes in the t'Hooft limit.}
\end{center}
\label{N2}
\end{figure}

We consider the t' Hooft limit~\cite{tHooft:1973alw} of a gauge
theory with $N_c$ colours, for a review see \cite{Manohar:1998xv}:
\beq \lb{thooft} \lim_{N_C\to \infty}  g_s^2 \,N_C =
\rm{constant}. \enq In Fig \ref{N2} we consider a two gluon
contribution to a hadron propagator. The thick line stands for a
hadron, the thin line for a (anti-)quark and the wavy line for a
gluon. The summation over all colours amounts to the following.
Since in a gauge theory with $N_C$ colours a colour-neutral meson
consists of  $N_C$ quark-antiquark  pairs, insert for  each
hadron-quark vertex the normalization factor $\frac{1}{N_C}$. For
colour summation, each  gluon line can in the large $N_c$
limit be replaced by a quark-antiquark double line. Colour
summation gives for each loop a factor $N_c$. This yields for the
graph a) the representation b) and the colour summation factor:
\beq \frac{1}{N_C} N_C^3 \,g_s^4 = (N_C\,g_s^2)^2  \enq which
survives in the large $N_C$ limit. In graph c) the gluon lines
cross (aplanar diagram) and the representation is d). Here we have
only 2 loops, and therefore this diagram contributes like \beq
\frac{1}{N_C} N_C^2 \,g_s^4 = \frac{(N_C\,g_s^2)^2}{N_C} \to 0
\enq and does not contribute in the large $N_C$ limit. Generally
one can show that all planar diagrams survive in the large $N_C$
limit, all aplanar ones do not. In Fig. \ref{N2} e) we give the
first order contribution to a decay of a hadron into two hadrons.
The two-line representation is given in f) and yields the
colour factor: \beq \frac{1}{N_C^{3/2} }\,  N_c^2 \,g_s^2=
\frac{N_C\, g_s^2}{\sqrt{N_C}} \to 0 \enq and hence does not
contribute in the large $N_C$ limit. This is true for any order
and hence we obtain the important result that in the $N_C\to \infty$
limit all hadrons are stable.

One can also show easily that all interactions between hadrons
(colourless objects) vanish in the limit $N_C \to \infty$, that is
in this limit only the confining forces survive. Therefore we can
hope to get realistic results for spectra and form factors, but
cannot try to calculate scattering cross sections.

%%%%%%%%27.11.
%%%%%%%%%%%%%%%%%%%%%
\chapter{The AdS action and wave equations for a (pseudo-)scalar and a vector field}
\section{The (pseudo-)scalar field}

\subsection{Euclidean metric}
As mentioned, we can derive all the properties of the
4-dimensional quantum field theory  from the solutions of the
classical action of the higher dimensional theory with a
non-Euclidean metric. Therefore we construct now the AdS action
for a (pseudo-)scalar field $\Ph(x)=\Phi(\vec x, ct, z)$.

We start with  a more familiar case, the action of a free scalar
field in Minkoswski space. It is given by the integral over the
Lagrangian $\cL$:
 \beq A= \int d^4x
\underbrace{\frac{1}{2}(\eta^{\nu\rh}\,\pa_\nu\Phi\, \pa_\rh \Phi -
\mu^2\, \Phi^2)}_{\cL}\label{l0} \enq

The solutions of the classical equations of motion are  the
functions  $\Phi(x)$ for which the action is minimal. From this
follows that the equations of motion for classical fields are the
Euler-Lagrange equations:
  \beq \pa_\nu \frac{\pa \cL}{\pa(\pa_\nu \Phi)} -
  \frac{\pa\cL}{\pa\ph}=0\lb{el0}.\enq
 This leads to the wave equation for a free (pseudo-)scalar field, called the Klein-Gordon equation:
  \beq \eta^{\nu \rh } \pa_\nu \pa_\rh \Phi + \mu^2 \Phi =0.\lb{sw0}\enq

\subsection{AdS$_5$ metric \lb{adsmetric}}
 If we  go from Euclidean metric  to the non-Euclidean AdS$_5$ metric we have    in the action \req{l0}  to replace:
  \begin{itemize}
    \item $\eta^{\mu \nu } \to g^{MN}$
    \item $d^4 x \to d^4 x \,\sqrt{|g|}\, dz $, see \req{volume}
    \item
In principle we have also to replace the normal (Euclidean)
derivative $\pa_\mu$  by the so called covariant derivative
$D_\mu$ in AdS$_5$, but for a scalar field the covariant
derivative is equal to the the normal
one~\footnote{\lb{fn1} Since the displacement of a
vector or a tensor in non-Euclidean geometry depends on the
metric, this has also to be considered in the derivative. The
covariant derivative $D_M$ of a vector field $V_N(x)$ contains the
so called Christoffel symbol $\Ga^K_{MN}$: $D_M \,V_N(x) = \pa_M
V_N - \Ga^K_{MN} \,V_K$, the Christoffel symbol can be calculated
from the metric tensor $g_{MN}(x)$, see e.g.
\cite{Brodsky:2014yha}, App. A.}
\end{itemize}

 Instead of the action Euclidean (\ref{l0}) we obtain in
 AdS$_5$ geometry:
 \beq A= \int d^4x\, dz\,
\underbrace{\sqrt{|g|}\frac{1}{2}(g^{MN}\,\pa_M\Phi\pa_N\Phi -
\mu^2\Phi^2)}_{\cL}\label{l1} \enq
 and from (\ref{el0}) we obtain instead of the Euclidean equations of motion (\ref{sw0}) the equations  in non-Euclidean metric:
\beq \pa_R \bigg(\sqrt{|g|} g^{RN} \pa_N \Phi\bigg) +
\sqrt{|g|}\mu^2 \Phi =0.\lb{sw1ne}\enq or
 \beq
 g^{RN} \big(\pa_R\pa_N \Phi\big) +
\mu^2 \Phi = \frac{-1}{\sqrt{|g|}} \,\pa_R \big(\sqrt{|g|}
g^{RN}\big)\,  \pa_N \Phi \lb{sw1b}\enq

We see that  the interaction in  the non-Euclidean metric leads to
an interaction term namely  the r.h.s. of (\ref{sw1b}), this is
due to the term $\pa_R \big(\sqrt{|g|} g^{RN}\big)\,  \pa_N\Phi$. If the
metric is Euclidean, the elements of the metric tensor are
independent of the coordinates and the r.h.s of (\ref{sw1b})
vanishes in that case.

In AdS$_5$ we have, see sect. \ref{adsmath}
  \beq \{g_{AB}\} =
\frac{R^2}{z^2} \{\eta_{AB}\} \qquad  \{g_{AB}\}^{-1}  =
\{g^{AB}\}= \frac{z^2}{R^2} \{\eta^{AB}\} \qquad |g|=
\left(\frac{R^2}{z^2}\right)^5 \enq
 therefore the l.h.s. of \req{sw1b} depends only on the holographic variable $x^5=z$.

It is convenient for further calculations to write
  \beq \frac{R^2}{z^2} \equiv e^{ 2
A(z)} \quad {\rm with}\; A(z)=  -\log z + \log R \lb{expm}\enq

Then we have \beq g_{MN} = e^{2 A(z)} \eta_{MN};  \quad g^{MN} =
e^{-2 A(z)} \eta^{MN} \quad \sqrt{|g|} = e^{5 A(z)}\enq and obtain
\beqa \cL&=& \frac{1}{2} e^{5 A(z)}\left( e^{-2 A(z)}\eta^{MN}
\pa_M
\Ph\, \pa_N \Ph -\mu^2 \Ph^2\right) \\
  &=& \frac{1}{2} e^{\ka A(z)}\left( \eta^{MN} \pa_M
\Ph\, \pa_N \Ph - e^{2 A(z)} \mu^2 \Ph^2\right) \lb{LA} \enqa with
$\ka =3$.

The ingredients of the  Euler-Lagrange
 equations \beq \pa_A \frac{\pa \cL}{\pa(\pa_A \Phi)}-\frac{\pa \cL}{\pa \Phi} =0
 \enq
 are
 \beq
\pa_A \frac{\pa \cL}{\pa(\pa_A \Phi)}=\pa_A\bigg(e^{\ka
A(z)}\eta^{AB}\pa_B\Phi\bigg); \quad \frac{\pa \cL}{\pa \Phi} =
 - \mu^2 \Phi  e^{(\ka+2) A(z)}; \quad \ka=3 \enq

From that we obtain the  wave equation for the (pseudo)scalar
field in AdS$_5$:
 \beq
 e^{\ka A(z)}\bigg(\eta^{\al\be}\pa_\al\pa_\be \Phi -\pa_z^2 \Phi - \ka
 \pa_z A(z) \pa_z \ph +\mu^2 e^{2 A(z)}  \Phi\bigg) = 0. \lb{whw}\enq
  where according to the convention of sect. \ref{adsmath} the indices $\al,\be$ run from 1 to 4 (Minkowski space).

For most cases it is convenient to work  with the transformed
field  $\tilde \Phi(q,z) $ where the Minkowski coordinates of the
field $\Phi(x,z)$ are Fourier transformed. \beq \Phi(x,z) = \int
\frac{d^4q}{(2\pi)^4}\, e^{i q x}\, \tilde \Phi(q,z) \label{Fourier}\enq Then we
can replace
$$\eta^{\al\be}\pa_\al\pa_\be \Phi(x,z) \rightarrow -q^2\, \tilde \Phi(q,z)$$
and  obtain for  $ \tilde \Phi(q,z)$ the equation \beq \lb{specA}
\left( - \pa_z^2- \big( \ka \pa_z A(z)\big) \pa_z  -q^2 +
\frac{(\mu R)^2}{z^2}\right) \tilde \Phi(q,z) =0 \enq inserting
\req{expm} we arrive finally at the wave equation \beq
\label{specB} \left(-\pa_z^2 +  \frac{\ka }{z} \,\pa_z  +
\frac{(\mu R)^2 }{z^2}-q^2 \right) \tilde \Ph(q,z) =0 \enq with
$\ka=3$.

\subsection{Solution and transformation of the equation of motion}

The general solution of the differential equation \req{specB} can
be obtained with the  mathematica program: \beq \tilde \Ph(q,z)=
(qz)^{(1+\ka)/2} \big(A\, J_\nu(qz) + B\, Y_\nu(qz)\big) \mbox{ with }
\nu = \sqrt{(\mu R)^2 + (\ka + 1)^2/4} \enq {\small
\begin{quote} Mathematica: "BesselJ[n, z] gives the Bessel
function of the first kind $J(n, z)$.", "BesselY[n, z] gives the
Bessel function of the second kind $Y(n, z)$."
\end{quote}}

The general solution $\tilde \Ph(q,z)$ of \req{specB} is
therefore: \beq \lb{specgen} \tilde \Ph(q,z) = A \, z^2\,J_\nu
(qz) + B \, z^2\,Y_\nu(qz) \mbox{ with } \nu^2 = (\mu R)^2 + 4
\enq

In the next few chapters  we shall only consider the solution
which is regular at $z=0$, that is we put $B=0$. We can transform
(\ref{specB}) also into a Schr\"odinger-like equation by rescaling,
we introduce $\Phi_S(q,z)$: \beq \tilde \Phi(q,z) = z^{3/2}
\tilde \Phi_S(q,z) \enq

then the linear derivative in \req{specB} vanishes and we obtain:
\beq \lb{specC} \left( - \pa_z^2 + \frac{ (4 \mu
R)^2+15}{4z^2}\right)\tilde \Phi_S(q,z) =q^2 \tilde \Phi_S(q,z)
\enq This looks like a Schr\"odinger equation with a potential;
but this potential does not lead to the formation of bound states.

  We note the disappointing fact, that there are indeed nontrivial solutions to the  equation of motion \req{specB} or \req{specC} , but there is no sign of confinement, since for any value of $q^2$ we find a solution. This is, however, not astonishing. The AdS$_5$ has what is called maximal symmetry and this results in conformal symmetry in the corresponding quantum field theory in the 4-dimensional Minkowski space. We shall come to conformal symmetry later, here it is sufficient to say that conformal symmetry demands, that there is no mass scale in the theory. This applies also for classical QCD in the limit of massless quarks, but not to the quantized QCD. Here we have indeed scales (e.g. the nucleon mass).

\section{Modifications of the action: The hard- and soft-wall model}
\subsection{The hard wall model~\cite{Erlich:2005qh,Brodsky:2006uqa}}
A way to impose the existence of discrete solutions is the hard
wall model. Here it is assumed, that the Lagrangian $\cL$  in
\ref{l1} is  only valid for  values of $z\leq z_0$ (so to speak
inside a hard wall at position $z_0$), and one imposes that  the
solutions of \req{specB} or \req{specC} vanish  at that  boundary
$z_0$.

In the case of $ \Phi(q,z)= z^2 J_\nu(q z)$ this means that the
$q$ must satisfy $qJ_\nu(q\,  z_0) =0$, i.e. the values of the
hadron masses $M^2= q^2$ are determined by the zeros $j_{\nu s}$
of the Bessel functions, see Fig. \ref{bessel} \beq \sqrt{q^2} \,
z_0 = j_{\nu s} \enq .

\begin{figure}
\bec
\includegraphics*[width=7cm]{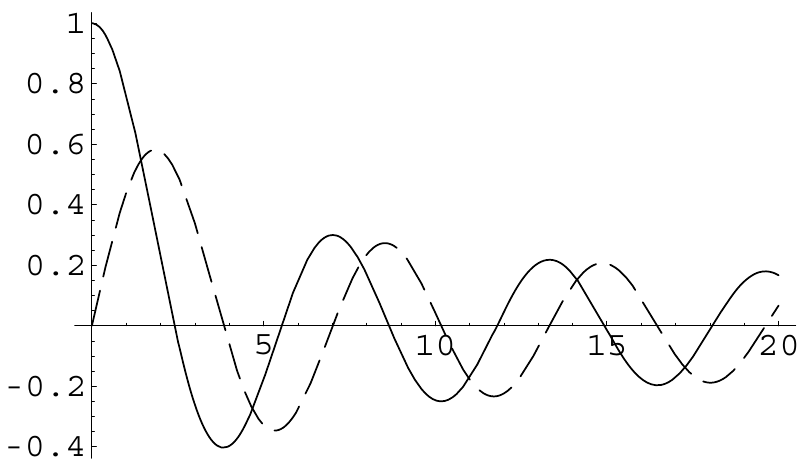}
\enc
\begin{picture}(0,0)
\setlength{\unitlength}{1mm} \put(60,11){$j_{0 1}$}
\put(63,13){\vector(0,1){8}} \put(70,11){$j_{0 2}$}
\put(72.5,13){\vector(0,1){8}}
\end{picture}
\caption{\lb{bessel} The Bessel functions $J_0(z)$ (solid) and
$J_1(z)$ (dashed).  The first and second zero of the Bessel
function $J_0$, which determine the mass of the ground state and
the first excitation,  are indicated.}
\end{figure}
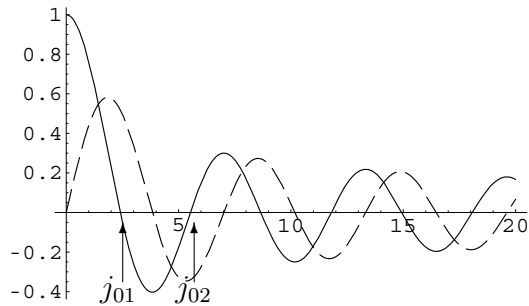

The zeros of the Bessel functions can for not so high $\nu$
 quite well be approximated by
  \beq j_{\nu,s} \approx (s + \frac{\nu}{2} - \frac{1}{4})\pi -
  \frac{4 \nu^2-1}{8 \pi(s + \frac{\nu}{2} - \frac{1}{4})} -
  \dots. \lb{approx}\enq
Exact values are $j_{01}= 2.405,\; j_{02}= 5.520,\;j_{03}= 8.653$.
\begin{table}
\begin{center}
\begin{tabular}{|c|c|c|c|} \hline
Experiment& $\pi \quad  140$MeV &  $\pi \quad  1300 \pm 100 $MeV &
$\pi \quad
1812 \pm 14 $MeV \\
$\nu=0; \quad z_0^{-1}=207$ MeV& 501 & 1140& 1797\\
\hline
\end{tabular}
\end{center}
\vspace{-6mm} \caption{Experimental and hard-wall masses of the
pion resonances based on equation \ref{specB}. \lb {hw1}}
\end{table}

Comparison of the very simple model based on the wave equation
(\ref{specB}) with the data is given in Tab. \ref{hw1}. Much better
agreement with the data can be obtained if one takes into account
additional symmetry breaking fields~\cite{Erlich:2005qh}, see also
\cite{Brodsky:2014yha}, chapter 7 and literature quoted there.

\subsection{The soft wall model \cite{Karch:2006pv}\lb{swm}}
In this model a scale is introduced by multiplying the Lagrangian
\req{LA} with a Dilaton term $e^{\vph(z)}$. This yields the new
Lagrangian ($\ka=3$) \beq \cL = \frac{1}{2} e^{\ka A(z)+\vph(z)}\left(
\eta^{MN} \pa_M \Ph\, \pa_N \Ph - e^{2 A(z)} \mu^2 \Ph^2\right)
\lb{LAs} \enq and the Euler Lagrange  equation becomes, see
\req{whw}
 \beq
 e^{\ka A(z)+\phi(z)}\bigg(\eta^{\al\be}\pa_\al\pa_\be \Phi -\pa_z^2 \Phi - \big(\ka
 \pa_z A(z)+\pa_z \vph(z)\big) \pa_z \Ph +\frac{(\mu R)^2}{z^2} \Phi\bigg) = 0. \lb{whws}\enq

The most popular choice for the dilaton term is: \beq
 \vp(z) = \la z^2
\lb{svm} \enq After Fourier transformation, see \req{Fourier} we
obtain \beq
 \bigg(-q^2 -\pa_z^2  + \big(\frac{\ka}{z} - 2 \la\,z\big)\pa_z
 +\frac{(\mu R)^2}{z^2} \bigg) \tilde \Phi(q,z)= 0.
\lb{whwsf}\enq

Here the programm mathematica gives no useful solution  and we
better transform the equation into a more useful form. With the
rescaling
 \beq \lb{rescale1s} \tilde \Phi(q,z) = z^{(\ka+1)/2 }e^{-\la z^2/2}  \tilde \Phi_N(q,z)\enq
we obtain

\beq \Big( -\pa_z^2 - \frac{1}{z} \pa_z + \frac{L_{AdS}^2 }{z^2} +
\la^2 z^2-(\ka -1) \la   \Big) \tilde  \Phi_N(q,z)= q^2 \,
\tilde  \Phi_N(q,z) \lb{sv1} \enq with
   \beq
L_{AdS}^2=(\ka+1)^2/4
+(\mu R)^2.\enq
 As is shown below this differential equation is
closely related to the harmonic oscillator wave equation and  has
normalized eigenstates with eigenvalues \beq q^2= M_{nL}^2 = (4 n
+ 2L_{AdS} +2)|\la|  - (\ka-1) \la \lb{sv3} \enq and the
eigenfunction for the eigenvalue specified by $L_{AdS}$ and $n$ is
\beq \tilde  \Phi_{N,nL_{AdS}}(z)= z^{L_{AdS}} \, e^{-
|\la|z^2/2}\,  L_n^{L_{AdS}}(|\la| z^2) \lb{sv2} \enq where
$L^L_n(x)$ are the associated Laguerre polynomials. In
mathematica: {\tt LaguerreL[n,x]}; examples are : $L^L_0 (x) =
1,\;  L^L_1 (x) = 1 + L x$.

For the solution of \req{whwsf} we obtain, according to
\req{rescale1s} \beq
 \tilde\Phi_{nL_{AdS}}(z)= z^{(\ka+1)/2} \,e^{- \la z^2}\,\tilde  \Phi_{N,nL_{AdS}}(z)=
z^{L_{AdS}+(\ka+1)/2} \, e^{- (|\la|+\la)z^2/2}\,
L_n^{L_{AdS}}(|\la| z^2) \lb{sv7} \enq

{\small
\begin{quote}
{\bf Derivation of \req{sv2}-\req{sv7}:} The differential operator
\beq \Big( -\pa_z^2 - \frac{1}{z} \pa_z + \frac{L^2 }{z^2} + \la^2
z^2  \Big)= 2 H_{ho} \enq is twice the radial differential
operator for a two dimensional harmonic oscillator  with angular
momentum $L$.

The  the eigenvalues are \beq H_{ho}\Phi_{nL} (q,z)=  E_{nL}
\Phi_{nL} \enq with \beq E_{nL}=(2 n + L +1)|\la| \enq where $L$
is the angular momentum and $n$ is the radial excitation number.
Hence we obtain for the spectrum of the eigenvalues of \req{sv1}
as $q^2 = 2 E_{nL}-(\ka-1) \la$ and the eigenfunctions are
directly those of the 2 dimensional harmonic oscillator.
\end{quote}}

For the scalar field we have $\ka=3$ and with $L=0$ we thus obtain
: \beq M^2_n= (4n +2)| \la| -2 \la \enq With $\la >0$ the lowest
(pseudo)scalar particle has mass 0, which is indeed  the  expected
value in the limit of massless quarks (chiral limit).

By rescaling \beq \tilde \Phi(q,z) = z^{\ka/2} e^{-\vp(z)/2}
\tilde \phi(q,z) \enq we obtain  the Schr\"odinger-like  form
\beq \Big( -\pa_z^2 +  \frac{4L_{AdS}^2-1 }{4z^2} + \la^2 z^2  -(\ka-1)
\la \Big) \tilde   \phi(q,z)= q^2  \tilde  \phi(q,z) \lb{sv4}
\enq where $L_{AdS}^2= (\ka+1)^2/4+(\mu R)^2$.
%%%%%%%%%%%%%%%%%%%%%%%%%
\section{Vector Field \lb{vector}}
{ Here we proceed similarly as in electrodynamics in 4
space-time dimensions. We start from a vector field $A_M$,
corresponding to the electromagnetic potential and construct the
tensor field \beq F_{MN}= \pa_M A_N - \pa_N A_M \enq which
corresponds to the electromagnetic field tensor.

We can use the normal derivatives, since the additional
contributions due to the non-Euclidean metric vanish due to the
antisymmetric construction. We start from the Lagrangian
 \beqa
 \cL &=& \sqrt{|g|} \big(\frac{1}{4} g^{MM'}\,g^{NN'} F_{MN}\,F_{M'N'}- \half \mu^2 \, g^{MM'}A_M\,A_{M'}  \big)\\
&=& \frac{R}{z}\left(  \frac{1}{4}\, \et^{MM'}\,\et^{NN'}
F_{MN}\,F_{M'N'} - \left(\frac{R}{z}\right)^2 \frac{\mu^2}{2}
\,\eta^{MM'}A_M A_{M'} \right) \lb{lgvm}\enqa In contrast to
electrodynamics we have added a mass term which breaks gauge
invariance in the AdS$_5$. (This is a so called Proca-Lagrangian).
For the soft wall model this Lagrangian is multiplied by a factor
$e^{\vp(z)}$ and thus our starting point is the Lagrangian:
  \beq
  \cL= e^{A(z)+\vp(z)}\left(\frac{1}{4}\, \et^{MM'}\,\et^{NN'} F_{MN}\,F_{M'N'} -  \frac{\mu^2R^2}{2z^2} \,\eta^{MM'}A_M A_{M'} \right)
\enq where, as in (\ref{expm})
 \beqa A(z)&=& - \log z + \log R \qquad
\mbox{and} \nn\\
\vph(x) &=& \left\{ \begin{array}{l l} 0 &\quad \mbox{ for the hard wall model}\\
\la^2 z^2 &\quad \mbox{ for the soft wall model}\end{array}\right.
\enqa In  a gauge theory the 5-mass $\mu =0$.

The 5 equations of motion are:
 \beq
  \pa_K\frac{\pa}{\pa(\pa_K A_L)}\cL = \frac{\de \cL}{\de A_L} , \quad L=1,\cdots 5\enq
From the expression
 \beq
 \frac{\pa}{\pa(\pa_K A_L)}\cL = e^{A(z) + \vp(z)}\et^{MK} \et^{NL} (\pa_M
 A_N - \pa_N A_M) \enq
  we obtain the Euler-Lagrange equations:
 \beqa
 \pa_K\frac{\pa}{\pa(\pa_K A_L)}\cL &-& \frac{\de \cL}{\de A_L}\\
&=& e^{A(z) +\vp(z)}\bigg( \de^5_K (\pa_z A+ \pa_z\vp) \et^{MK}
\et^{NL} (\pa_M A_N - \pa_N
 A_M)\nn \\
 && \qquad+\et^{MK} \et^{NL} (\pa_K\pa_M A_N -\pa_K \pa_N A_M) +\frac{(\mu R)^2}{z^2}\,\eta^{LL'}A_{L'}\bigg) \nn \\
 &=&e^{A(z) +\vp(z)}\bigg( (\pa_z A+\pa_z\vp) \et^{M5} \et^{NL} (\pa_M A_N - \pa_N
 A_M)\nn \\
 && \qquad+\et^{MK} \et^{NL} (\pa_K\pa_M A_N -\pa_K \pa_N A_M) +\frac{(\mu R)^2}{z^2}\,\eta^{LL'}A_{L'}\bigg) \lb{el1}\enqa

 Since we have in Minkowski space three vector particles (spin components)
 and 5 components of the potential $A_K$, we
can eliminate two components. We choose \beq A_5 = 0 \qquad {\rm
and} \quad \et^{MK} \pa_ K A_M = 0 \quad( {\rm  Lorenz~gauge})\enq
This simplifies (\ref{el1}) to:
    \beq
 \pa_K\frac{\pa}{\pa(\pa_K A_L)}\cL = e^{A(z)
-\vp(z)}\bigg((\pa_z A+ \pa_z\vp) \et^{M5}\et^{NL}\pa_5 A_N +
\eta^{MK}\et^{NL} \pa_K\pa_M A_N\bigg) \enq
     from which we obtain
in our notation, where Greek indices run from 1 to 4 and
$\pa_5\equiv \pa_z$:
 \beq
\eta^{\nu \la} \bigg(\eta^{\mu \ka} \pa_\mu \pa_\ka A_\nu -\pa_z^2
A_\nu -(\pa_z A+ \pa_z\vp)  \pa_z A_\nu +\frac{(\mu
R)^2}{z^2}\,A_\nu\bigg) =0\enq

We make the ansatz for the Fourier transform
 \beq A_\la(q,z)  =  \ep_\la(q) \tilde \Phi(q,z)\lb{ansatzv}\enq
where $\ep(q)$ is the polarization vector of a transverse vector
field, i.e. $\ep \cdot q=0$,
 and we obtain for $\tilde \Ph(z)$ the wave equation:
 \beq \big(-\pa_z^2 -(\ka \pa_z A+\pa_z\vph)\pa_z +\frac{(\mu R)^2}{z^2}\big )
  \tilde \Phi(q,z)=q^2 \tilde \Phi(q,z) \lb{vmw1}
\enq For the hard wall model we have $\vp=0$, that is \req{vmw1}
is like \req{specB} with $\ka=1$. For the soft wall model we have
$\vp(z) = \la\,z^2$, that is  \req{vmw1}
 is like  \req{whwsf} with $\ka=1$.

We can use equations \req{bessel} and  \req{sv1} $-$ \req{sv4}
also for the vector field, just using for $\ka$ the value $\ka=1$,
notably we have for the vector particle
 \beq L^2_{AdS} = (\mu R)^2+1.\enq}

%%%%%%%%%%%%%%%%%%%%%%%%%%%%%%%%%%%%%%%
%  corrected 3-12-17:
%asso1: (L+n)! (Fakultaet)   resc1 and HSX: \Phi_S replaced by \phi arround asso2:

 %The  solution for the equation \req{HEL} of the unmodified field $\Ph$ is,  see \req{resc1},\beq
 %\lb{asso2}\Ph_{nL_{AdS}}(z) =  \frac{1}{N} z^{2 +L_{AdS}-J} L_n^{(L_{AdS})}(|\la| z^2)e^{-(|\la|+\la) z^2/2}  %\enq
%t is normalized as:\beq  \lb{normHEL} \int_0\infty dz\, e^{\la z^2} z^{2 J-3} \Ph_{nL_{AdS}}(z)^2 =1\enq

\chapter{Light front holographic QCD}
Before we proceed further in the holographic appoach, we shortly
present the kinematical scheme, which is most adequate for a
relativistic semiclassical treatment of a quantum field theory,
the Light Front Quantization.
%%%%%%%%%%%%%%%%%%%%%%%%%%%%%%
\section{Wave functions in light front (LF) quantization \lb{wflfq}}
%%%%%%%%%%%%%%%%%%%%%%%%%%%%%%%%%%%%5
There are several schemes on which one can formulate the
quantization rules. The most usual is the instantaneous one, which
is based on correlators at the same time $x^4$. The light front
(LF) quantization~\cite{Dirac:1949cp} is based on quantization
rules at equal light front time $x^+=x^4+x^3 $. For a review of
applications in QCD see e.g. \cite{Brodsky:1997de}. In the limit
of the 3-component of the hadron going to infinity  the usual
frame based on equal time quantization approaches the light front
quantization frame.

{In light front quantization we have the variables
$x_+=x^4+x^3, \, x_-=x^4-x^3$ and $\vec x_\perp =(x^1,x^2)$. A
wave function in transverse position space with two constituents
depends on the the
following three variables, see also Fig. \ref{lf-scheme}:\\
 $\bullet$   The  longitudinal momentum fractions of the
 constituents~\footnote{The notation $x_i$ for the longitudinal momentum fraction is
 commonly used, it is not to be confused with the space-time  coordinates.},{ $x_i$}, with $\sum_i \,x_i=1.$ If the longitudinal momentum of the hadron is $P$, the longitudinal momentum of the constituent $i$ is $x_i\,P$.\\
$\bullet$ The two dimensional vector of transverse separation of
the two constituents, $\vec b_{\perp }= x^{(1)}_\perp
-x^{(2)}_\perp$ or, in polar coordinates,  on $b_{\perp }= |\vec
b_{\perp}|$ and the polar angle $\theta$. The LF angular momentum $L$ is given by $L=i \frac{\pa}{\pa \theta}$ }

%%%%%%%%%%%%%%%%%%%%%%%%%%%%%
\begin{figure}
\setlength{\unitlength}{1mm}
\begin{center}
\begin{picture}(160,40)(-15,0)
\put(0,28){Two partons} \put(2,10){\oval(8,25)} \put(1.5,0)
{\circle*{2}} \put(1.5,20) {\circle*{2}}
\put(2,0){\vector(2,0){30}} \put(2,20){\vector(2,0){20}}
\put(10,1){${ x_2} P$} \put(10,21){${ x_1 }P$} \put(0,0){
\line(0,1){20}} \put(1,10){ $b_\perp$}
%%%%%%%%%%%%%%%%%%%%%%
\put(48,28){Three partons} \put(51.5,10){\oval(13,25)}
\put(50.5,0) {\circle*{2}} \put(50.5,20) {\circle*{2}}
\put(50.5,8) {\circle*{2}}

\put(50,0){\vector(1,0){25}} \put(50,20){\vector(1,0){20}}
\put(51,8){\vector(1,0){12}}

\put(60,1){${ x_2} P$} \put(59,9.5){${ x_3} P$} \put(60,21){${ x_1
}P$}

\put(49,0){ \line(0,1){20}} \put(651,8){ \line(0,1){12}}
\put(50,3){ $b_{\perp 2}$} \put(50,12){ $b_{\perp 3}$}

%%%%%%%%%%%%%%%%%%%%%%%%
\put(97,28){Two parton clusters} \put(102,10){\oval(13,25)}
 \put(101,2) {\circle*{7}}
 \put(101,18) {\circle*{7}}
\put(100,2){\vector(1,0){30}} \put(100,18){\vector(1,0){20}}
\put(110,3){${ x_b} P$} \put(110,20){${ x_a }P$} \put(100,0){
\line(0,1){20}} \put(101,10){ $b_\perp^{\it eff}$}

%%%%%%%%%%%%%%%%%
\end{picture}
\end{center}
\caption{ \lb{lf-scheme} The LF variables for states
with two or more constituents.}
\end{figure}
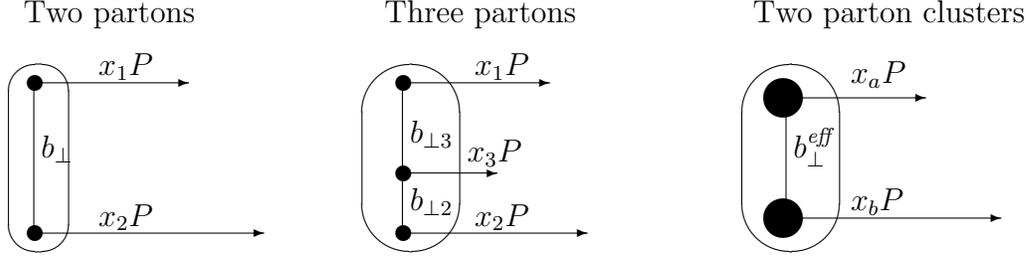

~

The mass for a Hadron with two constituents is in the LF form in
momentum space given \beq M^2= \int_0^1 dx \int d^2 k_\perp
\,\tilde \ph^{LF *}(x, \vec k_\perp)  \left(   \frac{1}{x(1-x)}
{\vec k_\perp}^2 + \frac{m_1^2}{x}+ \frac{m_2^2}{1-x} \right)
\tilde \ph^{LF}(x, \vec k_\perp)+\mbox{interaction} \lb{lfmk} \enq
where $\ph^{LF }(x, \vec k_\perp) $ is the LF wave function of two
constituents with relative momentum $\vec k$ and longitudinal
momentum fractions $x_1=x, \; x_2=(1-x)$. By Fourier
transformation we obtain: \beq M^2= \int_0^1 dx \int d^2 b_\perp
\,\ph^{LF *}(x, \vec b_\perp)  \left( -  \frac{1}{x(1-x)} {\vec
\pa_{b_\perp}}^2 + \frac{m_1^2}{x}+ \frac{m_2^2}{1-x} \right)
\ph^{LF}(x, \vec b_\perp)+\mbox{interaction} \enq For vanishing
constituent  masses, $m_i = 0$, the $x$ and $ \vec b_\perp$
dependence can be expressed by the LF variable \beq \lb{zedef}
\vec \ze = \sqrt{x(1-x)\,} \, \vec b_{\perp i}, \; \ze= |\vec
\ze\,| \enq and we construct the LF Hamiltonian
  \beq \lb{LFH}
   H=  \left( -
\vec \pa_\ze^2 + U(\ze) \right)=
 \left( - \pa_\ze^2 - \frac{1}{\ze} \pa_\ze - \frac{1}{\ze^2} \pa_\varphi^2+ U(\ze) \right),
\enq with $H  \ph^{LF }= M^2\,  \ph^{LF }$.

The complicated interaction is here approximated by the LF
potential $U(\ze)$. By separating the variables  $\ph^{LF }(\vec
\ze) = e^{i L \varphi}  \ph^{LF }(\ze) $ and by rescaling
  \beq \lb{rsx} \phi^{LF}(\ze) =\mathcal{\chi}(x) (2\pi\ze)^{-1/2} \phi(\ze) \enq
   we obtain for $ \phi(\ze) $ the Schr\"odinger-like equation:
\beq \label{lfhs4} H  \phi(\ze) =\left( -\pa_\ze^2  +
\frac{4L^2-1}{4\ze^2}+ U(\ze)\right)  \phi(\ze)  = M^2  \phi(\ze)
\enq

$L$ is the LF angular momentum. The LF potential {
$U(\zeta)$} is   in principle determined by  QCD it (hopefully)
contains also some part of the influence of higher Fock states,
that is states with more than two constituents.

In the following we shall mainly work with this form \req{lfhs} ,
but we should keep in mind that the LF wave function is obtained
from the solution $\phi(\ze)$ of the Schr\"odinger like equation
\req{lfhs4}  by dividing through $\sqrt{\ze}$.

{ We obtain an expression for the up to now arbitrary
function $\mathcal{\chi}(x) $  by the normalization conditions of
the two wave functions. }If we normalize $\phi(\ze)$,
{the  solution of \req{lfhs4}
 by
\beq \int dz |\ph(z)|^2 =1 \enq and the LF wave function
$\phi^{LF}$
 by
\beq \int_0^\infty  dx \, \int d^2 b_\perp |\phi^{LF}[x,b_\perp]|^2 =1 \enq
Then we get the relation \beq \label{slf} \phi^{LF}(x, b_\perp)
=\sqrt{\frac{x(1-x)}{2 \pi \ze}}\, \ph(\ze). \enq }
\paragraph{More than two constituents}
Since in the holographic correspondence  there is only one
variable   to describe the internal structure of hadrons, namely
the coordinate of the 5th dimension, hadrons with more than two
constituents have to be treated as consisting of clusters. In that case one
introduces the effective longitudinal momentum fraction of a
cluster $a$: \beq{x_a}^{\it eff} = \sum_{i=1}^{N_a} x_i, \enq
where $N_a$ is the number of constituents in the cluster $a$, and
correspondingly one introduces effective transverse and
longitudinal coordinates: \beq \vec{ b}_{\perp , a}^{\it eff}=
(\sum_{i=1}^{N_a} x_i\,\vec{ b}_{\perp ,i})/{ {x_a}^{\it eff}},
\qquad
 \zeta= \sqrt{{x_a}^{\it eff}\, {x_b}^{\it eff}} \,|\vec{ b}_{\perp ,b}^{\it eff}-\vec{ b}_{\perp ,a}^{\it eff}|.
\enq There is no theoretical limit for $N_a$. For the nucleon with
three constituents we have: $\vec{b}^{\it eff}_{\perp,d} =
(x_2\,\vec{ b}_{\perp ,2}+x_3\,\vec{ b}_{\perp
,3})/(x_2+x_3)$,\;\; $\ze_{\it eff}=\sqrt{(x_2+x_3) x_1} \;
\vec{b}^{\it eff}_{\perp,d}= \sqrt{\frac{x_1}{x_2+x_3}}
(x_2\,\vec{ b}_{\perp ,2}+x_3\,\vec{ b}_{\perp ,3})$.

The introduction of a  cluster is purely kinematical and necessary
in order to apply the the holographic approach to  hadrons with
more than two constituents,  since there is only one variable --
the holographic variable of the fifth dimension -- which describes
the internal structure.  This identification does not imply that
the cluster is a tightly bound system; it only requires that
essential dynamical features can be described in terms of  the
holographic variable.  This assumption is supported by the
observed similarity between the baryon and meson spectra.

The  cluster occurring in this approach cannot be considered as a
dynamical diquark and our approach is essentially different from a
dynamical diquark picture. In the chiral limit the cluster does
not acquire a finite mass, since the nucleon and delta masses are
described well by the model without any additional mass terms in
the supersymmetric LF Hamiltonian.

In Light Front Holographic QCD (LFHQCD)  one identifies the
holographic variable $z$ with the LF variable $\ze$ introduced
above. The equal form of the LF Hamiltonian  and   the bound state
operators \req{lfmk} and \req{lfhs4} with the LF Hamiltonian of a
two particle (two cluster)  state with LF angular momentum $L$
makes it suggestive, to identify the holographic variable $x_5=z$
with the LF variable $\zeta$. The purely formal quantity
$L_{AdS}=\sqrt{(\ka+1)^2/4 +(\mu R)^2}$ of AdS/CFT becomes then a
physical quantity related to  the AdS-mass $\mu$, we shall discuss this in detail in sect. \ref{lfhqcd}.

\section{Bound state  equations for mesons with arbitrary spin \lb{bsar}}
For spin higher than 1 the situation becomes very involved, since
now we have to use covariant derivatives. A field with spin $J>1$
is a symmetric tensor of rank $J$, $\Phi_{N_1 \dots N_J}$. An
invariant  action, modified by a dilaton term $e^{\vp(z)}$  is
\beqa \label{action2}
S_{\it eff} &=& \int d^{d} x \,dz \,\sqrt{g}  \; e^{\vp(z)} \,g^{N_1 N_1'} \cdots  g^{N_J N_J'}   \Big(  g^{M M'} D_M \Phi^*_{N_1 \dots N_J}\, D_{M'} \Phi_{N_1 ' \dots N_J'}  \nn \\
&& \qquad - \mu_{\it eff}^2(z)  \, \Phi^*_{N_1 \dots N_J} \,
\Phi_{N_1 ' \dots N_J'} \Big). \enqa Here one has to take into
account that in non-Euclidean geometry the shift of a tensor is
not only a shift in the variables but generally also mixes the
components of the tensor see sect. \ref{adsmetric}, footnote
\ref{fn1}. Therefore one has to use covariant derivatives and due
to these the Lagrangian is very complicated and we refer to
\cite{deTeramond:2013it} for details. Here we only quote the
resulting bound state  equations for mesons with angular momentum
$L$ and total spin $J$.

 We Fourier transform the field and extract a $z$-independent polarization vector $\ep_{\nu_1 \dots \nu_J}$:
\beq \tilde \Phi_{\nu_1 \dots \nu_J}(q,z) = \ep_{\nu_1 \dots
\nu_J}\tilde \Phi_{L,J}(q,z) \enq This leads to the  equation of
motion: \beq \lb{HEL} \Big(-q^2 -\pa_z^2+\left( \frac{3-2 J}{z}-
\pa_z \vp\right)\,\pa_z  + \frac{(\mu R)^2}{ z^2}\Big) \tilde
\Phi_{L,J}(q,z) = 0 \enq Comparing this equation with \req{whwsf}
we see that we can use the solutions obtained in sect. \ref{swm}
by inserting $\ka = 3 - 2 J$. This leads to: \beq  \lb{muL1}
L^2_{AdS}  = (J - 2)^2 + (\mu R)^2
 \enq

We bring \req{HEL}  into a Schr\"odinger like form by rescaling
 \beq \lb{resc1}\phi(q^2,z) = z^{(2 J-3)/2 }\, e^{\vp(z)/2} \tilde \Phi_{L,J}(q,z)\enq  and obtain:
\beq \lb{HSX} \Big(-q^2 -\pa_z^2+\frac{4 \, L^2_{AdS}-1}{4 z^2}
+U_{AdS}(z) \Big) \phi(q,z) = 0 \enq with
   \beq \lb{uadsvoll}
U_{ADS}(z) = \frac{1}{4}(\pa_z \vp)^2 + \frac{1}{2} \pa^2_z \vp+
\frac{2 J-3}{2 z} \pa_z \vp \enq For the choice $\vp(z) = \la
\,z^2$ this simplifies to \beq U_{ADS}(z) = \la^2 z^2 +2(J-1)
\,\la \enq

{ Equation \req{HSX} has  normalizable eigenfunctions
$\ph_{nL_{AdS}}(z) $ for the discrete values
   \beq \lb{assp} q^2=
M_{nL_{AdS}J}^2= (4 n + 2 L_{AdS} + 2)| \la| + 2 (J-1) \, \la \enq
   with eigenfunctions
   \beq \lb{asso1} \ph_{nL_{AdS}}(z) = 1/N
\,z^{L_{AdS}+1/2} L_n^{L_{AdS}}(|\la| z^2) e^{-|\la| z^2/2}\;
\mbox{ with }\, N=\sqrt{\frac{(n+L) !}{2 n!}}\;|\la|^{-(L+1)/2} \enq
    They are normalized to  $\int_0^\infty dz \, (\ph_{nL_{AdS}}(z) )^2=1$

 The  solution for the equation \req{HEL} of the unmodified field $\Ph$ is,  see \req{resc1},
 \beq
 \lb{asso2}
\Ph_{nL_{AdS}}(z) =  \frac{1}{N} z^{2 +L_{AdS}-J} L_n^{(L_{AdS})}(|\la| z^2)
e^{-(|\la|+\la) z^2/2}  \enq
it is normalized as:
\beq  \lb{normHEL}
\int_0^\infty dz\, e^{\la z^2} z^{2 J-3} \Ph_{nL_{AdS}}(z)^2 =1
\enq}

\section{Light Front Holographic QCD (LFHQCD) \lb{lfhqcd}}

We compare the soft wall {result \req{HSX}} with the
general LF Hamiltonian: \beq \label{lfhs} H  \phi(z) =\left(
-\pa_\ze^2  + \frac{4L^2-1}{4\ze^2}+ U(\ze)\right)  \phi(z)  = M^2
\phi(z) \enq where the LF potential $U(\ze)$ is not determined,
and we see the structural identity, if we identify the holographic
variable $x_5=z$ with the LF variable $\ze$ and the  AdS quantity
$L_{AdS}  = \sqrt{(J-2)^2 + (\mu R)^2}$ with the the LF angular
momentum $L$.  The light front potential is then the potential
$U_{ADS}$ derived from the (modified) AdS action. Altogether we
obtain  the following dictionary between the AdS result and the LF
Hamiltonian: \beqa
x_5=z &\Leftrightarrow&  \ze\\
L_{AdS} = \sqrt{(J-2)^2 + (\mu R)^2} &\Leftrightarrow& L\\
U_{ADS}(z) &\Leftrightarrow& U(\ze) \enqa

The quantity $L_{AdS} $ in the formula for the spectrum \req{assp}
and the wave functions \req{asso1} are identified with the LF
angular momentum. The potential is related to the dilaton
modification of the AdS$_5$ action. The final bound state equation
for a meson with LF orbital angular momentum $L$ and total angular
momentum $J$ is for a dilaton $\vp(z) = \la _M z^2$: \beq \left(
-\pa_\ze^2  + \frac{4L^2-1}{4\ze^2}+ \la_M^2 \ze^2
+2(J-1)\,\la_M\right)  \phi(\ze)  = M^2  \phi(\ze), \lb{bsff} \enq
with the spectrum of LFHQCD
\beq q^2= M^2_{nLJ}= (4n + 2L + 2)
\,|\la_M|+2(J-1) \la_M \lb{spff}
 \enq
  { and the
solution \req{asso1}}.
%%%%%%%%%%%%%%%%%%%%%%%%%%%%%%%%%%%%%%%%%%%%%
In the limit of massless quarks there is only one parameter
$\la_M$  in the theory, which sets the scale, which can be fixed
e.g. by the $\rh$ mass $2 \la_M =m_\rh^2$. In this respect LFHQCD
is like lattice gauge theory. It should be noted that in LFHQCD
the parameter $\la$ is always positive, whereas in the original
paper of \cite{Karch:2006pv} its value has to be negative, see
\cite{Brodsky:2014yha}, sect. 5.1.2. for a discussion of that
point.

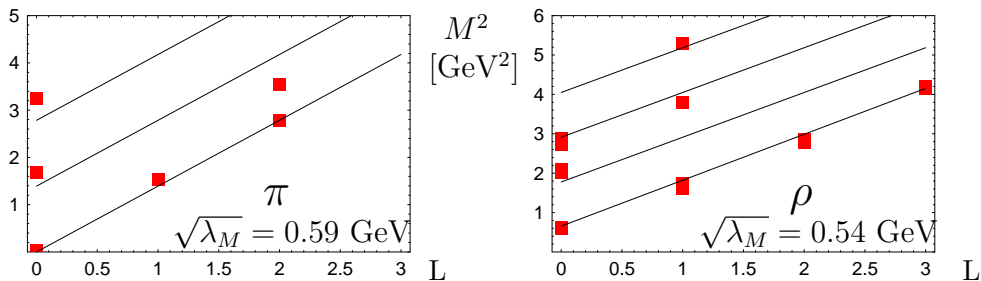
\begin{figure}
\setlength{\unitlength}{1mm}
\begin{center}

\includegraphics[width=6cm]{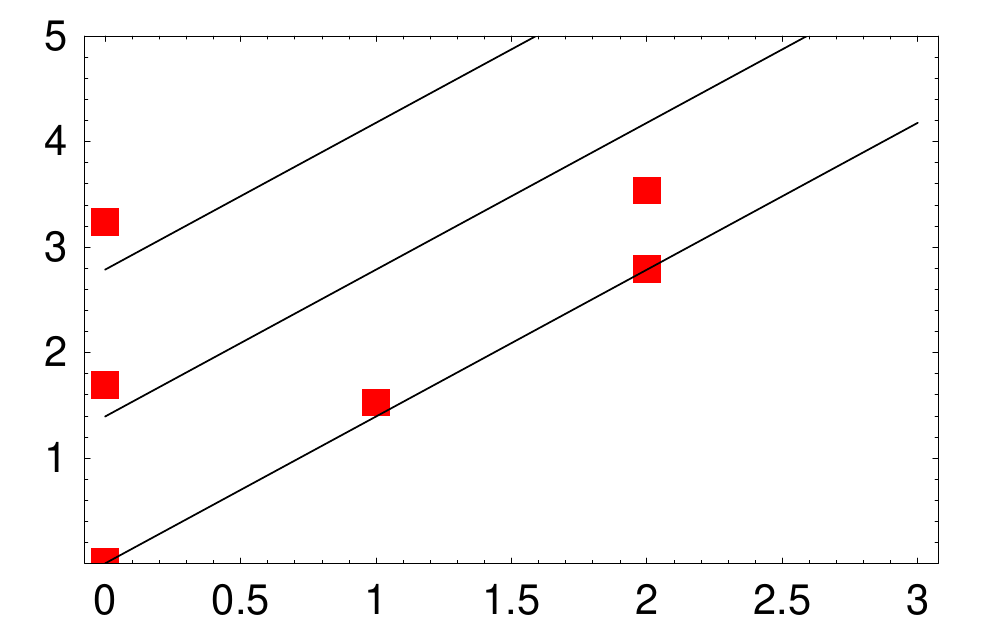} \hspace{7mm}
\includegraphics[width=6cm]{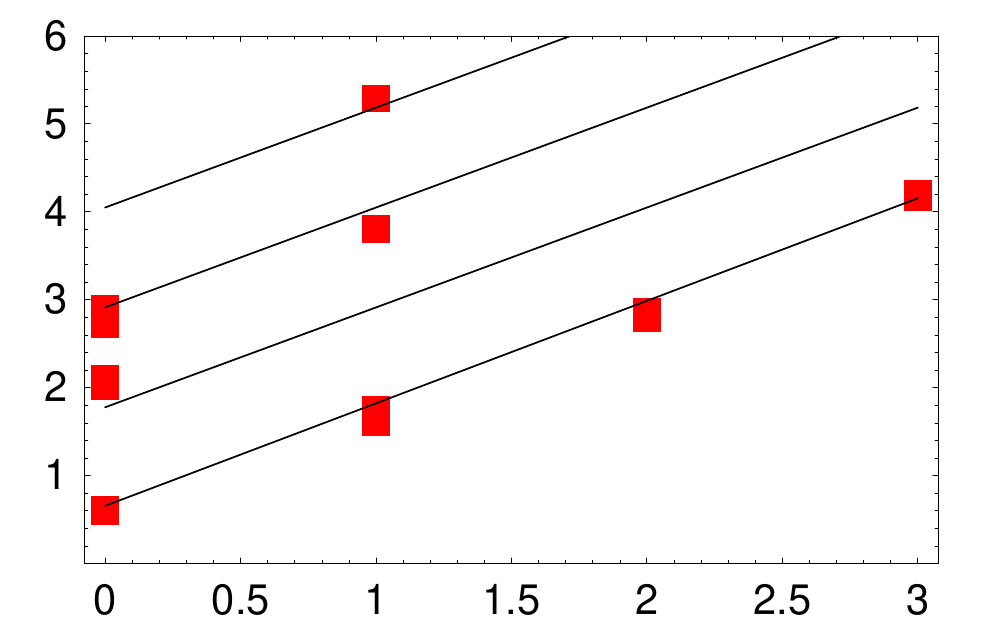}\\

\end{center}
\begin{picture}(30,20)
\put(78,57){$M^2$} \put(76,52){[GeV$^2$]}
%\put(65,50){$M^2$}
%\put(62,45){[GeV$^2$]}
%\put(62,62){\small L}
%\put(132,62){\small L}
\put(76,25){\small L} \put(147,25){\small L} \put(54,35){\Large
$\pi$} \put(42,30){$\sqrt{\la_M} = 0.59$ GeV} \put(124,35){\Large
$\rho$} \put(112,30){$\sqrt{\la_M} = 0.54$ GeV}
%\put(40,29){\large \tcb $N$}
%\put(28,25){$\sqrt{\la_B} = 0.50$ GeV}
%\put(110,29){\large \tcb $\De$}
%\put(98,25){$\sqrt{\la_B} = 0.52$ GeV}
\end{picture}
\vspace{-1cm} \caption{\lb{pirho} The theoretical spectra from
\req{spff} together with the data for the $\pi$ and $\rh$
trajectories and their daughters, from~\cite{Brodsky:2014yha}. }
\end{figure}

If $\la_M$ is fitted independently for the $\pi$  and $\rh$
trajectories, as done for the Fig. \ref{pirho}, the values  agree
within the expected variation of $\approx \pm 10\%$. From the
$\pi$ one obtains $\sqrt{\la_M} = 0.59$ GeV, from the $\rh$ one
obtains $\sqrt{\la_M} =0.54$ GeV. The theoretical predictions
given by \req{spff} (black lines) together with the observed
resonances are displayed in Fig. \ref{pirho}.

%%%%%%%%%%%%%%%%%%%%%%%%%%%%%%%%%%%%%%%%%%%%%
\section{Bound state  equations for baryons with arbitrary spin \lb{bsebas}}
%%%%%%%%%%%%%%%%%%%%%%%%%%%%%%%%%%%

We present  only the starting point and the final result
and refer to \cite{deTeramond:2013it} for a detailed presentation.
Particles with half integer spin are generally described
\cite{Rarita:1941mf} by a spinor with additional tensor indices,
$\Psi_{N_1 \cdots N_T}$. In 4 and 5 dimensions such a relativistic
spinor has 4 components. The starting point is the invariant
Lagrangian for a spinor field. It turns out that a dilaton term,
that is a factor $e^{\la z^2}$ in the  the Lagrangian does not
lead to an interaction~\cite{Kirsch:2006he}, since it can be absorbed by the fermion
field, we therefore do not include it in the action. In order to get bound states one has to add a Yukawa
like term $\bar \Psi_{N_1 \cdots N_T}\,\la_B z^2\,\Psi_{N_1' \cdots N_T'}$ to
the Lagrangian~\cite{Kirsch:2006he}.

Our  starting point  is the action: \beqa\label{af}
S_{F \it  eff} &=&\half  \int  d^d x \,dz\, \sqrt{g}  \; g^{N_1\,N_1'} \cdots g^{N_T\,N_T'} \\
 &&\left[ \bar  \Psi_{N_1 \cdots N_T}  \Big( i \, \Ga^A\, e^M_A\, D_M -\mu-{\la_B\,z^2}
\Big) \Psi_{N_1' \cdots N_T'} + h.c. \right] \nn \enqa Here
$\Psi_{N_1 \cdots N_T} $ is a spinor field in AdS$_5$ with
$T=J-\half$ covariant indices, that is $T=0$ for the nucleon. The
spinor is symmetric in the tensor indices. $\Gamma^A\,e^M_A $ are
the Dirac matrices in AdS$_5$ metric, $e^M_A$ is the so called
4-bein of AdS, $e^M_A = \frac{z}{R} \de^M_A$. The matrices
$\Gamma^A$ are the  Dirac matrices of flat 5-dimensional space,
$\Gamma^A = \ga^A, \; \Ga^5 = i \ga_5$. $ D_M\Psi_{N_1 \cdots N_T}$ is the
covariant derivative of a spinor (it is even more complicated than
the covariant derivative of a tensor, since it contains also the
so called spin connection.). A symmetry breaking term has been
inserted: the Yukawa term $\bar \Psi_{N_1 \cdots N_T} \la_B z^2
 \Psi_{N_1' \cdots N_T'}  $. As mentioned above, a term $e^{\vp(z)} $ like in \req{action2} could also  be inserted, but it has no influence on the equations of motion.

The procedure to obtain the equations of motion is the following\\
1) We evaluate the Euler Lagrange equations: \beq
  \pa_K\frac{\pa}{\pa(\pa_K \Psi_{N_1 \cdots N_T} )}\cL = \frac{\de \cL}{\de \Psi_{N_1 \cdots N_T}}; \qquad
\pa_K\frac{\pa}{\pa(\pa_K \bar \Psi_{N_1 \cdots N_T} )}\cL =
\frac{\de \cL}{\de \bar  \Psi_{N_1 \cdots N_T}}
 \enq
and obtain  equations, which can be brought into the form: \beq
\label{DE2} \left[ i \left( z \eta^{M N} \Gamma_M
\partial_N + \frac{4 -2 T}{2} \Gamma_z \right) - \mu R - R \,\la z^2
%\rho(z)
\right]    \Psi_{\nu_1 \dots \nu_T}=0.
 \enq

{2) We set all spinor tensors, which have at least one
index 5 to zero, that is all tensor-spinor fields have only
Minkowski indices$\Psi_{\nu_1 \cdots \nu_T}$. Then we  go to the
momentum space
 \beq \tilde \Psi_{\nu_1 \cdots \nu_T}(q,z) = \int
d^4x e^{i q x} \Psi_{\nu_1 \cdots \nu_T}(x,z) \enq
 and extract the spin content by spinors which satisfy the Dirac equation:}
 \beq
(\ga q  - M)\,u_{\nu_1 \cdots \nu_T}(q)=0, \mbox{ where } M^2=q^2.
\enq
 Then we define chiral spinors by $u^\pm_ {\nu_1 \cdots \nu_T}(q)$ with
\beq u^\pm_ {\nu_1 \cdots \nu_T}(q)= \half(1\pm \ga_5) u_ {\nu_1
\cdots \nu_T}(q) \enq
The original spinor field can be decomposed
into the chiral components: \beq \lb{rsbar} \tilde \Psi_{\nu_1
\cdots \nu_T}(q,z) =z^{2-T}\left(u^+_ {\nu_1 \cdots \nu_T}(q)
 \psi^+(q,z) +u^-_ {\nu_1 \cdots \nu_T}(q)
\psi^-(q,z)\right) \enq

From \req{DE2} one  obtains coupled first order differential
equations  for $\tilde  \psi^\pm(q,z)$, by reciprocal insertions
they can be made to decouple and we finally obtaib
    \beqa
\left( -\pa_z^2 + \frac{4 L_{AdS}^2-1}{4z^2} +\la_B^2 z^2 +2(L_{AdS}+1) \la_B\right) \psi^+(q,z)&=& M^2 \psi^+(q,z)  \lb{bsb}\\
\left( -\pa_z^2 + \frac{4 (L_{AdS}+1)^2-1}{4z^2} + \la_B^2 z^2
+2L_{AdS}\, \la_B\right) \psi^-(q,z)&=& M^2 \psi^-(q,z) \lb{bsb1}\enqa with
 \beq
L_{AdS} = |\mu R| - \half. \enq
We identify, as for the mesons, this
quantity $L_{AdS}$ with the LF angular momentum $L$ of the hadron,
strictly speaking the LF angular momentum between a quark and the
cluster.

These equations have the same structure as the one for bosons
\req{HEL}. The LF  potential for the two chirality components is:
\beq U^\pm (z)  = \la_B^2  z^2 + 2 (L+\half \pm \half) \la_B \enq
that is it contains a quadratic confining  term $\la_B^2
z^2$,  as the meson potential \req{uadsvoll}, but  different constant terms.

By comparing with the results obtained in sect. \ref{bsar} we
obtain the spectrum:
 \beq M^2_{nL} = \la_B \left(4n +2(L +\half \mp
\half) + 2 + 2 (L + \half \pm \half) \right)= 4 \la_B(n+L+1)
\lb{barsp1} \enq
and the same wave functions as the ones obtained
for the mesons, see \req{asso1}: \beq \lb{solbar} \ps^+_{nL}(q,z)
= \ph_{nL}(z), \quad  \ps^-_{nL}(q,z) = \ph_{nL+1}(z) \enq with
\beq
\ph_{nL}(z) = 1/N \,z^{L+1/2} L_n^{(L)}(|\la| z^2) e^{-|\la| z^2/2}.% \quad N=\sqrt{\frac{n+L}{2 n!}}.
\enq

Note that the two components of the baryon have different angular
momentum. In the LF form   the chirality $+$ component has the
spin aligned in $+x_3$ direction, the $-$ component in $- x_3$
direction. If we speak of a baryon with spin $L$, we always refer
to the $L$ of the positive chirality component.

The positive and negative chirality components of the original
field, which satisfies \req{DE2} are (see \req{solbar} and
\req{rsbar}) \beqa \lb{nwf3}
\tilde \Psi^+(z) &=& z^{2+L+1/2-T} L_n^{(L)}(|\la| z^2) e^{-|\la| z^2/2}\\
\tilde \Psi^-(z) &=& z^{3+L+1/2-T} L_n^{(L+1)}(|\la| z^2) e^{-|\la|
z^2/2} \nn \enqa where $T=J-\half$. Note that for baryons the
variable $\zeta$ corresponds to the  separation between one of the
quarks and a two quark cluster.

The spectrum is independent of $J$, it only depends on $L$. In
some way this is good, since many states with the same orbital
angular momentum $L$ but different total angular momentum $J$ have
the same mass. But the mass of the Delta  with $L=0,\;
J=\frac{3}{2}$ is different from that of  the nucleon, which has
$L=0,\; J=\frac{1}{2}$. Therefore we have to insert in
\req{barsp1} an additional term $+2 \la_B \cS$ , where  $\cS$  is
defined as
 the minimal spin of any 2-quark cluster, which can be formed in the baryon.
  For the $\De$ the spin $\cS = 1$, since only in that way we can obtain the
  total spin  $\frac{3}{2}$ , but in the nucleon
the spin $\half$ can be formed from a two quark system with spin
0. \beq \lb{barspin} M^2_{nL} = 4 (n+L+1) \, \la_B + 2 \cS\,
\la_B \enq This is in contrast to the mesons, where the mass
difference of the $\pi$ and the $\rho$ was a consequence of the
AdS action. As can be seen from Fig. \ref{lhfig}, the quality of
the results for baryons is  comparable to that of mesons. It is
remarkable, that the value of the scale $\la$ is very similar both
for mesons and baryons.

We shall come to a theoretical framework in which this must be the
case in the next chapter, but before we expand the applicability
of the model to a larger data basis by including the effects of
small quark masses.

\begin{figure}
\begin{center}
\includegraphics[width=10cm]{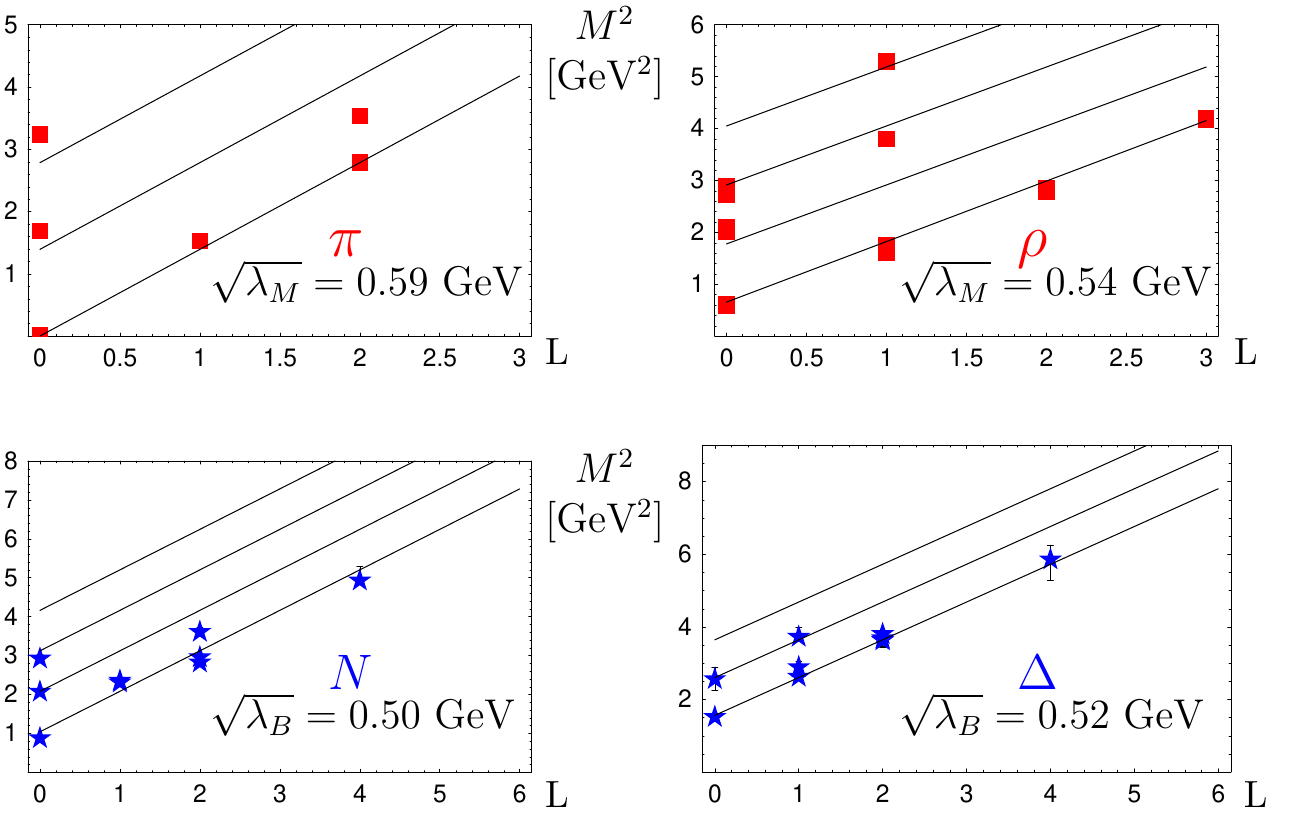}
 \caption{\lb{lhfig}The light hadron trajectories.}
\end{center}
\end{figure}

\section{Inclusion of small quark masses \lb{smqm}}
For small quark masses one expects that the mass effects can be
treated in a perturbative way. Therefore one first tries
perturbation theory. In the basic formula \req{lfmk} for the
construction of the Hamiltonian, \beq \lb{lfmk2} M^2= \int_0^1 dx
\int d^2 k_\perp \,\tilde \ph^{LF *}(x, \vec k_\perp)  \left( -
\frac{1}{x(1-x)} {\vec k_\perp}^2 + \frac{m_1^2}{x}+
\frac{m_2^2}{1-x} \right) \tilde \ph^{LF}(x, \vec
k_\perp)+\mbox{interaction} \enq the mass terms occur as $\sum
\frac{m_i}{x_i}$. Therefore a first guess for mass shift due to
the quark masses is: \beq \De M^2 = \int_0^1 \left(
\frac{m_1^2}{x} +\frac{m_2^2}{1-x}\right) \, \phi^{LF}(\vec
b_\perp, x)^2 \, b_\perp db_\perp \, d\phi \, dx \lb{DM1} \enq
where $\phi^{LF}$ is the normalized LF wave function, see sect.
\ref{wflfq}. Inserting the relation \req{slf} and using
$x(1-x)b_\perp db_\perp = \ze \,d\ze$ one obtains \beq \De M^2 =
\int dx \left( \frac{m_1^2}{x} +\frac{m_2^2}{1-x}\right) \,\int d
\ze \phi(\ze)^2  \lb{DM1} \enq where $\phi(\ze)$ is the normalized
wave function \req{asso1}. This expression for $\De M^2 $
diverges!

Therefore we have to look for more realistic wave functions: The
LF wave function \req{asso2} has the exponential behavior $\sim
e^{-\la x(1-x) b_\perp^2/2} $. Its Fourier transform is \beq \int
d^2 b \; e^{i \vec b_\perp \vec k_\perp} e^{-\la\, x(1-x) \, b_\perp^2/2}
= \frac{2 \pi}{x(1-x) } e^{-k_\perp^2/( 2 \la x(1-x))} \enq The
expression $\frac{k_\perp^2}{x(1-x)}$ in the wave function
describes the off-energy shell behaviour in LF form for massless
quarks. Including quark masses, makes this quantity, see
\req{lfmk} : \beq \frac{k_\perp^2}{x(1-x)}+\left( \frac{m_1^2}{x}
+\frac{m_2^2}{1-x}\right) \enq So it is plausible to include this
factor also in the wave function, that is replace: \beq
 e^{- \frac{1}{2 \la}\, \frac{ k_\perp^2}{ x(1-x)} }\to
 e^{- \frac{1}{2 \la}\left( \frac{k_\perp^2}{x(1-x)} + \frac{m_1^2}{x} +\frac{m_2^2}{1-x}\right)}
\enq This amounts to a multiplication of the wave functions with
the factor \beq
 e^{\frac{-1}{2 \la} ( \frac{m_1^2}{x} +\frac{m_2^2}{1-x})}
\enq The modified normalized  wave function is then: \beq \ph_m =
\frac{1}{N} e^{\frac{-1}{2 \la} ( \frac{m_1^2}{x}
+\frac{m_2^2}{1-x})} \,  \ph \enq
 with $N^2= \int_0^1dx\,
e^{\frac{-1}{ \la} ( \frac{m_1^2}{x} +\frac{m_2^2}{1-x})}$.

The expression for the mass correction becomes: \beq \De M^2 =
\frac{1}{N^2} \Bigg[ \int_0^1 dx\,  \left( \frac{m_1^2}{x}
+\frac{m_2^2}{1-x}\right)\; e^{\frac{-1}{ \la} ( \frac{m_1^2}{x}
+\frac{m_2^2}{1-x})}\Bigg] \enq This expression can be extended to
many quarks~\cite{Brodsky:2016yod}: \beq \De M_n^2(m_1, \cdots
m_n)=  (-2 \la^2)  \frac{\pa}{\pa \la}\log\Bigg[ \int_0^1 dx_1
\cdots dx_n \de(x_1 + \cdots x_n -1) \,  e^{\frac{-1}{\la} (
\frac{m_1^2}{x_1} +\cdots \frac{m_n^2}{x_n})} \Bigg]\enq

As can be seen from Tab. \ref{ground} and Figs. \ref{strmesfig}
and \ref{strbarfig} one can get reasonable fits to all hadron
trajectories  containing strange quarks. From $\pi$ and $K$ one
deduces the effective quark masses: $m_q=0.046 GeV, \; m_s=0.35$
GeV.  For the  ground states of the other hadrons, the resulting
masses agree very well with experiment, see Tab. \ref{ground}.
\begin{table}
\bec
\begin{tabular}{ccc}
&LFHQCD& Experiment\\
&GeV&GeV\\
$\Lambda$& 1.15 &1.116\\
$\Sigma^*$&  1.35  &1.385\\
$\Xi $&1.32 &1.314\\
$\Xi^*$& 1.50 &1.530\\
$\Om$& 1.68 &1.672\\
&&\\
$K^*$& 0.90 &0.892\\
$\Phi$& 1.08 &1.020\\
\end{tabular}
\enc \caption{\lb{ground}Masses of ground states with $m_q=0.046,
\; m_s=0.35$ GeV. The values for $\la$ are the same as for the
non-strange hadrons.}
\end{table}

\begin{figure}
\bec
\includegraphics[width=6cm]{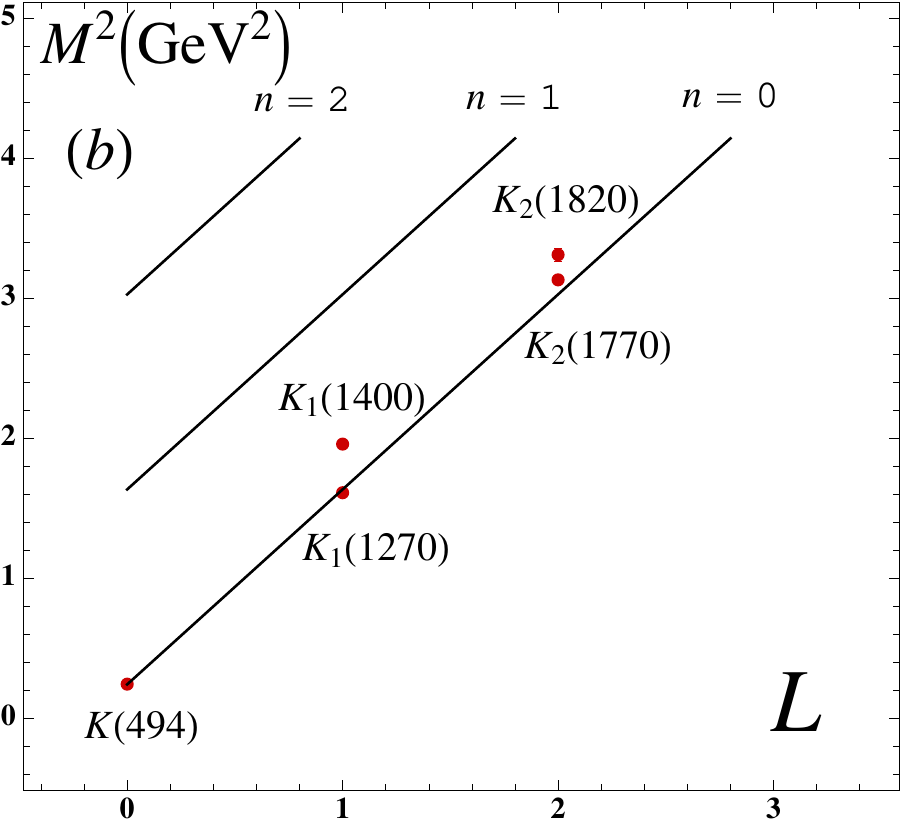}~~
\includegraphics[width=6cm]{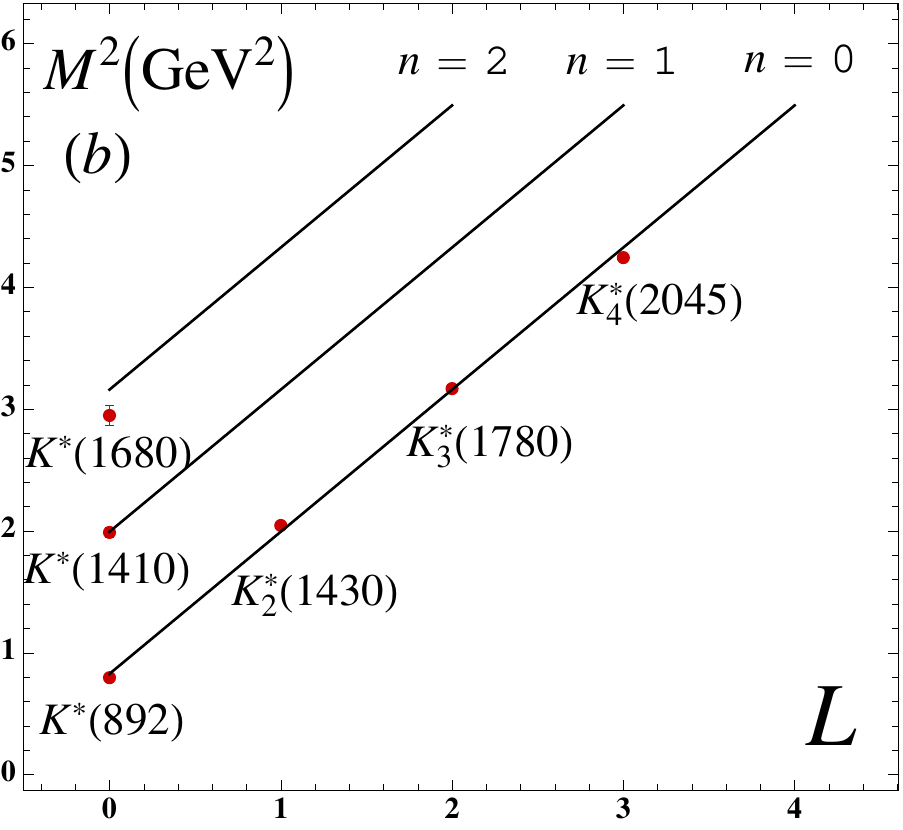}
\enc
\caption{\lb{strmesfig}Spectra of strange mesons with $m_q=0.046
GeV, \; m_s=0.35$ GeV. The values for $\la$ are the same as for
the non-strange hadrons.}
\end{figure}
\begin{figure}
\bec
\includegraphics*[width=8cm]{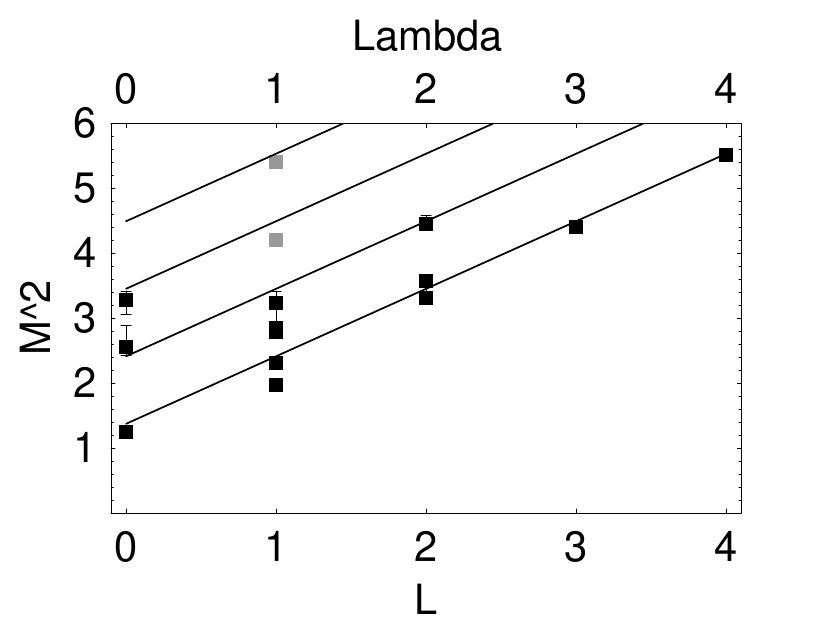}
\includegraphics*[width=8cm]{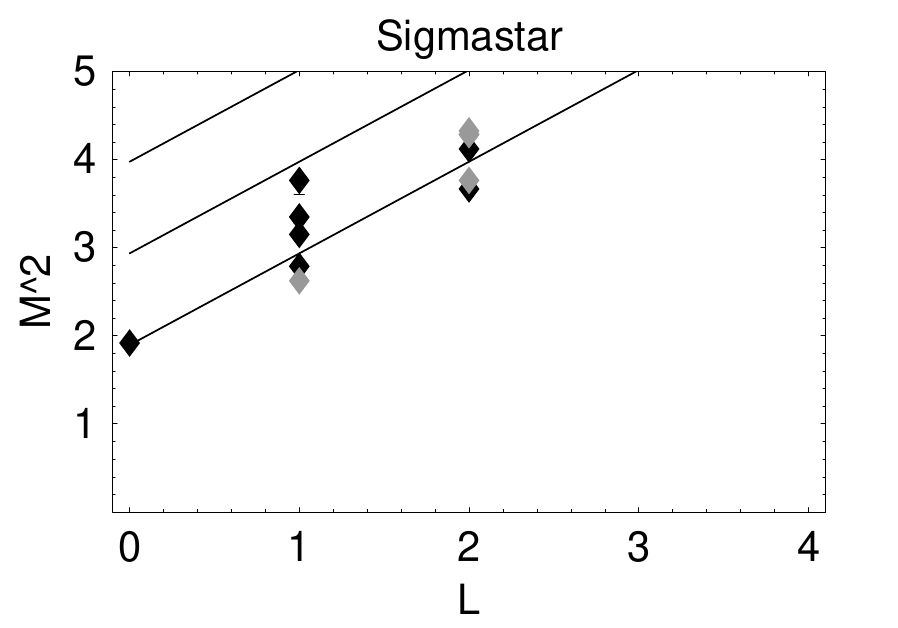}
\enc
\caption{\lb{strbarfig}Spectra of strange baryons with $m_q=0.046
GeV, \; m_s=0.35$ GeV. The values for $\la$ are the same as for
the non-strange hadrons.}

\end{figure}
%%%%%%%%%%%%%%%%%%%%
\clearpage
\section{Summary}
From the holographic principle one can derive wave functions for
hadrons. By modifying the action by suitable terms one can
reproduce the spectrum of all hadrons containing only light quarks
within the expected accuracy. The form of the wave functions
allows an interpretation of the quantities occurring in AdS$_5$.
The holographic variable can be identified with the LF variable
$\ze$, \req{zedef},  and the product of the AdS mass $\mu$  and the
curvature $R$ is related to the LF angular momentum, see
\req{muL1}.
 By a well justified modification of the wave function
due to finite but small quark masses, one can describe all light
hadrons in a satisfactory way, see Figs. \ref{lhfig} --
\ref{strbarfig}  .

There remain, however, two major unsatisfactory points: 1) The
modification of the invariant AdS$_5$ action was not determined by
some theoretical principles but completely arbitrary and only
chosen to satisfy the data. 2) The observed similarity between
meson and baryon spectra seems fortuitous, since the modification
of the meson and baryon action has nothing in common, for mesons
one multiplies the Lagrangian by a function, for baryons one has
to add a Yukawa-like term; the observed equality (within the
expected accuracy) of the values of the fundamental parameters
$\la_M$ and $\la_B$ seems also accidental. A third and minor point
is: that there was also no theoretical justification for adding
the spin term $2 \cS \la$  to the baryon spectrum in
\req{barspin}.

In the next section we shall see that there is indeed a
theoretical remedy for  all these unsatisfactory aspects.

%%%%%%%%%%%%
%\end{document}
%%%%%%%%%%%%%%

%%%%%%%%%%%%%%%%%%%%%%%%%%%%
\chapter{Supersymmetric light front
holographic QCD}

{\small The content of this chapter is based on the publications
\cite{deTeramond:2014asa,Dosch:2015bca,Brodsky:2016yod,Dosch:2016zdv}}.

The classical QCD Lagrangian contains in the limit of massless
quarks no scale and is therefore invariant under the conformal
group in 4 dimensions \footnote{Information on many concepts in
this section, as conformal group, Noether theorem etc. can be
found conveniently in Wikipedia or Wikischolars.}  . In the
AdS/CFT scheme, the action of the quantum gauge theory shows an
even larger symmetry, it is also invariant under supersymmetry.
Light Front Holographic QCD has approximated the quantum field
theory by a semiclassical theory (Quantum mechanics) by reducing
the dynamics to a Light Front Hamiltonian with two constituents
(or clusters of constituents) and a potential. This is feasible,
since in the limit of many colours in the gauge field theory all
its features can be obtained by the classical solutions of an
action in a 5 dimensional space. Quantum mechanics, and hence the
semiclassical approximation, can be viewed as a quantum field
theory in one dimension. It is therefore tempting to apply the
symmetry constraints of the 4 dimensional quantum field theory of
the AdS/CFT correspondence also to the semiclassical theory, which is a one-dimensional quantum
field theory. In short we will discuss
the consequences of an implementation of the superconformal
(graded) algebra on our approach of light front holographic QCD.
Fortunately superconformal quantum mechanics is much simpler than
general superconformal field theory and notably it is a symmetry
of wave functions and therefore it does not lead to the existence
of new (stable ) particles which are the superpartners of the
existing particles. For pedagogical reasons we start with
conformal symmetry, leaving the superconformal symmetry for the
following section.

\section{Constraints from conformal algebra \lb{confal}}
Our aim is to incorporate into a semiclassical effective theory
 conformal symmetry. We will require that the
corresponding one-dimensional effective action which encodes the
conformal symmetry of QCD remains conformally invariant.  De
Alfaro, Fubini and Furlan~\cite{deAlfaro:1976je} investigated in
detail the simplest scale-invariant one-dimensional field
theory, diven by the action
 \beq \label{A}
A[Q]= \half \int dt  \left(\dot Q^2 - \frac{g}{Q^2} \right)  =
\int dt \cL, \enq
 where  $\dot Q \equiv d Q / dt$. Since the action is dimensionless (in natural units),
 the dimension of the field $Q$ must be half the dimension of the ``time'' variable
$t$, dim[$Q$] = $\half$dim$[t]$,
 and  the constant $g$  is dimensionless.
 The translation operator in $t$, the Hamiltonian, is
 \beq \label{HtQ}
H =
 \half \left(\dot Q^2 + \frac{g}{Q ^2}\right),
 \enq
 where the field momentum operator is $P = \frac{\pa \cL}{\pa \dot Q} = \dot Q$,  and the quantum equal time commutation relation is
\beq \label{CCR} [Q(t), \dot Q(t)] =[Q(t), P(t)]= i. \enq The
equation of motion for the field operator $Q(t)$ is  given by the
usual quantum mechanical evolution \beq  \label{QME}
 i \left[H, Q(t) \right]  = \frac{d Q(t)}{d t}.
 \enq
Up to now we have worked in the Heisenberg picture: states are
time independent, but operators depend on time. Since we want to
obtain a Schr\"odinger-like equation we go to the Schr\"odinger
picture with time dependent states $\vert \psi(t)\rangle$ and time
independent operators. The time dependence of the states is
determined by the Hamiltonian:
    \beq  \label{QEf}
H  \vert \psi(t) \rangle = i \frac {d}{d t} \vert \psi(t)\rangle.
    \enq
We realize the states as elements of a Hilbert space of functions with one variable and therefore the fields  $Q$ and  $P = \dot
Q$ are operators  in that space. They are given by the
substitution \beq \lb{QC}
 Q(0) \to x,  \qquad  \dot Q(0) \to - i \frac{d}{d x}.\enq
Then we obtain  the usual quantum mechanical evolution \beq
\label{Sp} i \frac{\partial}{\partial t} \psi(x,t) = H \Big(x,
- i \frac{d}{d x} \Big) \psi(x, t), \enq Using \req{HtQ} and
\req{QC} we obtain the familiar form \beq  \label{Htxf} H  = \half
\left(- \frac{d^2}{dx^2} + \frac{g}{x^2} \right) . \enq It has the
same structure as the LF Hamiltonian \req{lfhs} with a vanishing
light-front potential,  as expected for a conformal theory. The
dimensionless constant $g$ in action \req{A} is now related to the
angular momentum  in the light front wave equation \req{asso2}.

As emphasized  in \cite{deAlfaro:1976je}, the absence of
dimensional constants in \req{A} implies that the action  $A[Q]$
is invariant under   the full conformal group in one dimension,
that is, under translations, dilatations, and special conformal
transformations. These can be easily expressed by the
infinitesimal transformations of the variable $t$ and the filed
$Q$: \beqa
 {\rm Translation}\quad  t &\to&  t'= t+\ep, \qquad Q\to Q'=Q \\
 {\rm Dilatation} \quad t &\to&  t'= t (1+\ep),  \qquad Q\to Q'=Q \,\sqrt{1+\ep} \\
 \mbox{\small spec. conf. transf.}\;  t &\to&  t'= \frac{t}{1-\ep t}, \qquad Q\to Q'=\frac{Q}{1-\ep t}
\enqa

One can convince oneself, that \req{A} is indeed invariant under
these transformations. In checking this, be aware that
$\epsilon$ is assumed to be infinitesimal, that is terms of
$O(\epsilon^2)$  must be neglected.

The constants of motion of the action are obtained by applying the
Noether theorem. These constants of motion are the generators of
the conformal group. They are

Translations : \beq \lb{tr} H(0) = \half \left( \dot Q^2 +
\frac{g}{Q^2} \right), \enq Dilatations: \beq \lb{dil} D(0) =  -
{\textstyle{\frac{1}{4} }}  \left( \dot Q Q + Q \dot Q \right)
\enq Special conformal transformations: \beq \lb{spec}
 K(0)= \half Q^2.
 \enq

Using the commutation relations \req{CCR} one can check that the
operators  $H,  D ~ {\rm and} ~ K$ do indeed fulfill the algebra
of the generators of the  one-dimensional conformal group ${\it
Conf}\!\left(R^1\right)$
 \beq
 [H,D]= i\,H,  \quad [H ,K]=2\, i \, D, \quad [K,D]=- i\,  K .
\enq

The conventional Hamiltonian $H$ is one of the generators of the
conformal group. One can extend the concept of a Hamiltonian by
forming a linear superposition of all three generators \beq
\label{G} G = u \, H + v\,D + w\,K. \enq
           The new Hamiltonian $G$ acts
on the state vector and its evolution involves a new time
variable $\tau$, which is related to $t$ by \beq   \label{dtau}
d\tau= \frac{d t} {u+v\,t + w\,t^2}. \enq

We now insert into  \req{G} the expressions (\ref{tr} --
\ref{spec}) and the Schr\"odinger picture substitutions \req{QC}
and obtain: \beq \label{Htaux} G = \frac{1}{2} u \left(-
\frac{d^2}{d x^2} + \frac{g}{x^2}\right)  + \frac{i}{4} v  \left(x
\, \frac{d}{d x}+ \frac{d}{d x} \, x  \right) +\frac{1}{2}  w x^2.
\enq

Comparing with the LF Hamiltonian: \beq \lb{LFH} H= - \pa_\ze^2+
\frac{4 L^2 - 1}{4
 \ze^2} +{  U(\zeta)};
\enq we see that the constraint to construct the Hamiltonian
inside the conformal algebra restricts the possible forms of the
potential considerably.

Comparing with the AdS-Hamiltonian for mesons \beq \Big(-\pa_z^2+
\frac{4L^2-1}{4 \ze^2} +\la^2 \ze^2 +2(J-1) \la\Big) \enq shows
that with $u=2,\; w= \la^2,\; v=0, \; g=L^2- \frac{1}{4}$ the
''conformal Hamiltonian" $G$, \req{Htaux}, reproduces the $\zeta$
dependent  terms of the AdS Hamiltonian for mesons.

This is very good, but not enough. We have also constant terms,
which are essential for the spectra, and, above all we have
baryons, therefore we also implement supersymmetry, which relates
meson wave functions with baryon wave functions, that is we extend the conformal algebra to
the superconformal algebra.
\section{Constraints from superconformal (graded) algebra}
\subsection{Supersymmetric QM}

As mentioned in the Introduction, Supersymmetry (SUSY) relates
particles of different spin. From a theorem of Coleman and Mandula
follows, that this is impossible for symmetry groups based on
algebras, like the  rotational group, the gauge group $SU(3)$ etc. It
has been shown that only symmetries based on generators which obey
commutation and anti-commutation rules can do that. Such an
extension of an algebra is called a graded algebra.

{\small \begin{quote} For a certain time, SUSY  was very popular.
Especially string theory is only fully consistent in a
supersymmetric world. Supersymmetric QFT is very complicated.
Moreover there is no sign in nature that it is realized, although
the hopes were very high that one would detect new particles which
are supersymmetric partners of known ones. One had even speculated
that supersymmetric partners of neutrinos might be good candidates
for dark matter in cosmology. But since -- at least up to now --
no new particles, which could be  supersymmetric partners, were
found at LHCb, SUSY  came in disrepute with phenomenologically
interested physicists.
\end{quote}

In contrast to SUSY Quantum Field theory, SUSY Quantum Mechanics
is very simple~\cite{Witten:1981nf}. It is based on a graded
algebra consisting of two ``supercharges'' (fermionic operators),
$Q$ and $Q^\dagger$,  and a bosonic operator (the Hamiltonian)
$H$. The two supercharges obey anticommutation relations,
$$\{Q,Q^\dagger\}= 2 H,  \; \{Q,Q\}=0 , \;\{Q^\dagger,Q^\dagger\}=0$$
a supercharge and a bosonic operator obey commutation relations.
$$[Q,H]=0, \;[Q^\dagger,H]=0. ~\footnote{Note that the operator $2 H$ corresponds to the Meson Hamiltonian to be compared with the LF Hamiltonian} $$

 It is easy to realize this graded algebra as matrices in a Hilbert space with two components:
%%%%%%%%%%%%%%%%%%%%%%%%%%%
\beq
 Q=\left(\begin{array}{cc} 0 & -\pa_x+\frac{f}{x}+V(x)\\
0&0\end{array}\right), \; Q^\dagger=\left(\begin{array}{cc} 0 &
0\\ \pa_x+\frac{f}{x}+V(x)&0\end{array}\right) \
\lb{q1}
\enq
\beq
2 H=\left(\begin{array}{cc}  -\pa^2_x+\frac{f^2+f}{x^2}+V^2(x)-\pa_xV(x) + \frac{2 f\,V(x)}{x} & 0\\
 0 & \hspace{-2cm}  -\pa^2_x+\frac{f^2-f}{x^2}+V^2(x)+\pa_xV(x) + \frac{2 f\,V(x)}{x}\\
\end{array}\right)
\enq where $f$ is a { dimensionless} constant and $V$ has
{$\lim_{x\to 0} x V(x) =0$, otherwise arbitrary}.

\subsection{Superconformal quantum mechanics}

If the superpotential $V$ vanishes,  there is no dimensionful
quantity in the game and we can extend the SUSY algebra to a
superconformal algebra~\cite{Akulov:1984uh,Fubini:1984hf}. For
$V(x)=0$ the operators \req{q1}  become \beq
Q=Q_0=\left(\begin{array}{cc} 0 &-\pa_x+\frac{f}{x}\\
0&0\end{array}\right), \;
 Q_0^\dagger=\left(\begin{array}{cc} 0 & 0\\\pa_x+\frac{f}{x}&0\end{array}\right)
\enq

The absence of a dimensionful constant allows to extend the
supersymmetric algebra to a superconformal one. This is done by
introducing a new supercharge $S$ and its Hermitian adjoint
$S^\dagger$ with the following anticommutation relations \beq
\{S,S^\dagger\}= 2 K, \quad \{Q_0,S^\dagger\}-\{Q_0^\dagger S\} = 4
i D,\quad \{Q_0,S^\dagger \}+\{Q_0^\dagger, S\} =2 f I +
\si^3\enq all other anticommutators vanish \beq \{Q_0,Q_0\}=
\{Q_0,S\}=\{S,S\}=
\{Q_0^\dagger,Q_0^\dagger\}=\{Q_0^\dagger,S^\dagger\}=\{S^\dagger,S^\dagger\}=0
\enq Here $K$ and $D$ are the conformal operators introduced
above. We see that the new supercharges $S,\, S^\dagger$ extend
the supersymmetric graded algebra to a superconformal one, which
contains also the generators of the conformal algebra, introduced
in the previous section. In matrix notation we have: \beq
S=\left(\begin{array}{cc} 0 & x\\ 0&0\end{array}\right), \;
S^\dagger=\left(\begin{array}{cc} 0 & 0\\x&0\end{array}\right) ,
\; I=\left(\begin{array}{cc} 1 & 0\\ 0&1\end{array}\right),
\;\si_3 = \left(\begin{array}{cc} 1 & 0\\ 0&-1\end{array}\right).
\enq

\subsection{Consequences of the superconformal algebra for dynamics \lb{consu}}
In  subsection \ref{confal} a new Hamiltonian $G$ was constructed
inside the conformal algebra. Here we construct a new Hamiltonian
inside the superconformal (graded) algebra. We now
follow~\cite{deTeramond:2014asa,Dosch:2015nwa} in applying the
 procedure of
\cite{deAlfaro:1976je}, which was extended to the superconformal
algebra by Fubini and Rabinovici~\cite{Fubini:1984hf}and construct
a generalized Hamiltonian.  For that we introduce the supercharge
$R_\la$ as linear combination of $Q$ and $S$: \beq R_\la= Q_0 +
\la\, S \enq that is \beq \lb{Rex} R_\la=\left(\begin{array}{cc} 0
& r_\la\\ 0&0\end{array}\right), \;
 R^\dagger=\left(\begin{array}{cc} 0 & 0\\r^\dagger _\la&0\end{array}\right)
\enq with: $r_\la =-\pa_x+\frac{f}{x}+\la x,\; r^\dagger _\la =
\pa_x+\frac{f}{x}+\la x$

The new generalized Hamiltonian is constructed analogously to the
original one, but now not as an anticommutator of the supercharges
$Q$ and $Q^\dagger$ , but as an anticommutator of the supercharges
$R_\la $ and $R_\la^\dagger$: \beq G= \{R_\la, R^\dagger_\la\}=
\{Q_0, Q_0^\dagger\} + \la^2 \{S, S^\dagger\} +\la \Big(
\{Q_0,S^\dagger\}+\{S, Q_0^\dagger\}\Big) \enq

In this generalization of the Hamiltonian  the dimensionful
quantity $\la$ appears naturally, since $S$  and $Q$ have
different physical  dimension,
 $Q$ has dimension $[1/l]$ (1 over length),  $S$ has dimension $[l]$, hence $\la$ must have $[l^{-2}]$.

The new Hamiltonian $G$ is diagonal: \beq
G=\left(\begin{array}{cc} r_\la \,r_\la^\dagger&0\\ 0&
r_\la^\dagger  \, r_\la\end{array}\right)=\left(\begin{array}{cc}
G_{11}&0\\ 0&G_{22}\end{array}\right) \enq

Supersymmetry implies that the two eigenvalues of $G_{11}$ and
$G_{22}$ are identical. This can easily checked:

Be $\phi=\left( \begin{array}{c} \phi_M\\ \phi_B
\end{array}\right)$ an eigenvector of $G$, then we have \beqa
\label{x1}
R_\lambda^\dagger \, G \phi &=&E\,  R_\lambda^\dagger\, \phi \\
&=& R_\lambda^\dagger(R_\lambda R_\lambda^\dagger +R_\lambda^\dagger R)\phi\\
&=& R^\dagger\,R\, R_\lambda^\dagger \phi \quad \mbox{since } R_\lambda ^\dagger\,R_\lambda ^\dagger =0\\
&=& (R_\lambda R_\lambda^\dagger +R_\lambda^\dagger R) R_\lambda^\dagger \,\phi\\
&=& G R_\lambda^\dagger \,\phi \enqa That is if $\phi$ is
eigenstate of $G$ , then $R_\lambda^\dagger \phi$ is also
eigenstate with the same eigenvalue. Let $\phi^{1} $ be the state
$ \phi^{1}=\left( \begin{array}{c} \phi_M\\ 0 \end{array}\right) $
and  $\phi^{2}=\left( \begin{array}{c}0\\
\phi_B\end{array}\right), $ then  \beq R_\lambda^\dagger \phi^1 =
\left( \begin{array}{c} 0\\ r_\la^\dagger \phi_M
 \end{array}\right) \enq
has a lower component.  Since it has the same eigenvalue as
$\phi^1$, the two components have the same eigenvalues.
%%%%%%%%%%%%%%%%%%%%%%%%%%%
%%%29.11.,

We now construct the new Hamiltonian explicitly. Inserting
\req{Rex} we obtain the explicit expressions
 \beqa G_{11}&=& -
\pa_x^2 + \frac{4 (f + \half)^2 - 1}{4
 x^2} +\underbrace{ \la^2 \,x^2+ 2 \la\,(f-\half)}_{U_-},\\
\label{Gsd} G_{22}&=& - \pa_x^2 + \frac{4 (f-\half)^2-1 }{4  x^2}+
\underbrace{\la^2 \, x^2+ 2
 \la\,(f+\half)}_{U_+}.
\enqa
  and compare this result with the Hamiltonians obtained from LFHQCD.

   For
the baryons we had obtained, see \req{bsb}: \beqa
\left( -\pa_z^2 + \frac{4 L^2-1}{4z^2} + \la_F^2 z^2  +2(L+1) \la_F\right) \Psi^+(q,z)&=& M^2 \Psi^+(q,z)\\
\left( -\pa_z^2 + \frac{4 (L+1)^2-1}{4z^2} +  \la_F^2 z^2  +2L \,
\la_F\right) \Psi^-(q,z)&=& M^2 \Psi^-(q,z) \enqa
   and for mesons
with $J=L+S$, see \req{bsff} \beq \lb{HMes} \Big(-\pa_z^2+
\frac{4L^2-1}{4 \ze^2} +\la^2 \ze^2 +2(J-1) \la\Big) \tilde
\Phi_{L,J}(q,z) = q^2 \tilde \Phi_{L,J}(q,z) \enq
we identify $f=L+\half$   and $x=\zeta$ and we notice two very nice features:

1)We recover the baryon equations \req{bsb}f.  The potentials were in LFHQCD
constructed with a modification of the AdS
Lagrangian~\cite{deTeramond:2014asa}, which was there chosen {\it
ad hoc} for purely phenomenological reasons. In superconformal
LFHQCD the potentials are a consequence of the algebra. The
chirality transformation, that is multiplication by $\ga_5$ acts
like a supercharge.

 2) $G_{11}$ is  the  Hamiltonian of a meson with  $L_M=J=f+\half$ and
 $G_{22}$ is the Hamiltonian  of the baryon component $\psi_{B}^+$ with $L_B=f-\half$.
 So we can put the meson wave function $\phi_{L_M}$ with
LF angular momentum $L_M$ and the positive chirality baryon wave
function $\psi_{B}^+$ with angular momentum $L_B = L_M - 1$ in a
supersymmetric doublet: \beq
\Phi=\left(\begin{array}{c}\ph_{L_M}\\ \psi^+_{L_M -1}\end{array}
\right) \enq
      The supercharge $R^\dagger_\la$  transforms the wave
function of a boson into that of a fermion~\cite{Dosch:2015nwa}. That is
Fermion and Boson wave functions  are transformed into each other
by a supercharge, namely $R_\la^\dagger$, as it should be in
supersymmetry !

The meson with angular momentum $L_M=f+\half$ is superpartner of
Baryon with  $L_B=f-\half$, therefore mesons with $L_M=0$ ($\pi$
{\it e.g.}) can have no superpartner since $L_B=-1$ is excluded.
{ One can easily check that the lowest eigenstate
$\phi_0$ of the Hamiltonian $G$ has the eigenvalue 0 and the form
\beq \Phi_0= \left( \begin{array}{c} \phi_{00}\\ 0 \end{array}
\right),
\enq
 where $\phi_{00}$ is given by \req{asso1}. This
implies that there exists  no nontrivial eigenstate of the baryon
with eigenvalue 0 which could be a partner of the lowest mesonic
state. This is due to the fact that $r_\la^\dagger \phi_{00} =0$.
}

Up to now we have only treated hadrons where the quark spin does
not enter explicitly, like in the pion and in the nucleon. in the
next subsection the treatment will be extended to other cases,
notably the rho-meson and the Delta resonance.

\subsection{Spin terms and small quark masses \lb{spsqm}}
{ For mesons the action AdS distinguishes between the
$\rho$ trajectory, where the total quark spin $\cS= 1$, and the
$\pi$ trajectory, where the quark spin $\cS=0$. Therefore the
Hamiltonian \req{bsff} depends both on the orbital angular $L$ and
on the total angular momentum.} For mesons with $J = L + S$, where
$S$ is the total quark spin, we can separate the Hamiltonian
\req{bsff} into a part containing only the angular momentum plus
the spin term. \beq H_M = H_{J=L} + 2 \cS \,\la = G_{11}+
2\cS\,\la \enq The Hamiltonian for a meson with $J=L$ is identical
with $G_{11}$. Therefore we obtain as final Hamiltonian \beq
G_{SUSY}=\{R_\la,R^\dagger_\la\} + \cS {\mathbf I}, \lb{hfin}
 \enq
where $\mathbf I$ is the unit operator; $\cS$ is for mesons the
total quark spin, and for baryons the  minimal possible quark spin
of two-quark clusters inside baryon.

The supercharges  $R_\la, \, R^\dagger_\la$ are given by \req{Rex}
with $f=L_B+ \half$,  they connect a baryon wave functions with
angular momentum $L_B$ and positive chirality  and diquark spin
$\cS$ with a meson wave function with angular momentum $L_M=L_B+1$
and total quark spin $\cS$.

In LFHQCD, sect. \ref{bsebas},
 the additional spin term in \req{barspin} had to be added for baryons  by hand in order to obtain
agreement with experiment. In \req{hfin} it is a consequence of
supersymmtry.

The corrections through finite quark masses will be the same as
discussed in sect. \ref{smqm} and will break the the
supersymmetry. The final formulae for spectra from AdS with
superconformal constraints are: \beqa \label{mesbarfin}
\mbox{Mesons } && M_M^2 = 4 \la (n+L_M)+ 2 \la \,  \cS +\De M_2^2(m_1,m_2),\\
\mbox{Baryons} && M_B^2=4 \la(n+L_B+1) + 2 \la \,  \cS   +\De
M_3^2(m_1,m_2,m_3).\nn \enqa

Note that in applications of LFHQCD before 2016, e.g. in \cite{Brodsky:2014yha}, for baryons with $\cS=1$ the spin effect was taken into account by formally using  half integer LF angular momentum; this leads to the same mass formul\ae.

\subsection{Comparison with experiment}
In Fig. \ref{pirhondel} we display the theoretical curves obtained
from \req{mesbarfin} and the experimental results for the
non-strange hadrons, in Fig. \ref{strangehad} for hadrons
containing 1 or 2 strange quarks. The result for the $\Omega^-$
mass comes out to 1760 MeV, within the expected accuracy
compatible with the experimental mass of 1672 MeV. In Fig.
\ref{lamsym} we display the values of $\sqrt{\la}$ obtained by
independent fits to the different chanels. Indicated at the
abscissa are the lowest $(L=0)$ state of the trajectory. Theory
and experiment agree with the same accuracy of $\approx \pm 100$
MeV, as expected from the model and also observed in the previous
chapter.

\begin{figure}
\begin{center}
\includegraphics[width=6.5cm]{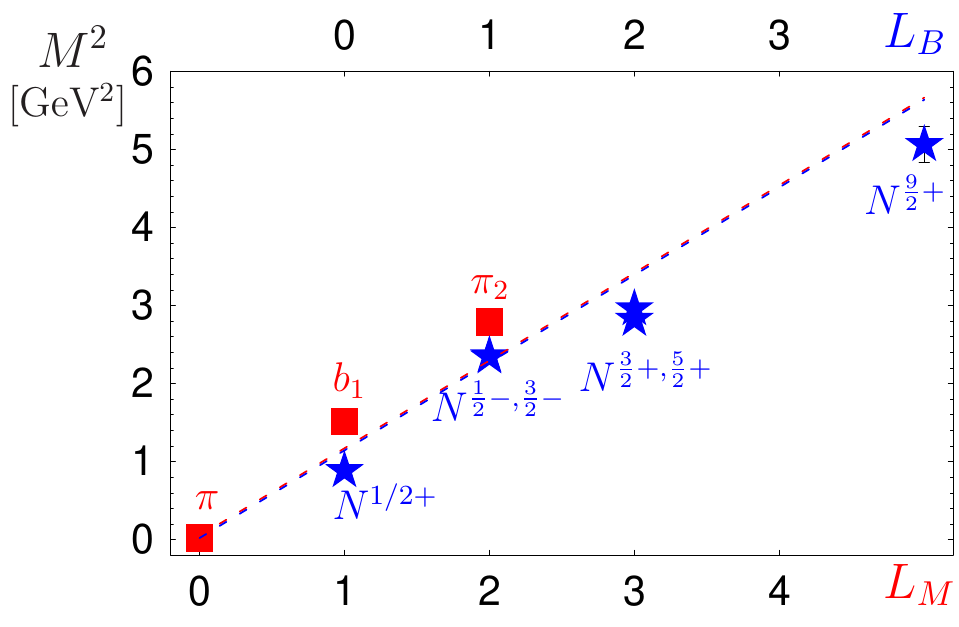}
\includegraphics[width=6.5cm]{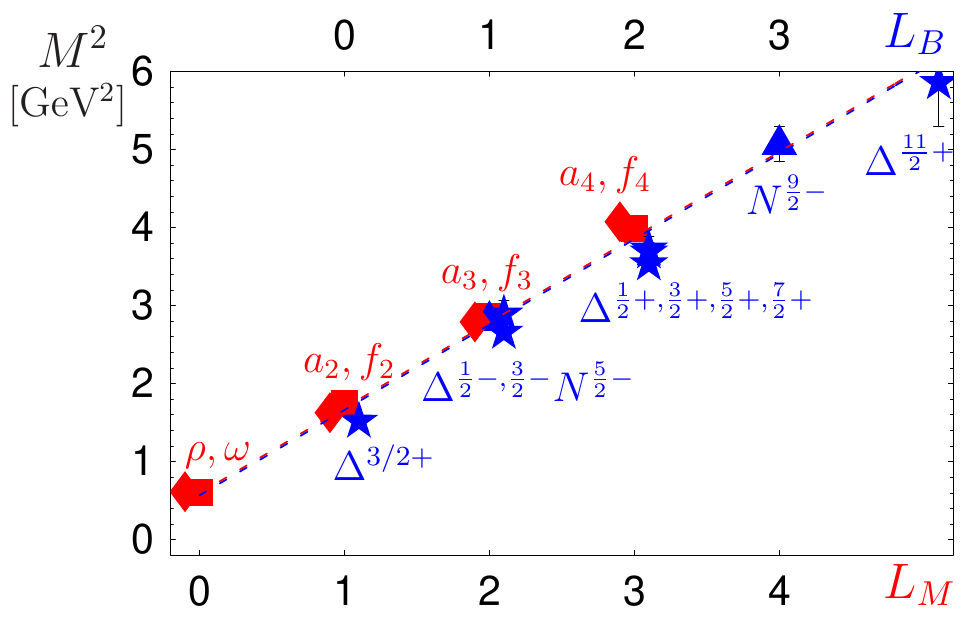}\\
\hspace{3cm}$\pi-N$   \hspace{4cm} $\rho-\De$ \caption{Theoretical
and experimental results for light hadrons. The dashed line is the
result of supersymmetric LFHQCD, \req{mesbarfin} , without mass
corrections), the red boxes denote the mesons, the blue stars the
baryons, adapted from ~\cite{Dosch:2015bca} \lb{pirhondel}.}
\end{center}
\end{figure}

\begin{figure}
\begin{center}
\includegraphics[width=14cm ]{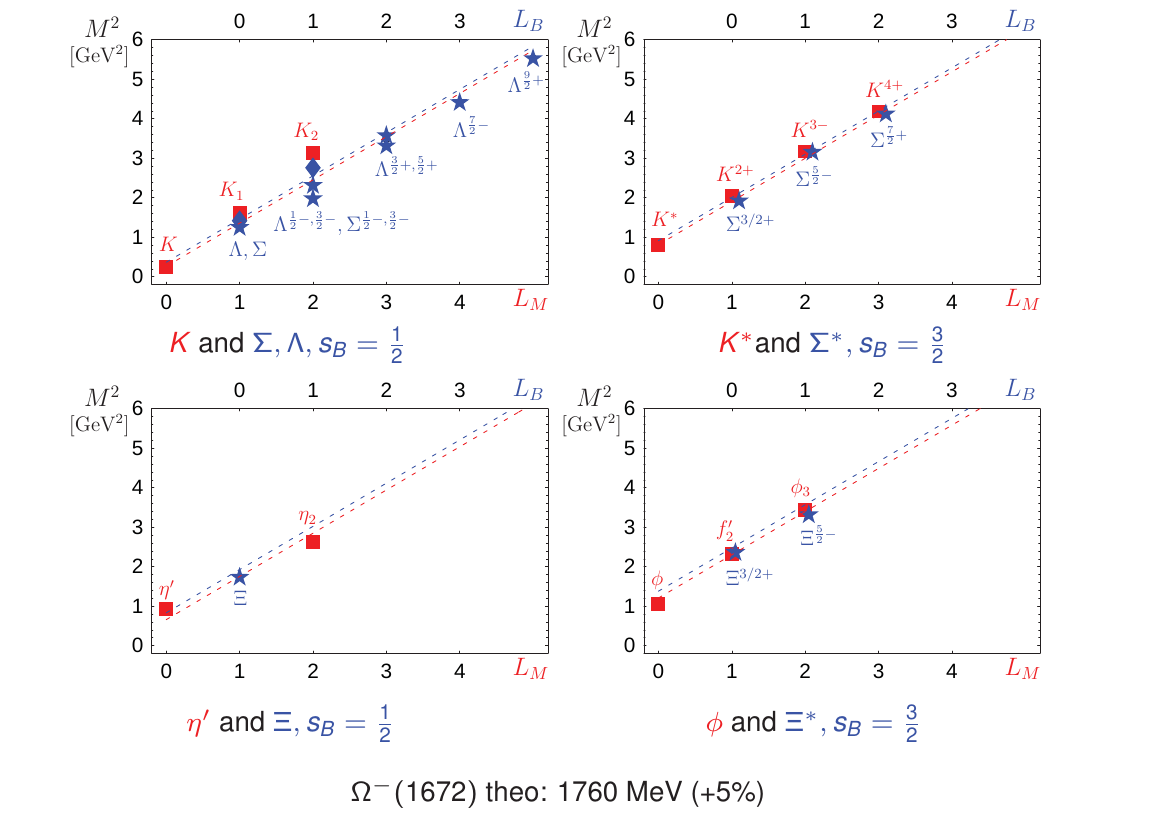}
\caption{Theoretical and experimental results for hadrons
containing one or two strange quarks. The dashed lines are the
result of supersymmetric LFHQCD, \req{mesbarfin}, including mass
corrections), the red boxes denote the mesons, the blue stars and
diamonds the baryons, adapted from ~\cite{Dosch:2015bca}
\lb{strangehad}.}
\end{center}
\end{figure}

\begin{figure}
\begin{center}
\includegraphics[width=10cm]{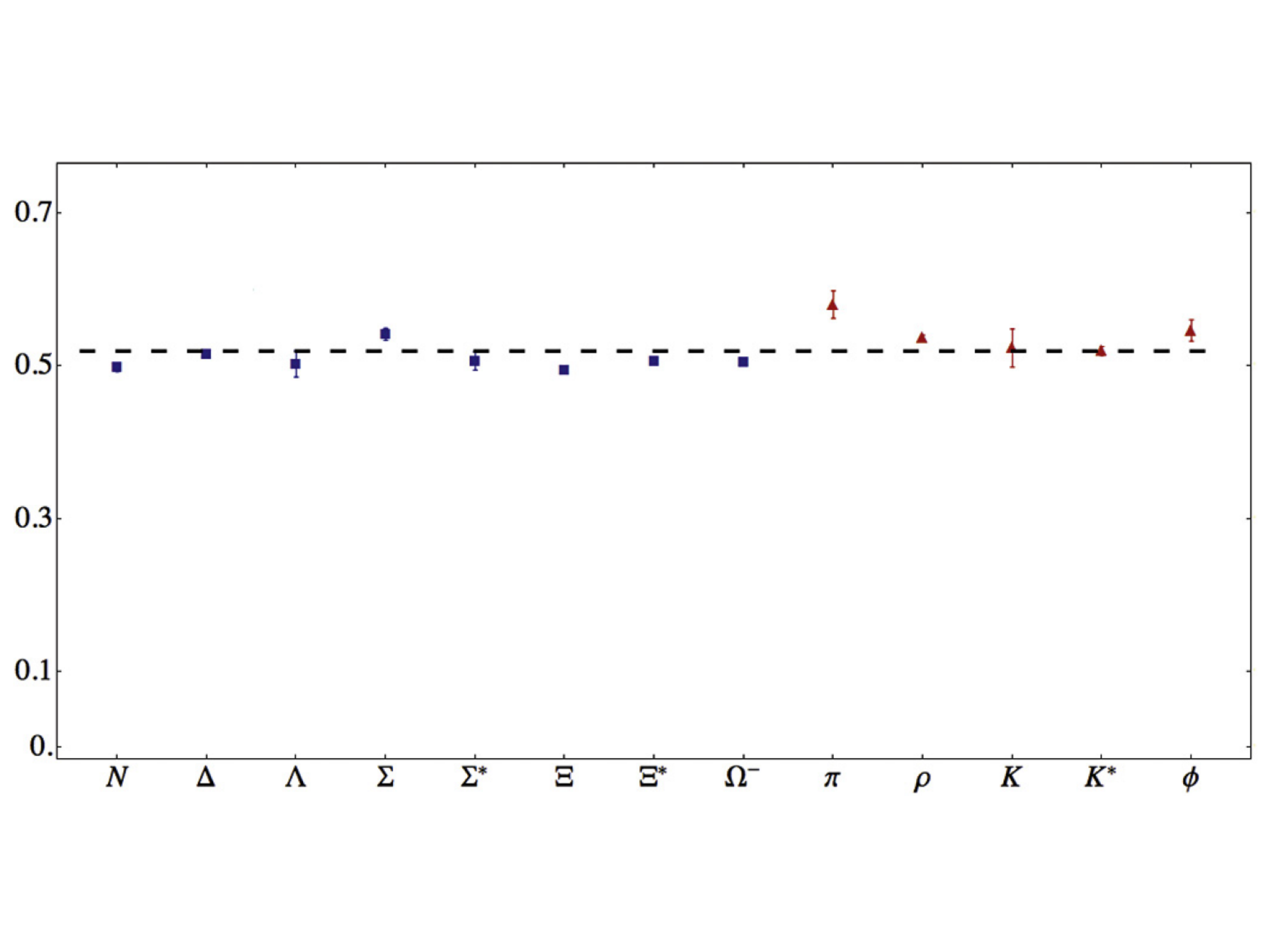}
\caption{The scale parameter $\la$ determined for the different
trajectories separately. Indicated at the abscissa are the lowest
$(L=0)$  states of the trajectory\lb{lamsym}.}
\end{center}
\end{figure}

\clearpage

\subsection{ Completing the supersymmetric multiplet - Tetraquarks \lb{tqs}}

Up to now we have only considered the supermultiplet of a meson
and the positive chirality component of the baryon, $\Psi^+$. The
negative chirality component must also have a superpartner in
order to complete the supersymmetric multiplet. The supercharge
$R^\dagger_\la$ transforms a meson wave function with LF angular
momentum $L_B + 1$ into a baryon wave function with angular
momentum $L_B$, it decreases the angular momentum by one unit and
changes the fermion number by ±1, since it is a fermionic
operator. \beq
 R^\dagger\left(\begin{array}{c}  \phi_{L_B+1}\\ 0
 \end{array}\right)=\left(\begin{array}{c} 0\\   \psi^+_{L_B} \end{array}\right)
\enq We can form a doublet containing the negative chirality
component of the baryon $\Psi^-_{L_B+1}$ (remember that the
negative chirality component has an angular momentum one unit
higher than the positive one, see \req{solbar}): \beq
\left(\begin{array}{c} \tilde \ps^-_{L_B+1}\\ \ph^T_{L_B}
\end{array}\right)\enq where  $ \ph^T_{L_B} $  is a bosonic wave
function with angular momentum $L_B$, and it must be in the same
radial excitation state as the meson and the baryon. This is the
only information we can draw from supersymmetric quantum
mechanics, since it does not contain any information on the quark
structure. One can only draw analog conclusions from the action
of the operator $R^\dagger_\la$ inside the first multiplet, where
the quark interpretation is fixed in LFHQCD.

In the original multiplet $\left(\begin{array}{c}\phi_{L_B+1} \\
\psi^+_{L_B} \end{array}\right)$ the operator $R^\dagger_\la$
 has transformed a two-quark state into a
three-quark state, that it is has increased the number of
constituents by 1. The two-quark state could have colour 3 or 6,
but it must be in a $\bar 3$ colour state, since the baryon is a
colour singlet. Therefore it is plausible to attribute to
$R^\dagger_\la$ in the quark configuration the property of
transforming an antiquark state into a two-quark state in the same
colour representation $\bar 3$, or correspondingly the
transformation of a quark into an two-antiquark state in the same
colour 3 representation. From that we infer that $R^\dagger_\la$
transforms a quark of the nucleon into an antiquark pair in colour
3 representation. Therefore we infer further that the superpartner
of the negative chirality component of the baryon is a tetraquark
state consisting of a two quark state in colour $\bar 3$
representation and a two quark state in colour 3 representation.
The previous considerations are graphically represented in Fig.
\ref{pict}.

\begin{figure}
\begin{center}
\setlength{\unitlength}{0.5pt}
\begin{picture}(350,250)(0,-50)
\put(25,35){\circle*{25}} \put(25,65){\circle{25}}
\put(25,-35){\circle{25}} \put(25,-65){\circle*{25}}

\redt{\put(75,85){$R^\dagger_\la$}}
\bluet{\put(75,-95){$R^\dagger_\la$}}

\redt{\put(230,85){$R^\dagger_\la$}}
\bluet{\put(230,-95){$R^\dagger_\la$}}

{\thicklines \redt{\put(15,65){\vector(1,0){100}}}} {\thicklines
\bluet{\put(15,-65){\vector(1,0){100}}}}

{\thicklines \redt{\put(165,35){\vector(1,0){100}}}} {\thicklines
\bluet{\put(160,-35){\vector(1,0){100}}}}

\put(120,65){\circle*{25}} \put(150,65){\circle*{25}}
\put(135,35){\circle*{25}}

\put(120,-65){\circle{25}} \put(150,-65){\circle{25}}
\put(135,-35){\circle{25}}

\put(280,65){\circle*{25}} \put(310,65){\circle*{25}}
\put(280,35){\circle{25}} \put(310,35){\circle{25}}

\put(280,-65){\circle{25}} \put(310,-65){\circle{25}}
\put(280,-35){\circle*{25}} \put(310,-35){\circle*{25}}

\end{picture}
\end{center}
\caption{\label{pict} Since particles and antiparticles are in
this semiclassical theory treated at the same footing, the
fermionic operator $R^\dagger_\la$ can be interpreted either as
transforming a quark into an antiquark pair (red arrows) or an
antiquark into a quark pair (blue arrows). In the transition from
the meson to the nucleon the two-quark state has to be in the
antisymmetric colour representation. It is natural to require this
also in the transition from the nucleon to the tetraquark.}
\end{figure}
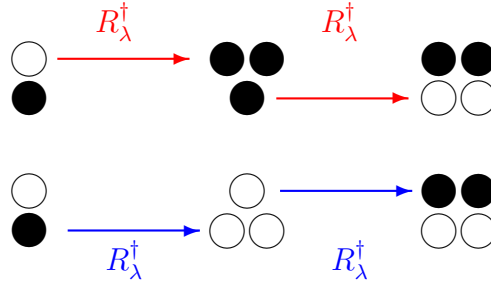

The complete supersymmetric quadruplet can be arranged into a
$2\times 2$ matrix, \beq
 \left(\begin{array} {cc} \phi_{L_B+1}&\psi^-_{L_B+1}\\ \psi_{L_B}^+ & \phi^T_{L_B} \end{array} \right)
\enq with \beq G  \, \left(\begin{array} {cc}  \phi_{L_B+1}&
\psi_{L_B+1}^-\\  \psi_{L_B}^+ & \phi^T_{L_B} \end{array} \right)=
 M^2 \,\left(\begin{array} {cc}  \phi_{L_B+1}  & \psi^-_{L_B+1}\\  \psi_{L_B}^+ & \phi^T_{L_B} \end{array} \right)
\enq

It is important to note that the two-constituent clustering has
only to be considered as a kinematical, but not as a dynamical
grouping. We shall see later in treating the form factors that
there is indeed no indication for a tightly bound diquark state.
Therefore we must assume that there is no excitation of the two
constituent cluster. The Pauli principle implies then that the
two-constituent cluster must be in an Isospin $I = 1$ and total
angular $J = 1$ state or an $I = 0, J = 0$ state, since it is
antisymmetric in colour and must also be totally antisymmetric. If
the  clusters are in a relative $S$ state, colour can rearrange
and the $3\bar 3$ state can change into a $0 − 0$ state since
\beq \sum_{i=1}^3\ep_{ijk}\,\ep^{i\ell m} = \de_j^\ell\, \de_k^m -
\de_k^\ell\, \de_j^m \enq This is a meson-molecule and can decay
easily without further involvement of strong interactions. The
status of tetraquarks is therefore rather uncertain, unless for
some reason it is stable under strong interactions, when the decay
threshold of the decay into to mesons is higher than he mass of
the tetraquark. Also higher orbital expiations make the colour
rearrangement more difficult, because of the spatial separations
of the two clusters by the centrifugal barrier. We consider
nevertheless in Table \ref{tabtet} two candidates for complete
super-quadruplets in the lowest possible angular momentum state:
Though the agreement - always in the limits of the expected
accuracy –- is very satisfactory, one should take into account
that there are also conventional interpretations for the states
listed in the table under tetraquarks. We therefore do not pretend
that these states are pure tetraquark states, but that the
tetraquark states should be taken into account if a detailed
analysis of states with similar masses and the same quantum
numbers is performed.
\begin{table}
\caption{\label{tabtet} Two candidates in the lowest possible
angular momentum state.}
\begin{center}
\begin{tabular}{ccc}
Meson $I(J^P)$&Baryon $I(J^P)$&Tetraquark $I(J^P)$\\
\hline
$b_1(1235)\,1(1^+)$&$ N_{+,-}(940)\,\half(\half^+)$ &$f_0(980)\,0(0^+)$\\
$a_2(1320)\,1(2^+)$&$\Delta_{+,-}(1230)\,\frac{3}{2}(\frac{3}{2}^+)$&$a_1(1260)1(1^+)$
\end{tabular}
\end{center}
\end{table}

%%%%%%%%%%%%%
%%Correctur13.12
\section{Implications of supersymmetry on Hadrons containing heavy quarks  \lb{ssh}}
This section is based on \cite{Dosch:2015bca,Dosch:2016zdv}.

We have seen in Fig. \ref{strangehad} that breaking of the
conformal symmetry by  small quark masses does not invalidate the
principal results of superconformal LFHQCD, if the effect of the
quark masses is taken into account perturbatively, see sects
\ref{smqm} and \ref{spsqm}. In this section we investigate the
inclusion of one heavy quark ($c$ or $b$ quark). The question is: Does the
supersymmetric part  of  the superconformal algebra  survive?  The fact
that supersymmetry played an essential role to fix the exact form
of the potentials gives us some hope that it plays generally a
fundamental role in the AdS/CFT correspondence and might even be
present, if conformal symmetry is broken strongly by heavy quark
masses.

\subsection{The experimental situation \lb{hqe}}
We have a look at the corresponding meson and baryon spectra where
we plot both spectra in the same graph and use the relation that
the meson LF angular momentum is by one unit larger than that of
its baryonic partner,  $L_M = L_B + 1$. The results, displayed in
Fig. \ref{c-hybr} and Fig. \ref{b-hybr} show that supersymmetry is realized
to a similar degree as for light quarks, and as far as one can
see,  there is even an indication for linear trajectories.
\begin{figure}
\bec
\includegraphics[width=8cm]{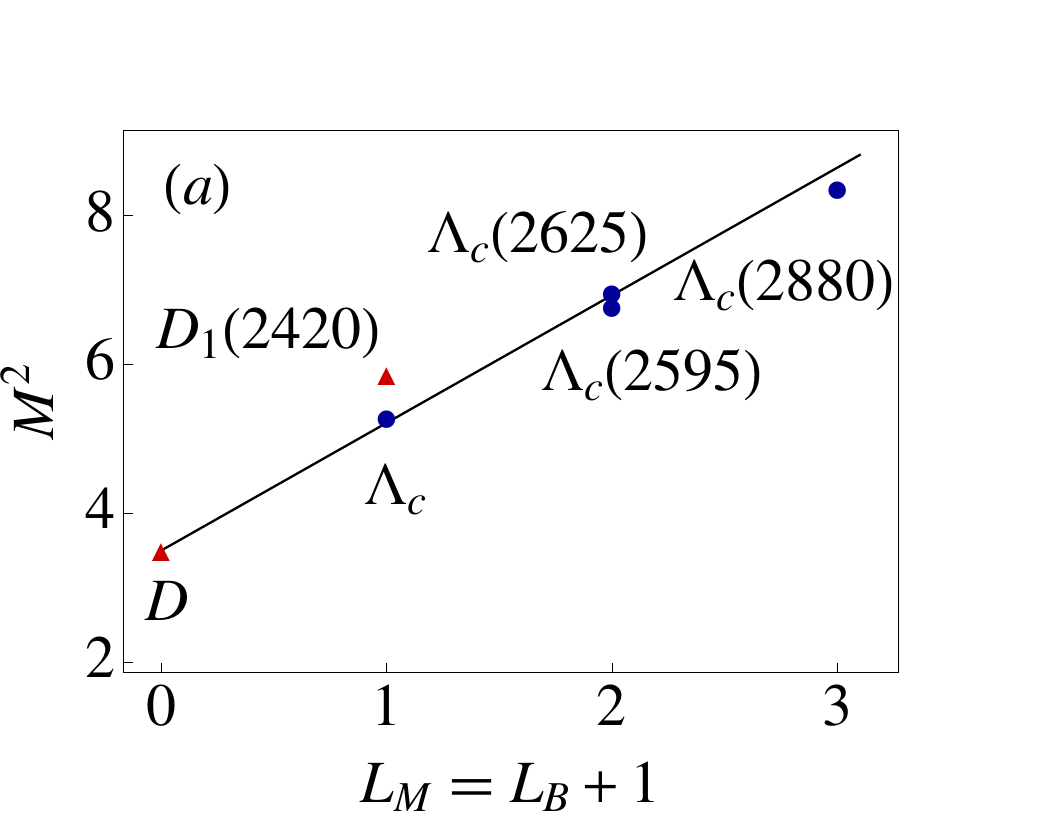}
\includegraphics[width=7cm]{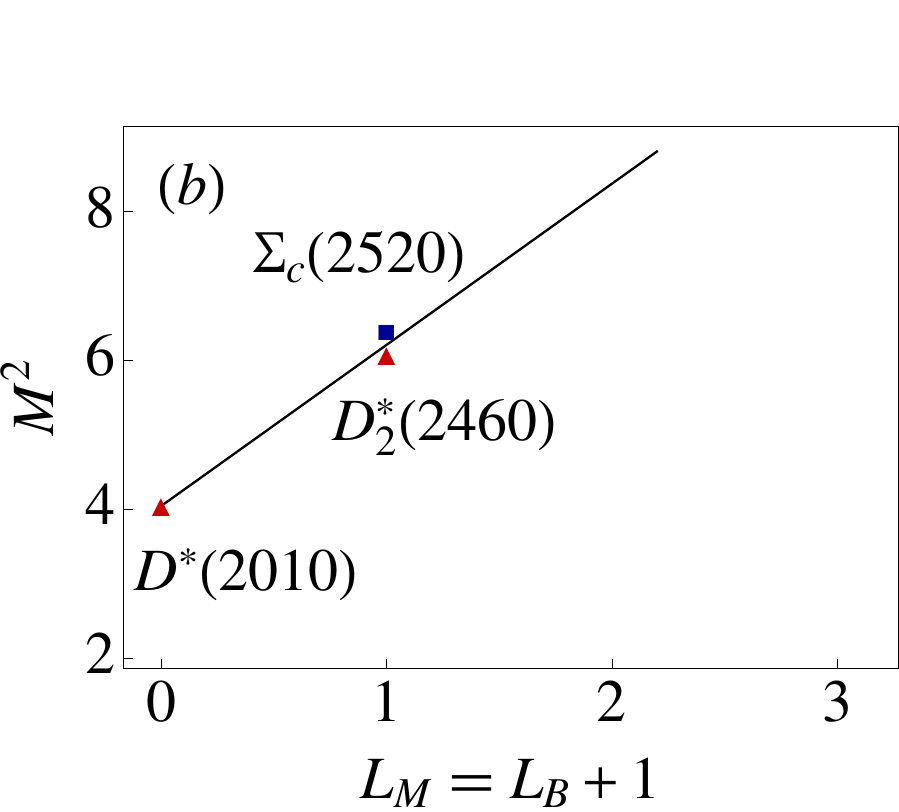}
\includegraphics[width=8cm]{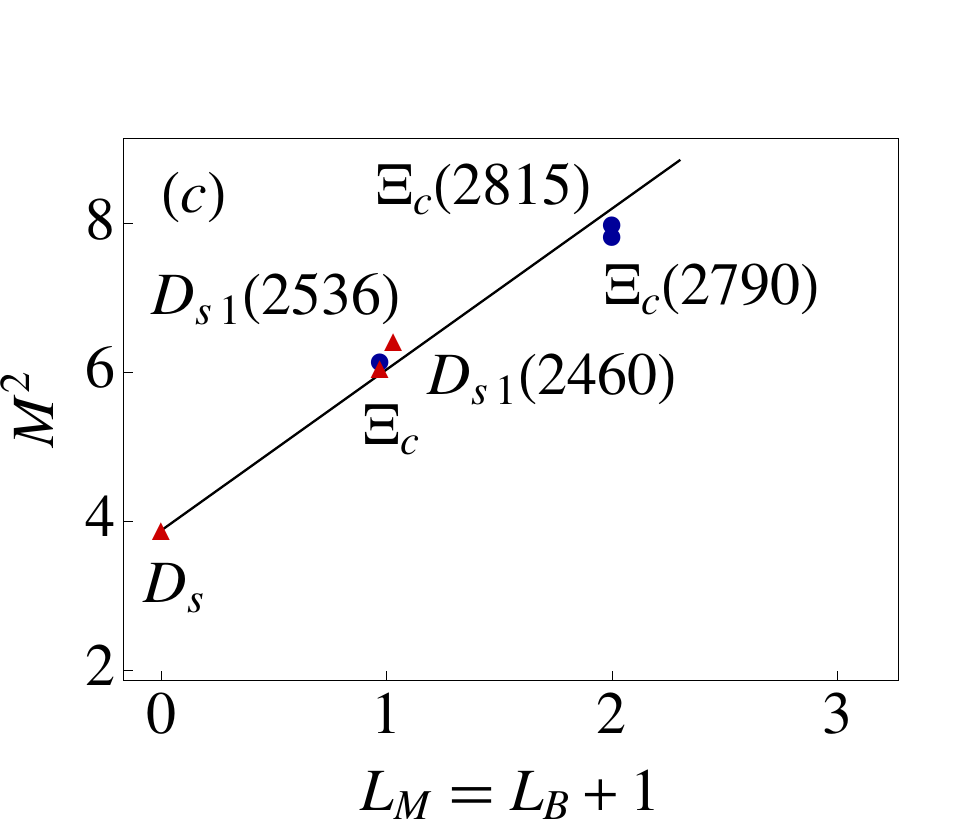}
\includegraphics[width=7cm]{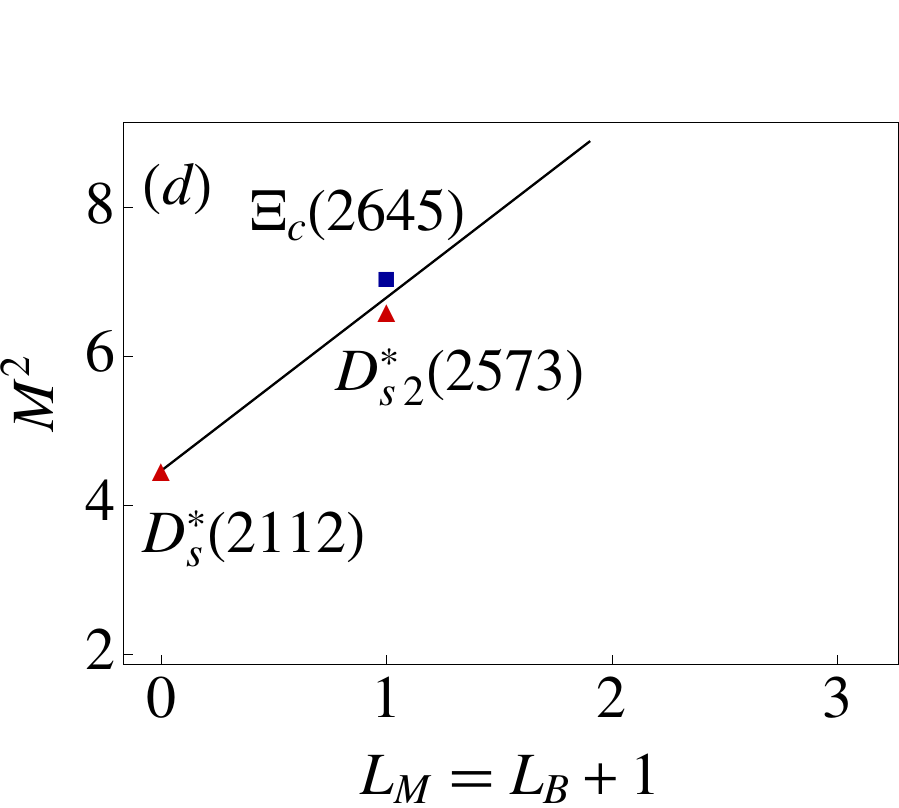}
\enc \caption{\lb{c-hybr} Supersymmetry for charmed hadrons.}
\end{figure}

\begin{figure}
\bec
\includegraphics[width=7cm]{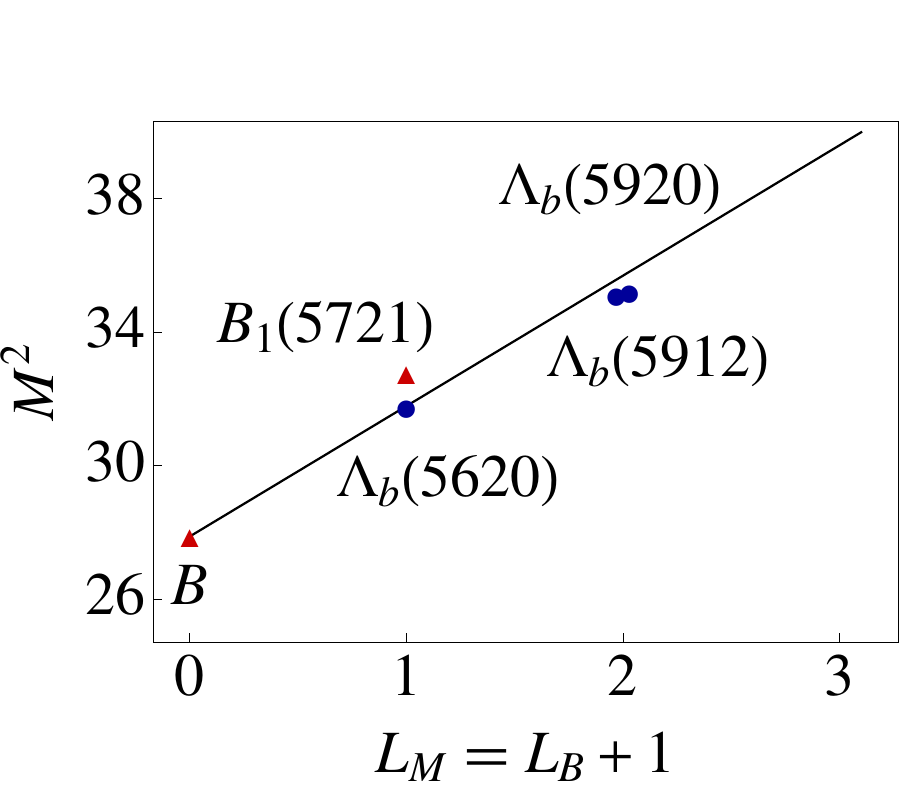}
\includegraphics[width=7cm]{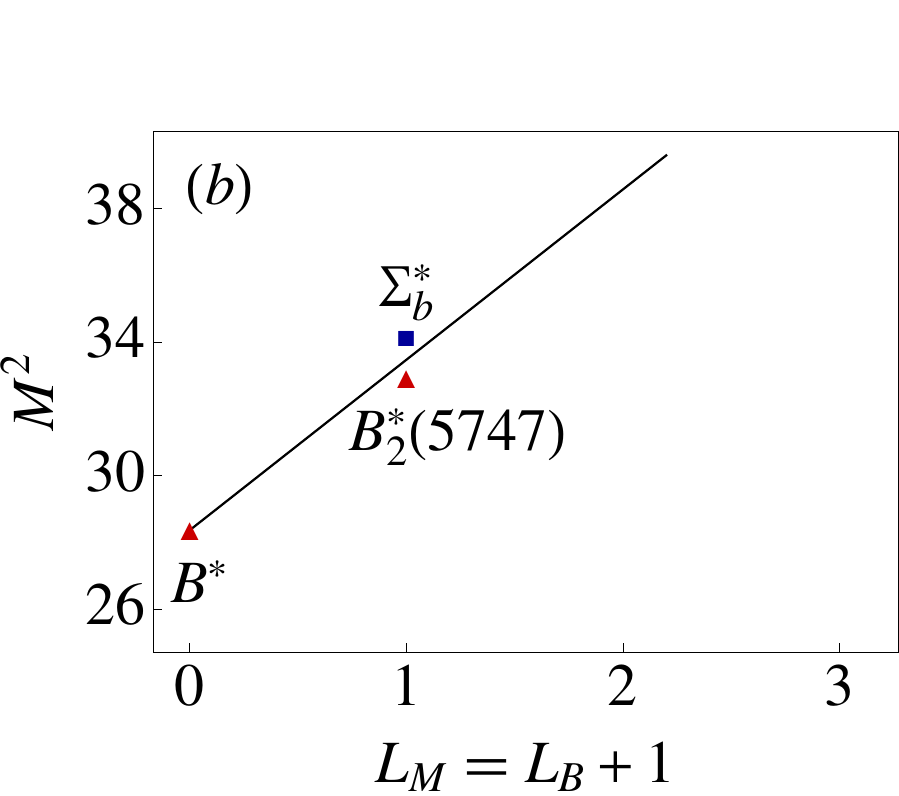}
\includegraphics[width=7cm]{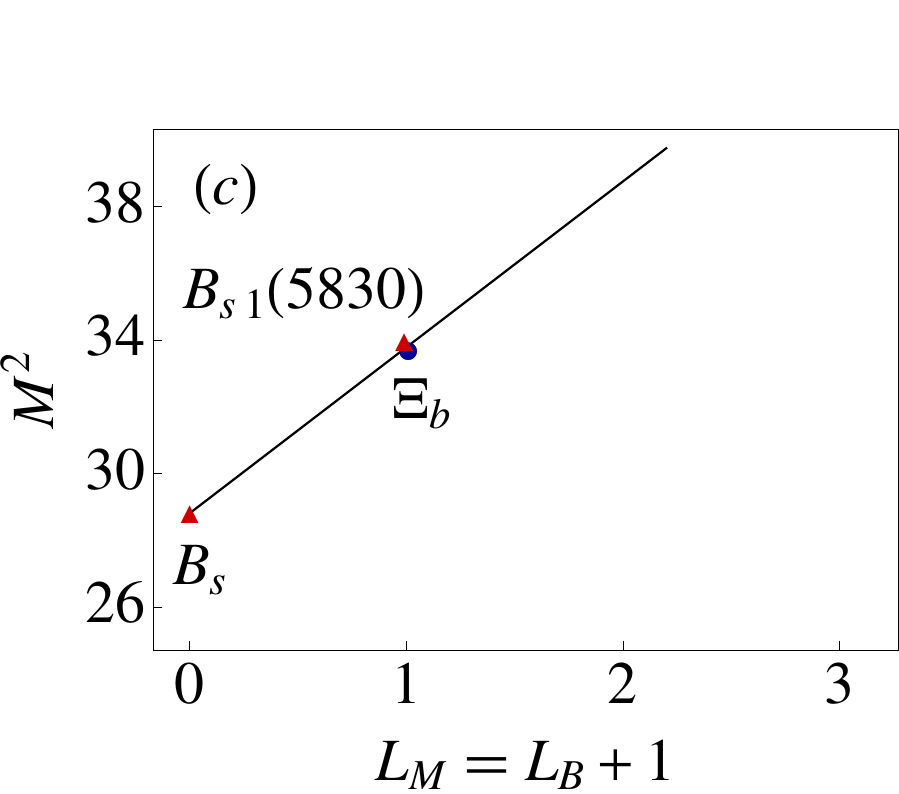}
\includegraphics[width=7cm]{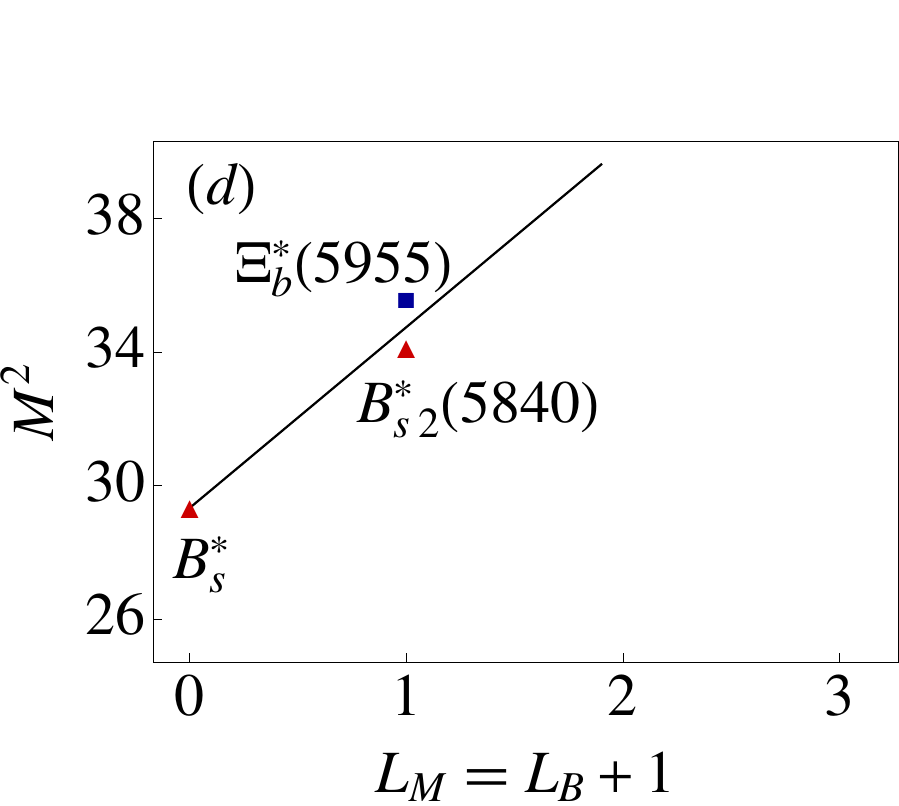}
\enc \caption{\lb{b-hybr} Supersymmetry for bottom hadrons.}
\end{figure}
\clearpage
%%%%%%%%%%%%%%%%%%%%%%%%%%%%%%%%%%%%%%%%%%%5

\subsection{Linear trajectories \lb{lt}}
In this subsection we show that linear trajectories are a
consequence of supersymmetry, if one demands that in the AdS/CFT
correspondence the modification of the action is realized by a
dilaton term $e^{\vp(z)}$, even if  the functional form of
$\vp(z)$  is not fixed. We go back to supersymmetric quantum
mechanics \cite{Witten:1981nf}. There the supercharge Q is
% (see sect. \ref{Witten:1981nf})
%. There the supercharge Q is (see sect. \ref{susyqm}
\beq Q = \left(\begin{array}{cc}
0 & q\\
0 & 0\\
\end{array}
\right) , \quad  \quad Q^\dagger=\left(\begin{array}{cc}
0 & 0\\
q^\dagger & 0\\
\end{array} \right),
\enq
 and
\beq \lb{H} H_M= \left(\begin{array}{cc}
q \, q^\dagger &  0\\
0 & q^\dagger q \\
\end{array}
\right) , \enq
 with
 \beqa \label{qdag}
q &=&-\frac{d}{d z} + \frac{f}{z} + V(z),\\
q^\dagger &=& \frac{d}{dz}  + \frac{f}{z} + V(z),
 \enqa
where  $f$ is a dimensionless constant. Without breaking
supersymmetry one  can add to the Hamiltonian \req{H} a constant
term proportional to a multiple of the unit matrix, $\mu^2 I$ \beq
\label{Hmu} H_\mu =  \{Q,Q^\dagger\} + \mu^2 I \enq where the
constant $\mu$ has the dimension of a mass; thus we obtain the
general supersymmetric light-front  Hamiltonian
 \beq \label{hmmu}
 H_\mu
  =     \left(\begin{array}{cc} - \frac{d^2}{d z^2}+\frac{4 L_M^2-1}{4z^2}+U_M(z)& \hspace{-1cm} 0 \\
0 & \hspace{-1cm} - \frac{d^2}{d z ^2} +\frac{4 L_B^2-1}{4z
^2}+U_B(z )
\end{array}\right) + \mu^2 \,\mathbf{I}
\enq where $L_B + \half = L_M - \half = f$ and $U_M$ is the meson
potential for a meson with $J=L_M$ and $U_B$ a baryon potential. They can be obtained from the Hamiltonian \req{H}
in terms of the superpotential $V$. \beqa
U_M(z) &=&    V^2(z) - V'(z) + \frac{2L_M - 1}{z} V(z),  \label{HM} \\
U_B(z) &=&    V^2(z) + V'(z) + \frac{2L_B+1}{z} V(z) . \label{HB}
\enqa The superpotential $V$ is only constrained by the
requirement that it is regular at the origin.

In LF holographic QCD  the confinement potential for  mesons
$U_M$~\req{HM} is due to  the dilaton term $e^{\vp(x)}$ in the
AdS$_5$ action, see \req{svm}.  It leads to, see \req{uadsvoll}
   \beq \label{dilU}
 U_{\rm dil}(z)=  \frac{1}{4}(\vp'(z))^2  + \frac{1}{2} \vp''(z)
+ \frac{2L_M -3}{2 z} \vp'(z) \enq for $J_M= L_M$.
  In the conformal limit the potential is harmonic and this is only compatible with a quadratic dilaton profile, $\varphi
= \lambda z^2$.

But since heavy quark masses break superconformal symmetry
strongly, the quadratic form $\vp = \la z^2$ cannot longer be
derived from symmetry arguments as in section \ref{consu}.
Additional constraints do appear, however, by  the holographic
embedding  of supersymmetry. To see that, we equate the potential
\req{dilU}, given in terms of the dilaton profile $\vp$, with the
meson potential \req{HM} written in terms of the superpotential
$V$:
   \beq \lb{susyacs}
\frac{1}{4}(\vp')^2 + \frac{1}{2} \vp''  + \frac{2 L-1 }{2 z} \vp'
= V^2  - V' +\frac{2 L +1}{z}V, \enq where $L = L_M -1$.

A simple calculation shows that for the ansatz  $\vp(z) = \la\,
z^n$ only the power n = 2 is compatible with \req{susyacs}.
Therefore we make the ansatz:
 \beqa
 \label{a1}
 \vp'(z) &=& 2 \la z
\,\al(z), \\ \label{a2} V(z) &=& \la z \,\be(z) . \enqa
 Then we obtain from  \req{susyacs}
 \beq \label{constraint} \la^2 z^2
(\al^2-\be^2) + 2 L  \la (\al-\be) + \la z (\al' + \be')=0 . \enq
  Introducing the linear combination \beqa
\si(z) &=& \al(z) + \be(z), \nn  \\
\de(z) &=& \al(z) - \be(z), \enqa \req{constraint} yields
 \beq \de(z) = - \frac{\la z\,
\si'(z)}{\la^2 z^2 \, \si(z) + 2 L \la} , \enq and therefore:
\beqa \label{a3}
\al(z) &=& \frac{1}{2} \left( \si(z)-  \frac{\la z\,  \si'(z)}{\la^2 z^2 \, \si(z) + 2 L \la}\right), \\
\label{b3} \be(z) &=& \frac{1}{2} \left( \si(z)+  \frac{\la z\,
\si'(z)}{\la^2 z^2 \, \si(z) + 2 L \la}\right) .\enqa

Using \req{a1}  and \req{a3}   we obtain  after an  integration the
condition for a dilaton profile for a meson with angular momentum
$L_M=L+1$ \beq \label{dil}
 \vp(z) = \int^z dz'\, \left(\la z'\, \si(z') - \frac{ \la^2 {z'}^2\,  \si'(z')}{\la^2 {z'}^2 \, \si(z') + 2 (L_M-1) \la}\right).
\enq

The modification of the general AdS action should be independent
of the angular momentum of a peculiar state, that is we must have
$\si'(z) =0$ thus \beq \si(z)= A~~ \mbox{ with $A$  an arbitrary
constant}. \enq
 From  \req{dil} and \req{a1}
it follows  that
 \beq
  \label{phiV}
  \vp(z) = \half  \la A \,z^2 + B,
  \enq
and
\beq  \al=\be = \half \si=\half A \enq
  from which follows:
  \beq
  V(z) = \half \la A \, z
 \enq

This result implies that the LF potential even for strongly broken
conformal invariance has the same quadratic form as the one
dictated by the conformal algebra. The constant $A$, however, is
arbitrary, so the strength of the potential is not determined.
Notice that the interaction potential \req{dilU} is unchanged by
adding a constant to the dilaton profile, thus we can set $B =0$
in \req{phiV} without modifying the equations of motion.

The LF eigenvalue equation  $H \vert \phi\rangle =  M^2 \vert \phi
\rangle$ from the supersymmetric Hamiltonian \req{hmmu} leads to
the hadronic spectrum
 \beq \label{mass-formulae}
\begin{tabular}{ll}
\mbox{Mesons:}   ~~~ &  $M^2 = 4 \la_Q \, (n+L_M) +  \mu^2$, \\
\mbox{Baryons:}  ~~~ &  $M^2 = 4 \la_Q \, (n+L_B+1) + \mu^2$,
\end{tabular}
\enq here $L_M$ and $L_B$ are the LF angular momenta of the meson and baryon respectively, the slope constant $\la_Q =
\half  \la \, A$ can depend on the mass of the heavy quark.  The
constant   term   $\mu$  contains the effects of spin coupling and
quark masses.

The fitted values of the slopes $\la$ for the different channels
are shown in Fig. \ref{lambda-channel}.
 There is no perfect agrement between different channels, but distinctly 3 groups are observed for the conformal case, for the $c$-channel,  and for the $b$-channel.
%%%%%%%%%%%%%%%%
\begin{figure}
\begin{center}
\includegraphics*[width=12cm]{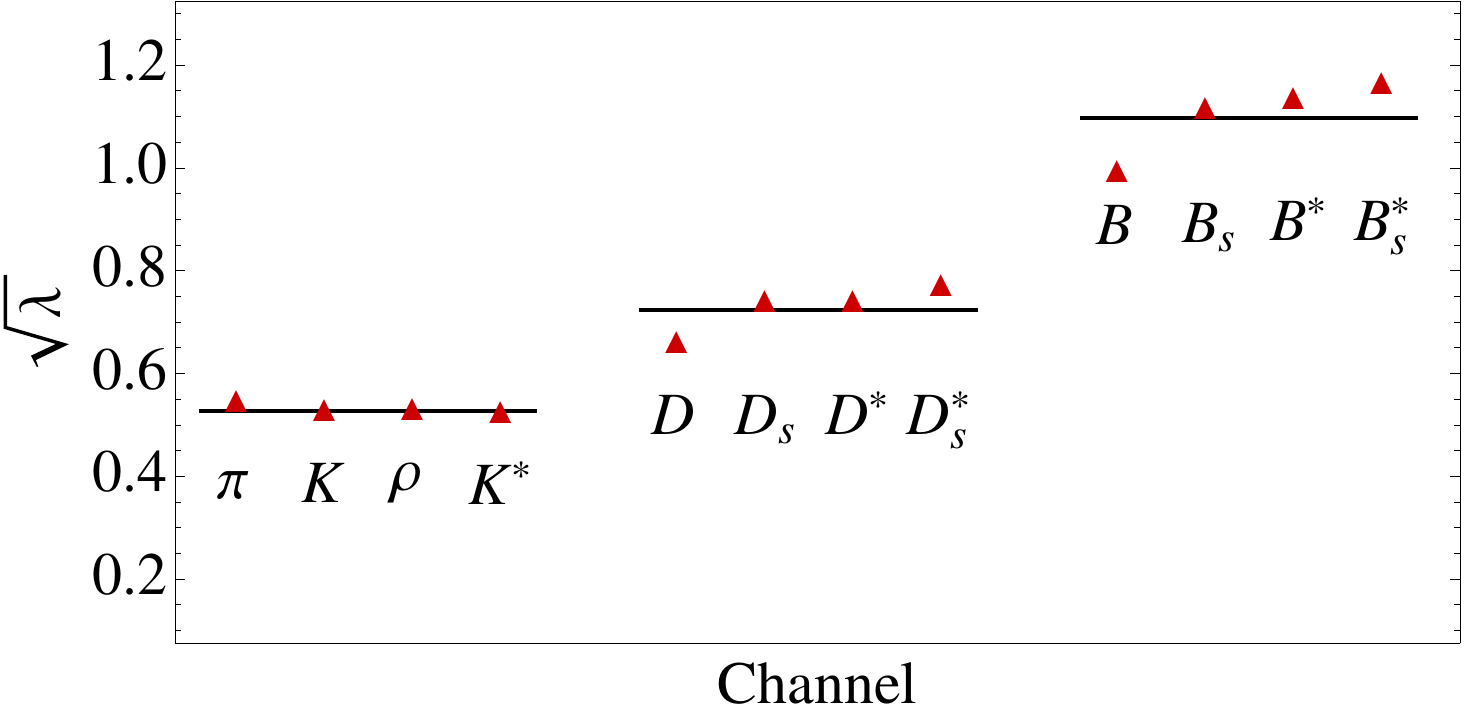}
\end{center}
\caption{\label{lambda-channel} The fitted value of $\sqrt{\la_Q}$
for different meson-baryon trajectories, indicated by the lowest
meson state on that trajectory.}
\end{figure}
%%%%%%%%%%%%%%%%
In tables \ref{tabcharm} and \ref{tabbottom}  results and
predictions of the model are shown together with the deviation
$\De M$ between theory and  experiment, which is typically $\De M <\sim 100 MeV$.

\begin{table}
\caption{\label{tabcharm}  Charmed Hadrons. The quark spin $s$ is
the total quark spin of the meson or the diquark cluster, $\la_Q$
is the fitted value for the  trajectory and $\Delta M$ is the
difference between the observed and the theoretical value
according to \req{mass-formulae}. The lowest lowest lying meson
mass determines de value of  $\mu^2$ in \req{mass-formulae} for
each trajectory. We have added predictions, if only one
superpartner has been observed and for $L_M\leq 2,\;L_B\leq 1$.}
\begin{center}
{\renewcommand{\arraystretch}{0.7}
\begin{tabular}{cccccccc}
status&particle&$I(J^P)$&quark&$s$&$n,L$&$\sqrt{\la_Q}$&$\Delta M$\\
&&& content&&&[GeV]&[MeV]\\
\hline
obs& $D(1869)$&$\half(0^-)$&$c \bar q$&0&$0,0$&0.655& 0\\
obs&$D_1(2400)$&$\half(1^+)$&$c \bar q$&0&$0,1$&0.655&139 \\
obs&$\La_c(2286)$&$0(\half^+)$&$c q q$&$0$&$0,0$&0.655&4\\
%obs&$\Si_c(2453)$&$1(\half^+)$&$c q q$&$0$&$0,0$&0.655&171\\
obs&$\La_c(2595)$&$0(\half^-)$&$c q q$&$0$&$0,1$&0.655&-36\\
obs&$\La_c(2625)$&$0(\thalf^-)$&$c q q$&$0$&$0,1$&0.655&-6\\
obs&$\La_c(2880)$&$0(\fhalf^+)$&$c q q$&$0$&$0,2$&0.655&-59\\
pred& $D_2({\it 2630})$&$\half(2^-)$&$c \bar q$&0&$0,2$&0.655&?\\
pred& $D_2({\it2940})$&$\half(3^+)$&$c \bar q$&0&$0,3$&0.655& ?\\
\hline
obs& $D^*(2007)$&$\half(1^-)$&$c \bar q$&1&$0,0$&0.736& 0\\
obs& $D_2^*(2460)$&$\half(2^+)$&$c \bar q$&1&$0,1$&0.736& -29\\
obs&$\Si_c(2520)$&$1(\thalf^+)$&$c q q$&$1$&$0,0$&0.736&28\\
pred& $D_3^*({\it 2890})$&$\half(3^-)$&$c \bar q$&1&$0,2$&0.736& ?\\
pred&$\Si_c({\it 2890})$&$1(\fhalf^-)$&$c q q$&$1$&$0,1$&0.736&?\\
pred&$\Si_c({\it 2890})$&$1(\thalf^-)$&$c q q$&$1$&$0,1$&0.736&?\\
pred&$\Si_c({\it 2890})$&$1(\half^-)$&$c q q$&$1$&$0,1$&0.736&?\\
\hline
obs& $D_s(1958)$&$0(0^-)$&$c \bar s$&0&$0,0$&0.735& 0\\
obs&$D_{s1}(2460)$&$0(1^+)$&$c \bar s$&0&$0,1$&0.735&23\\
obs&$D_{s1}(2536)$&$0(1^+)$&$c \bar s$&0&$0,1$&0.735&73\\
obs&$\Xi_c(2467)$&$\half(\half^+)$&$c s q$&$0$&$0,0$&0.735&31\\
obs&$\Xi_c(2575)$&$\half(\half^+)$&$c s q$&$0$&$0,0$&0.735&113\\
obs&$\Xi_c(2790)$&$\half(\half^-)$&$c s q$&$0$&$0,1$&0.735&-67\\
obs&$\Xi_c(2815)$&$\half(\thalf^-)$&$c s q$&$0$&$0,1$&0.735&-41\\
pred& $D_{s2}({\it 2856})$&$0(2^-)$&$c \bar s$&0&$0,2$&0.735& ?\\
\hline
obs&$D^*_s(2112)$&$0(1^-) ?$&$c \bar s$&1&$0,0$&0.766& 0\\
obs&$D^*_{s2}(2573)$&$0(2^+) ?$&$c \bar s$&1&$0,1$&0.766& -29\\
obs&$\Xi_c(2646)$&$\half(\thalf^+)$&$c s q$&$1$&$0,0$& 0.766&28\\
obs&$D^*_{s3}({\it 3030})$&$0(3^-) ?$&$c \bar s$&1&$0,2$&0.766& 0\\
pred&$\Xi_c({\it 3030})$&$\half(\fhalf^-)$&$c s q$&$1$&$0,1$& 0.766&?\\
pred&$\Xi_c({\it 3030})$&$\half(\thalf^-)$&$c s q$&$1$&$0,1$& 0.766&?\\
pred&$\Xi_c({\it 3030})$&$\half(\half^-)$&$c s q$&$1$&$0,1$& 0.766&?\\
\end{tabular}}
\end{center}
\end{table}

\begin{table}
\caption{\label{tabbottom}  Bottom Hadrons. The notation is the
same as for Table. \ref{tabcharm}.}

\begin{center}
{\renewcommand{\arraystretch}{0.7}
\begin{tabular}{cccccccc}
status&particle&$I(J^P)$&quark&spin&$n,L$&$\sqrt{\la_Q}$&$\Delta M$\\
&&& content&&&[GeV]&[MeV]\\
\hline
obs& $B(5279)$&$\half(0^-)$&$b \bar q$&0&$0,0$&0.963& 0\\
obs&$B_1(5721)$&$\half(1^+)$&$b \bar q$&0&$0,1$&0.963&101 \\
obs&$\La_b(5620)$&$0(\half^+)$&$b q q$&$0$&$0,0$&0.963&1\\
obs&$\La_b(5912)$&$0(\half^-)$&$b q q$&$0$&$0,1$&0.963&-28\\
obs&$\La_c(5920)$&$0(\thalf^-)$&$b q q$&$0$&$0,1$&0.963&-20\\
pred& $B_2({\it 5940})$&$\half(2^-)$&$c \bar q$&0&$0,2$&0.963&?\\
\hline
obs& $B^*(5325)$&$\half(1^-)$&$b \bar q$&1&$0,0$&1.13& 0\\
obs& $B_2^*(5747)$&$\half(2^+)$&$b \bar q$&1&$0,1$&1.13& -45\\
obs&$\Si^*_b(5833)$&$1(\thalf^+)$&$b q q$&$1$&$0,0$&1.13&44\\
pred& $B_3^*({\it 6216})$&$\half(3^-)$&$c \bar q$&1&$0,2$&1.13& ?\\
pred&$\Si_b({\it 6216})$&$1(\fhalf^-)$&$c q q$&$1$&$0,1$&1.13&?\\
pred&$\Si_b({\it 6216})$&$1(\thalf^-)$&$c q q$&$1$&$0,1$&1.13&?\\
pred&$\Si_b({\it 6216})$&$1(\half^-)$&$c q q$&$1$&$0,1$&1.13&?\\
\hline
obs& $B_s(5367)$&$0(0^-)$&$b \bar s$&0&$0,0$&1.11& 0\\
obs&$B_{s1}(5830)$&$0(1^+)$&$b \bar s$&0&$0,1$&1.11&16\\
obs&$\Xi_b(5795)$&$\half(\half^+)$&$b s q$&$0$&$0,0$&1.11&-16\\
pred& $B_{s2}({\it 6224})$&$0(2^-)$&$b \bar s$&0&$0,2$&1.11& ?\\
pred &$\Xi_b(\it 6224)$&$\half(\half^-)$&$b s q$&$0$&$0,1$&1.11&?\\
pred &$\Xi_b(\it 6224)$&$\half(\thalf^-)$&$b s q$&$0$&$0,1$&1.11&?\\
\hline
obs&$B^*_s(5415)$&$0(1^-) ?$&$b \bar s$&1&$0,0$&1.16& 0\\
obs&$B^*_{s2}(5840)$&$0(2^+) ?$&$b \bar s$&1&$0,1$&1.16& -55\\
obs&$\Xi_b(5945)$&$\half(\thalf^+)$&$b s q$&$1$&$0,0$& 1.16&55\\
pred&$B^*_{s3}(6337)$&$0(3^-) ?$&$b \bar s$&1&$0,2$&1.16& ?\\
pred&$\Xi_b({\it 6337})$&$\half(\fhalf^-)$&$b s q$&$1$&$0,1$&1.16& ?\\
pred&$\Xi_b({\it 6337})$&$\half(\thalf^-)$&$b s q$&$1$&$0,1$&1.16& ?\\
pred&$\Xi_b({\it 6337})$&$\half(\half^-)$&$b s q$&$1$&$0,1$&1.16& ?\\
\end{tabular}}
\end{center}

\end{table}

\clearpage

%%%%%%%%%%%%%%%%%%%%%%%%%%

\subsection{Consequences of heavy quark symmetry (HQS) \lb{hqs}}
The decay constant $f_M$ of a meson is the coupling of the hadron
to its current. For the pion the decay constant $f_\pi$ is defined
as: \beq \langle 0|A_\mu|\pi(p)\rangle = i p_\mu \,f_\pi, \enq
where $A_\mu$ is the axial vector current.
In a bound state model for mesons it is related to the value of
the LF wave function at the origin~\cite{Brodsky:2007hb}.
 \beq \label{fm2} f_M=\sqrt{\frac{2
N_C}{\pi}}\int_0^1dx\, \ph^{LF}(x, \mbf{b}_\perp = 0). \enq which
is identical with the result first obtained by van Royen and
Weisskopf~\cite{VanRoyen:1967nq}.

It has been known for a long time \cite{Shuryak:1981fza}, and has
been formally proved in HQET \cite{Isgur:1991wq}, that for the
masses of heavy mesons with mass $M_M$ and a decay constant $f_M$
the product $\sqrt{M_M\,}\, f_M$ approaches, up to logarithmic
terms,  a finite value \beq \label{hqet} \sqrt{M_M\,} \, f_M \to
C. \enq

The wave function and hence $f_M$ depends on the scale $\la_Q$
and so we can, by \req{hqet}
 relate
the scale with the heavy quark mass (in the limit of large masses
the hadron mass equals the quark mass). We shall here not go into
details of the calculation but just quote the result of the
analysis in \cite{Dosch:2016zdv} The dependence of $f_M$ on the
scale $\la_Q$ and the quark mass $m_Q$ comes out to be \beq f_M
\sim\frac{\la_Q^{3/2}}{m_Q^2} \enq From that and \req{hqet} we
obtain: \beq \sqrt{\la_Q} \sim \sqrt{m_Q}. \enq In the limit of
heavy quarks the meson mass $M_M$  equals the quark mass $m_Q$:  $M_M=m_Q$ and
therefore \beq \lb{hq} \sqrt{\la_Q}\sim \sqrt{M_M} \enq

This corroborates our statement that the increase of $\la_Q$ with
increasing quark mass is dynamically necessary. In Fig.
\ref{lambdamass} we show the value of $\lambda_Q$ for the
$\pi,\,K,\,D,$ and $B$ meson  as function  of the meson mass
$M_M$. For the light quarks we are of course far away from the
heavy quark limit result \req{hq}, but it is remarkable  that the
simple functional dependence \req{hq} derived in the heavy quark
limit predicts for the $c$ quark a value $\sqrt{\la_c}=  0.653$
-- after fixing the proportionality constant in \req{hq} at the B
meson mass -- which is indeed at the lower edge of the values
obtained from the fit to the trajectories (0.655 to 0.766).

%%%%%%%%%%%%%%%%
\begin{figure}
\includegraphics[width=12cm]{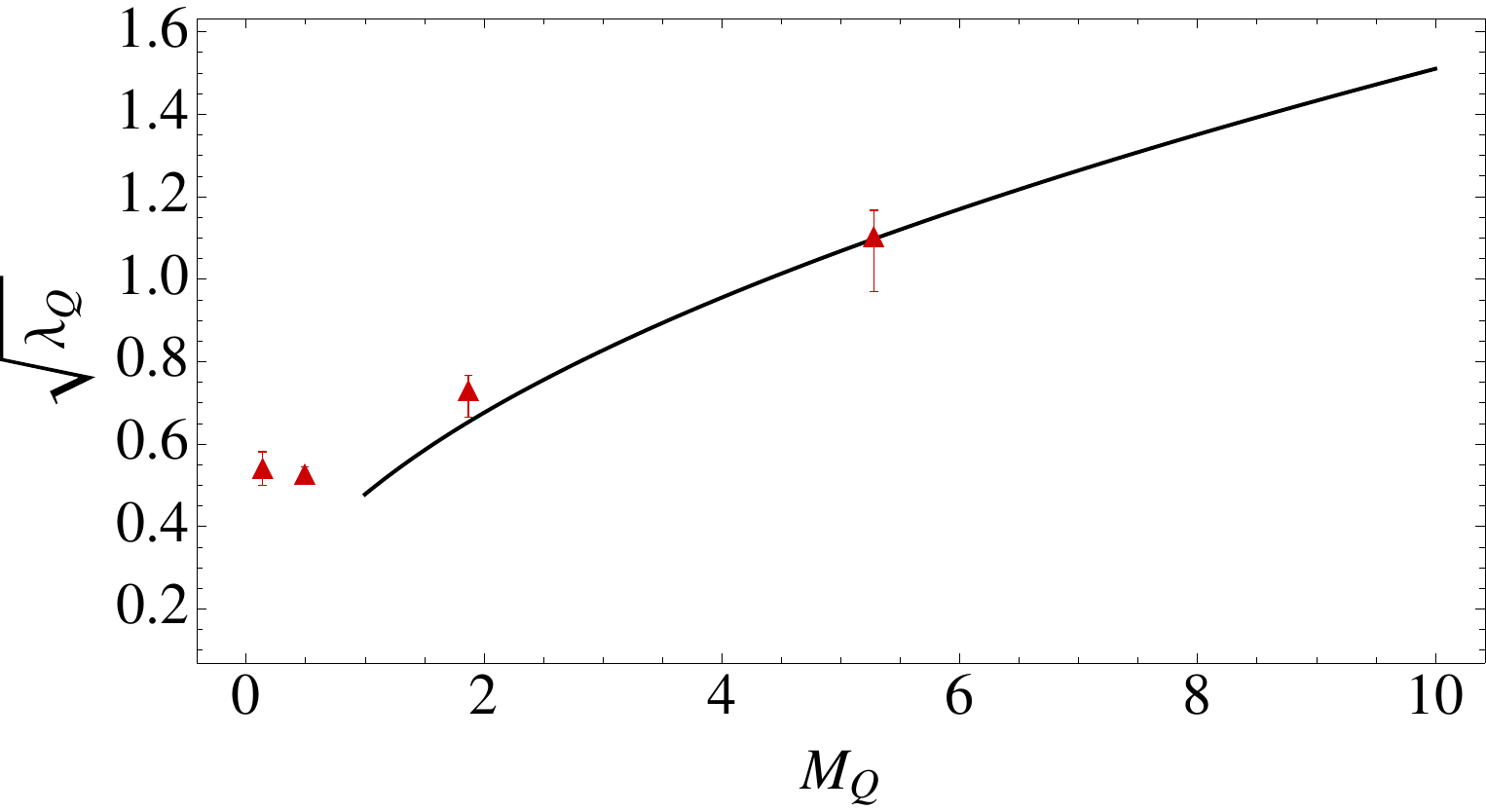}
\caption{\label{lambdamass} The fitted value of $\la_Q$ vs. the
meson mass $M_M$. The solid line is the square root dependence
\req{hq} predicted by HQET.}
\end{figure}
%%%%%%%%%%%%%%%%
%%%%%%%%%%%%%%%%%%%%%%%
%%%%%%%%%%%%%%%%%%%%%%%
\section{Extension to two heavy quarks}

The extension of the meson baryon supersymmetry to hadrons
containing two heavy quarks is very speculative and can by no
means inferred from the results of the previous investigations. A
system consisting of two light quarks and one consisting of a
light and a heavy quark are both ultra-relativistic, whereas a
system consisting of two heavy quarks is closer to nonrelativistic
dynamics. Therefore the statements made in this subsection are
taken only as propositions worth testing, but not as predictions
of the model.

The most obvious consequence of supersymmetry in this double-heavy
sector is the existence of double charmed baryon states
Ξ$\Xi_{cc}$ with a mass of approximately 3550 GeV, that is
approximately the same mass~\cite{Brodsky:2016yod} as the mesons $
h_c(1P)(3525)$ and the $\ch_{c2}(1P)(3556)$. Indeed a weakly
decaying doubly charmed baryon  has been found at
LHCb~\cite{Aaij:2017ueg} with a mass of 3614 GeV, within the
expected accuracy very well compatible with the that of the
expected superpartner $ h_c(1P)(3525)$  or  $\ch_{c2}(1P)(3556)$.
This gives some weight to the prediction of a double-bottom baryon
$\Xi_{bb}$ with a mass of ca 9900 MeV, that of  the mesonic
superpartners
 $h_b(1P)(9899)$ or  $\ch_{c2}(1P)(9912)$. Since one expects P-wave excitations of the $B_c(6276)$ at $\approx 6300$ MeV, supersymmery applied to this sector does predict  $\Xi_{bc}$ states at  $\approx 6300$ MeV.

%%%%%%%%%%%%%%%%%%%%
%%%%%%%%%%%%%%%%%%%%%
\subsection{Completing he supermultiplet in the heavy hadron sector \lb{smh}}
In sect. \ref{tqs} we have seen that the supersymmetric partner of
the negative chirality component of the baryon is naturally
interpreted as a tetraquark. We shall not repeat all the arguments
brought there in favour of that interpretation. We shall only
focus on the new situation which occurs if one or two quarks are
heavy and therefore the constituents cannot be treated on equal
footing. The situation is graphically represented in Fig.
\ref{scheme}. In A) we show the situation  discussed in sect.
\ref{tqs}, in B) the situation  where one quark is heavy, it is
not essentially different from A). A new element comes in for the
case of two heavy quarks, displayed under  C). Here we can
construct a tetraquark with hidden charm or beauty, or one with
double open charm or beauty. The latter case is very interesting,
since the predicted double charmed or double-bottom tertraquarks
could be stable against strong interactions. Their expected mass,
which is equal to that of the $L=1$ mesons, would be below the
threshold of a decay into two  mesons with open charm or beauty.
The main argument against the observation of tetraquarks, namely
the easy hadronization, is in these cases not applicable. Such a
situation has also been predicted for a double-bottom tetraquark
by Karliner and Rosner~\cite{Karliner:2017qjm}.
\begin{figure}
\includegraphics*[width=10cm]{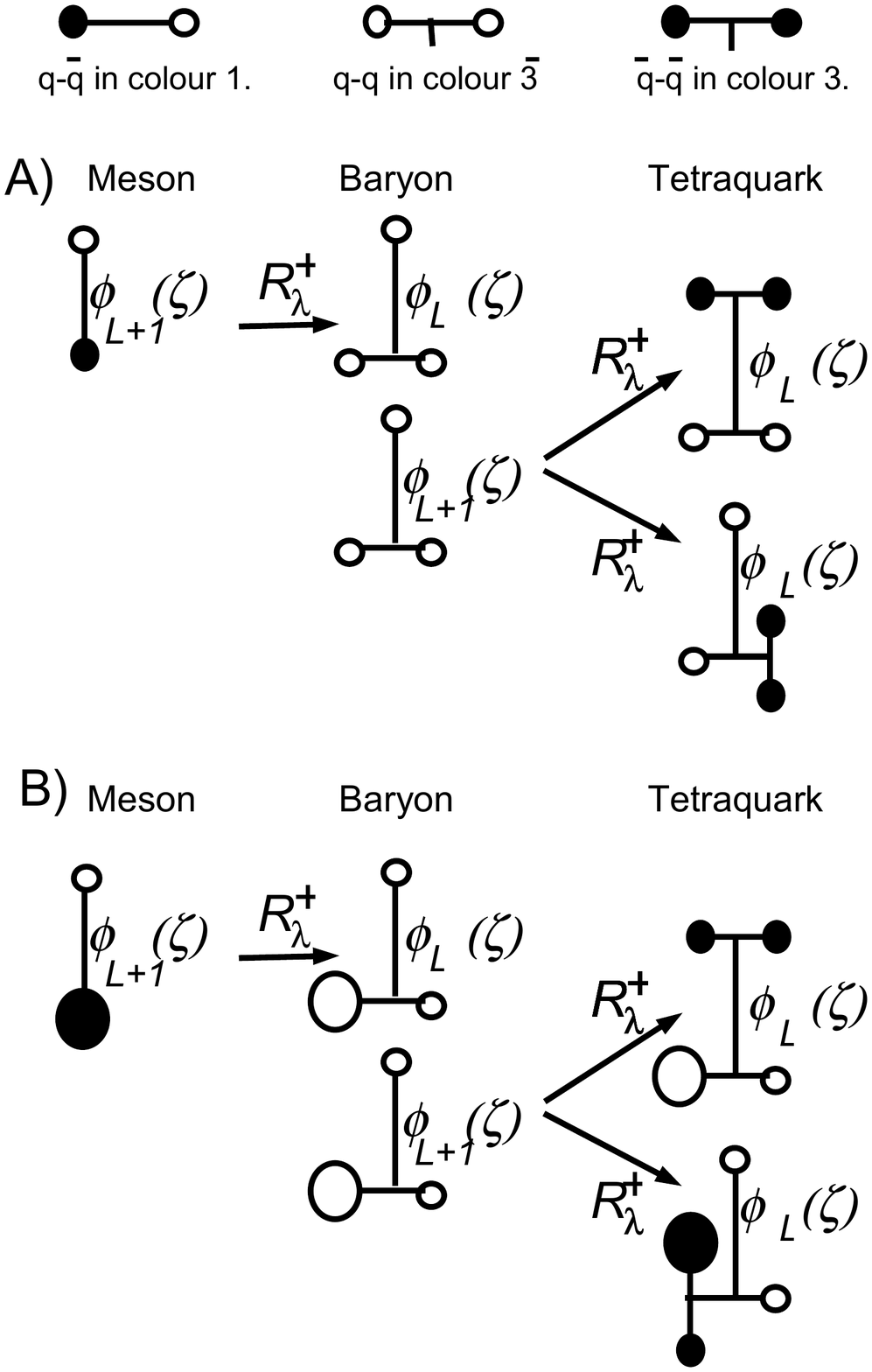} \hspace{-2 cm}
\includegraphics*[width=10cm]{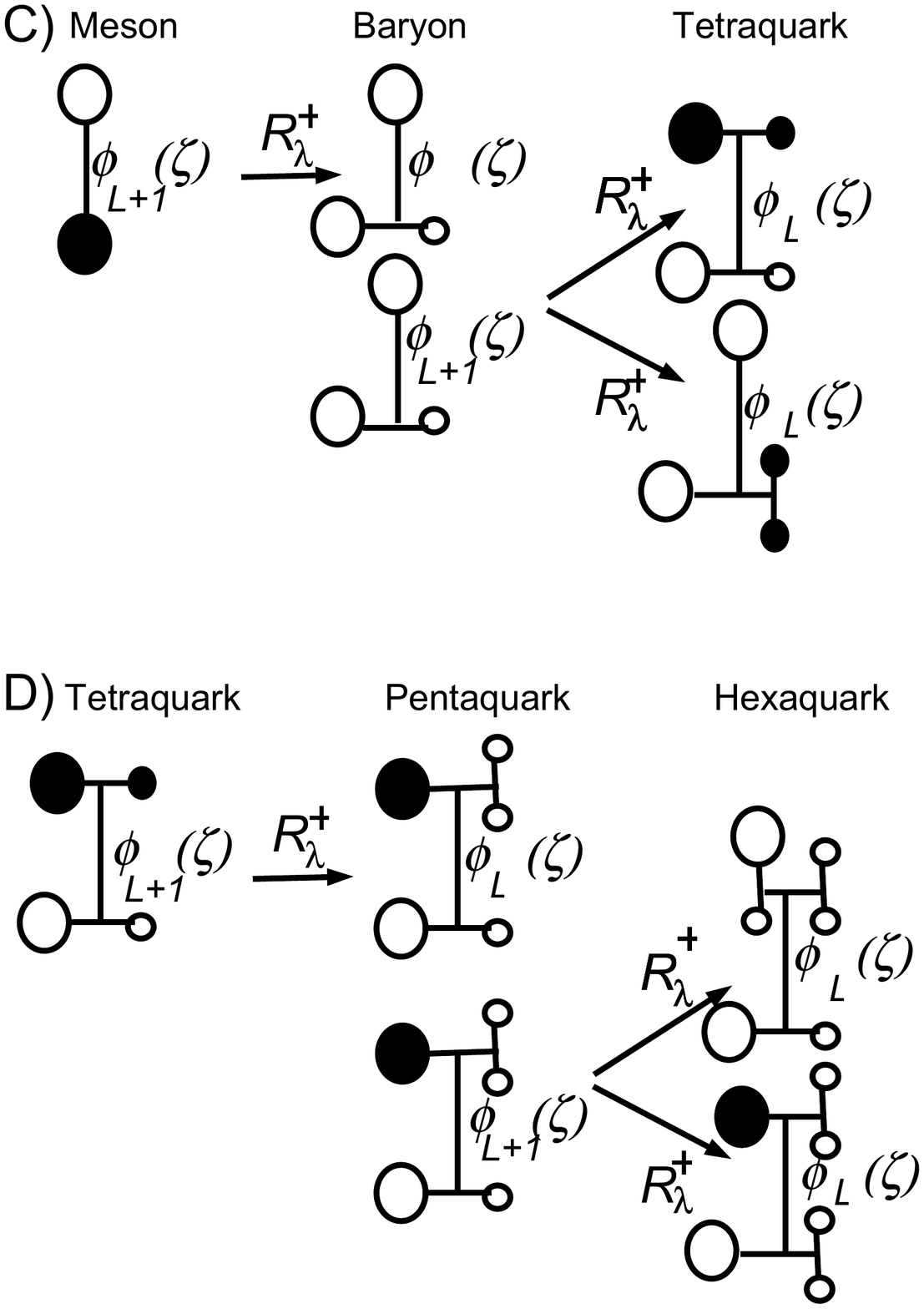}
\caption{\lb{scheme} The supersymmetric multiplets in SUSY LFHQCD.
Open circles denote quarks, filled circles antiquarks, small
circles light quarks, large circles heavy quarks. The supercharge
$R_\la^\dagger$, \req{Rex}, changes the fermion number by $\pm 1$
and lowers the angular momentum of the wave function by one unit.
A) and B) depict the case with none or one heavy quark, C) the
case with two heavy quarks; in that case either a light quark is
transformed into a light antiquark pair, or a heavy quark into a
heavy antiquark pair. In D) a possible scenario is indicated,
where the lowest member of a multiplet is a tetraquark, from which
a penta- and hexa-quark can be constructed. A quark and an
antiquark connected by a straight line are in a colour singlet
state; two quarks or two antiquarks connected by a line with
intersection are in a colour $\bar 3$ or 3 representation,
respectively.}
\end{figure}

\section{Summary}
If one imposes superconformal symmetry, more precisely expressed:
if one demands that the generalized LF Hamiltonian is constructed
from the generators of the (graded) superconformal algebra, all
the problems mentioned in the summary of the last chapter are
resolved. The LF potentials  for mesons and baryons are
unequivocally determined and we predict supermultiplets consisting
of a meson with angular momentum $L_M$, a baryon with angular
momentum $L_B=L_M-1$, and a bosonic state, which is plausibly
interpreted as tetraquark, with angular momentum $L_B$; the scale
parameter $\la$ is the same for mesons and baryons. The relations
in the meson-baryon sector, encoded in the mass formulae
\req{mesbarfin} are satisfied by the data  within the expected
accuracy of $\approx 100$ MeV, see Figs. \ref{pirhondel} and
\ref{strangehad}. Supersymmetry is also satisfied for hadrons
containing one heavy quark and perhaps even for hadrons with two
heavy quarks, see sect. \ref{ssh}.  Combining the conditions of
supersymmetry with those of AdS one concludes that  also hadrons
containing a heavy quark lie on linear trajectories, see sect.
\ref{lt},   Fig. \ref{c-hybr} and \ref{b-hybr}. Heavy quark symmetry
allows to estimate the relation between the $\la$-value for a
system containing a $c$-quark and that for a system containing a
$b$-quark, see Fig. \ref{lambdamass}. We again emphasize that the
supersymmetry relates wave functions of observed mesons and
observed  baryons. There is no need nor motivation  to introduce
new particles like squarks or gluinos.

%%%%%%%%%%%%%
%\end{document}
%%%%%%%%%%%%%%%%%

%%%%%%%%%%%%%%%%%%%%%%%%%%%%
\chapter{The propagator from AdS\lb{c6}}

\section{The two point function in Holographic QCD}
\subsection{The generating functional}
As shortly mentioned in the introduction, the generating
functional (partition function in statistical mechanics)  allows
 to calculate all possible matrix
elements and therefore all possible observables.

The generating functional for a quantum field $\phi$ is defined as

\beq Z[j] = \int \cD \phi \, e^{i\int d x j(x) \,\phi(x)} \,
e^{iS[\phi]} \lb{genfun}\enq Here $S[\phi]$ is the AdS  action
and $\cD \phi$ denotes the functional integration over all fields
$\phi$.

Unfortunately the functional integration can be performed
analytically only for a Gaussian integrand. This  is  the basis
for perturbation theory, but for more general cases ones has to
use numerical methods (lattice calculations).

From the generating functional one can obtain all n-point
functions by functional derivation. The latter -- in contrast to
functional integration -- is elementary and can in general be
performed analytically. Functional derivatives  occur also in
classical field theory, for instance for the derivation of the
classical equations of motion, see Chapt. 3.1. Indeed one can
often ignore that a derivative  is a functional derivative.  One considers the function, after which is derived,  as a variable and use the
rules of ordinary differentiation.

The expression for the n-point function is: \beq \langle \cT
\phi(x) \phi(y) \phi(z) \dots \rangle =(-i)^n\,\left[ \left(\frac{\de}{\de
j(x)}   \frac{\de}{\de j(y)}  \frac{\de}{\de j(z)} \dots
\right)Z[j] \right]_{/ j=0} \enq

Normally we are interested only in the connected n-point
functions, corresponding to connected Feynman diagrams. These are
obtained from functional differentiation of the connected
generating functional $W[j]$ which is the logarithm of $Z[j]$:
\beq  W[j]= \log Z[j]. \enq

Ads/CFT allows to calculate the generating functional of the full
quantum field theory (in the $N_c\to \infty$ limit) by solving the
classical equations of motion of the 5-dimensional gravitational
theory~\cite{Gubser:1998bc,Witten:1998qj}. One has to insert the
classical solution, $\phi(x,z=\ep)$ into the action and perform the
limit $\ep \to 0$. \beq \lb{fun} W[j] =i\,  \lim_{\ep\to 0}
S[\ph_{\it cl}]_{/ \lim_{\ep \to 0} \ph_{\it cl}(\ep)= j} \enq
This essentially states that in the functional integral
\req{genfun} only the fields which minimize the action, that is
the classical solutions, contribute. We cannot insert directly the
classical solution at $z=0$ into \req{fun} since it contains
infinities,
 as it is to be expected in a quantum field theory. We have to insert the
  classical solution evaluated a finite value  $z=\ep$ and the limit $\ep \to 0$
   corresponds to the renormalization procedure. Generally it can be performed analytically
   as we  shall see later.

\subsection{The classical action}
The classical action for mesons with arbitrary spin is (see sect.
\ref{bsar}, \req{action2})
 \beqa \lb{action2-b} S_{\it eff} &=&
\int d^{d} x \,dz \,\sqrt{g}  \; e^{\vp(z)} \, \,g^{N_1 N_1'}
\cdots  g^{N_J N_J'}
\Big(  g^{M M'} D_M \Phi^*_{N_1 \dots N_J}\, D_{M'} \Phi_{N_1 ' \dots N_J'}  \nn \\
&& \qquad - \mu_{\it eff}^2(z)  \, \Phi^*_{N_1 \dots N_J} \,
\Phi_{N_1 ' \dots N_J'} \Big). \enqa

The covariant derivatives $D_M$ of higher rank tensors are very
complicated (see sect. \ref{adsmetric},footnote \ref{fn1}) but, as
has been shown in \cite{deTeramond:2013it}, it turns out,   that
in the end one can largely ignore the complications and work as if
the covariant derivatives were usual partial derivatives.

We proceed in the usual way,% see sect \ref{bsa}. :

1) we Fourier transform the solution of the Euler-Lagrange equations over the 4 Minkowski variables:\\$ \tilde
\Phi_{N_1  \dots N_J}(q,z) = \int d^4x e^{-iqx} \Phi_{N_1 ' \dots
N_J'}(x,z)$

2) we extract the spin content into a spin tensor:\\
$\tilde \Ph(q,z)  = \vep_{N_1 \cdots N_J} \tilde \Phi(q,z)$ where
the spin tensor is totally symmetric in all indices and fulfills
the conditions: \beq \lb{60} q^{N_1} \, \vep_{N_1 \cdots
N_J}(q)=0, \quad \eta^{N_1 N_2}\vep_{N_1 N_2\cdots N_J}(q)=0 \enq
For the classical solutions in the 4 dimensional space (i.e. $z =
0$) we have $ \vep_{5,N_2,  \cdots N_J}=0$, that is the
polarization tensor vanishes if at least one of the indices ${N_i}
=5$ . The factors in the integrand of \req{action2-b} are
$\sqrt{g} = \left(\frac{R}{z}\right)^5,\,\;g^{N_1
N_1'}=\left(\frac{z}{R}\right)^2 \, \et^{N_1 N_1'}$ and we obtain
from \req{action2-b}: \beq  S = \int \frac{d^4q}{(2 \pi)^4}
S(q)\enq with \beq \lb{61} S(q)= X_{\it spin}  \int_\ep^\infty
dz\, \left(\frac{R}{z}\right)^{3-2J}  e^{\vp(z)} \Big[-\pa_z\tilde
\Phi^*\pa_z\tilde \Phi -\left(-q^2 + \frac{(\mu R)^2}{z^2}\right)
\tilde \Phi^*\tilde \Phi \Big] \enq With the spin term \beq X_{\it
spin} = \si^{\nu_1 \cdots \nu_J, \nu'_1 \cdots \nu'_J} \ep_{\nu_1
\cdots \nu_J} \ep_{\nu'_1 \cdots \nu'_J}\enq The term $\si^{\nu_1
\cdots \nu_J, \nu'_1 \cdots \nu'_J}$ is due to the conditions
\req{60}. For $J=1$ it is \beq \lb{spf1} \si^{\nu\nu'} =
\et^{\nu\nu'} - \frac{q^{\nu}\,q^{\nu'}}{q^2} \enq We abbreviate:
\beq A(z) =\left(\frac{R}{z}\right)^{3-2J}  e^{\vp(z)}; \;B(z)
=\left(-q^2 + \frac{(\mu R)^2}{z^2}\right) \enq then the  the
action reads: \beq\lb{62} S(q)= X_{\it spin}\int_\ep^\infty dz\,
\left(-A(z)\pa_z\tilde \Phi^*\pa_z\tilde \Phi -B(z) A(z)
\tilde \Phi^*\tilde \Phi \right) \enq
 The Euler Lagrange equations, of which the classical field $\tilde \Phi_{\it cl}$ is a solution, are
\beq \pa_z \frac{\pa \cL}{\pa(\pa_z\tilde \Phi^*})=\frac{\pa
\cL}{\pa \tilde \Phi^*} \lb{63-a} \enq For solutions of the Euler
Lagrange equations, that is the classical solutions $\tilde
\Phi_{\it cl}(q,z)$ we have therefore \beq -\pa_z\Big(A\,\pa_z
\tilde \Phi_{\it cl}\Big)=-A \,B \,\tilde \Phi_{\it cl}\lb{63-b}
\enq We perform a partial integration for the first term of
\req{62}:

\beq - \int_\ep^\infty dz\, A(z) \pa_z\tilde \Phi^*\pa_z\tilde
\Phi =  \int_\ep^\infty dz\, \tilde \Phi^*\pa_z(A \,\pa_z \tilde
\Phi ) - \Big[\tilde \Phi^* A \; \pa_z \tilde \Phi
\Big]^\infty_\ep \enq Inserting  \req{63-b} into \req{62} we
obtain: \beq S(q) =   \int_\ep^\infty dz\, \tilde \Phi_{\it
cl}^*(A B - AB )\tilde \Phi_{\it cl} -\Big[ \tilde \Phi_{\it
cl}^*\,A \; \pa_z \tilde \Phi_{\it cl} \Big]^\infty_\ep \enq The
classical action is therefore reduced to a surface term.

Collecting everything we we obtain for the full action with the
classical solution inserted:
     \beq S[\tilde \Phi_{\it cl}]= -
X_{\it spin} \int \frac{d^4q}{(2\pi)^4}
\Big[\left(\frac{R}{z}\right)^{3-2J}  e^{\vp(z)} \tilde \Phi_{\it
cl}^*(q,z)\pa_z \tilde \Phi_{\it cl}(q,z) \Big]^\infty_\ep \enq
      We
consider, as usual in field theory, only  solutions which vanish
at $z\to \infty$ and therefore only the lower surface term
contributes and we end up with
\beq S[\tilde \Phi_{\it cl}]=
X_{\it spin} \int \frac{d^4q}{(2\pi)^4}
\Big[\left(\frac{R}{z}\right)^{3-2J}  e^{\vp(z)} \tilde \Phi_{\it
cl}^*(q,z)\pa_z \tilde \Phi_{\it cl}(q,z)_{\big/ z= \ep} \enq

In order to fulfill the condition $\lim_{\ep \to 0} \ph_{\it
cl}(q,\ep)= j(q)$ in \req{fun} we choose the normalization \beq
\tilde \Phi_{\it cl,\nu_1 \cdots \nu_J}(q,z) =\frac{ \tilde
\Phi_{\it cl}(q,z)}{ \tilde \Phi_{\it cl}(q,\ep)}\; j_{\nu_1
\cdots \nu_J}(q) \enq

So our final result is: \beqa
W[j] &=& i\, \lim_{\ep\to 0}  S[\ph_{\it cl}]_{/ \lim_{\ep \to 0} \ph_{\it cl}(\ep)= j}\\
&=& i\, \lim_{z\to \ep \to 0}\Bigg[ \si^{\nu_1 \cdots \nu_J, \nu'_1 \cdots \nu'_J} \times \nn \\
&& \int \frac{d^4q}{(2 \pi)^4} \left(\frac{R}{z}\right)^{3-2J}
e^{\vp(z)} j^*(q)_{\nu_1 \cdots \nu_J} \frac{ \tilde \Phi^*_{\it
cl}(q,z)}{ \tilde \Phi^*_{\it cl}(q,\ep)}
j(q)_{\nu_1' \cdots \nu_J'}\, \frac{\pa_z \tilde \Phi(q,z)}{\tilde \Phi(q,\ep)}\Bigg]\nn\\
&=& i\, \si^{\nu_1 \cdots \nu_J, \nu'_1 \cdots \nu'_J} \, \int
\frac{d^4q}{(2 \pi)^4} e^{\vp(z)} j^*(q)_{\nu_1 \cdots \nu_J}
j(q)_{\nu_1' \cdots \nu_J'}\, \lim_{z\to \ep \to 0}
\left(\frac{R}{z}\right)^{3-2J}  \frac{\pa_z \tilde
\Phi(q,z)}{\tilde \Phi(q,\ep)}\nn \enqa

The connected two point function is given by \beqa  \lb{final1}
\Si_{\nu_1 \cdots \nu_J,\nu_1' \cdots \nu_J'} (q) &=&
(-i)^2\, \frac{\de}{\de j^*_{\nu_1 \cdots \nu_J}(q)}
\frac{\de}{\de j_{\nu_1 \cdots \nu_J}(q)}W[j]\nn \\
&=& \frac{-i}{(2\pi)^4} \, \si^{\nu_1\nu_1' \cdots \nu_J\nu_J'} \left(\frac{R}{z}\right)^{3-2J}  e^{\vp(z)}\left(\frac{\pa_z \tilde \Phi(q,z)}{\tilde \Phi(q,\ep)}\right)\nn \\
&=&  \frac{-i}{(2\pi)^4} \, \si^{\nu_1\nu_1' \cdots \nu_J\nu_J'} \Si(q^2)
\enqa with \beq \lb{final2}
\Si(q^2)=\left(\frac{R}{\ep}\right)^{3-2J}
e^{\vp(\ep)}\left(\frac{\pa_z \tilde \Phi(q,\ep)}{\tilde
\Phi(q,\ep)}\right) \enq It is always understood that $\ep \to 0$.

As to be expected in QFT there will be infinities occurring in the
limit $\ep \to 0$. These have to be treated by some
renormalization (this is not special to perturbation theory, but
also has to be done in lattice regularization, e.g.). If the
infinities are independent of $q$, they are not dynamically
relevant and can be discarded. This will be done in the following.
%%%%%%%%%%%%%%%%%%%%%%%%%%%%%%%%%%%%
%%%%%%%%%%%%%%%%%%%%%%%%%%%%%%%%%%%%%%
\section{Soft wall model}
\subsection{Solutions for soft wall model}
Here we take $\vp(z) = \la z^2$, that is the starting point is the
same as in sect. \ref{bsar}.  The Euler Lagrange equation for the
Lagrangian \req{action2-b} is: \beq \lb{ELG}
 \Big[ \pa_z^2 + \frac{1}{z}(2 J-3 + 2 \la z^2) \pa_z +q^2-\frac{(\mu R)^2}{z^2}\Big] \tilde \Phi(q,z)= 0
\enq where in LFHQCD the AdS mass $\mu$ is related to the LF
angular momentum $L$ by (see \req{muL1})
 \beq  L^2= (\mu
R)^2+(J-2)^2 \lb{muL}.
 \enq
In sect. \ref{bsar} we were interested
in bound state solutions, which  must be normalizable. This is only
possible for discrete values of $q^2$ which led us to the hadronic
spectra. If we give up the normalizability condition, we shall
find a larger set of solutions for every value of $q^2$.

Mathematica finds a rather complicated solution for \req{ELG}. So
it is better to rescale $\tilde \Phi$  and bring \req{action2-b}
into a form which has relatively simple and mathematically well
investigated  solutions.

A rather general and well investigated form for a differential
equation with first and second order derivatives is Kummer's
equation \cite{AS}, 13.1.1: \beq \lb{kummer} y \, w''(y) +(b-y)
w'(y) - a\, w =0 \enq with well studied  solutions. This form can
be  obtained from \req{ELG} by rescaling.

We rescale \beq \tilde \Phi(q,z)=z^\be  \tilde \Ph_n(q,z) \enq and
obtain \beq \lb{ELGn} \Big[ \pa_z^2 + \frac{1}{z}(2 J-3 +
2\be+2\lambda z^2) \pa_z +q^2+C \Big] \tilde \Phi_n(q,z)= 0 \enq
where \beq C= \frac{\be^2+ 2 \be (J-2) - (\mu R)^2}{z^2} +2 \be\la
+ q^2 \enq We can make the term $\sim 1/z^2$ vanish by choosing:
\beq \lb{LLmu}
 \be = 2 + L- J, \quad  L= \sqrt{(J-2)^2 + (\mu R)^2}
\enq We  introduce the dimensionless variable \beq y= |\la| z^2
\enq and use: $\pa_z f(y) = \pa_y f(y) 2 |\la| z, \quad \pa_z^2
f(y) = \pa_y^2 f(y) 4 \la^2 z^2  + \pa_y f(y) 2 |\la| $

Then we obtain with  choice $\be=2+L-J$ from  \req{ELGn} the
differential equation: \beqa \lb{EL-final}
y \,\pa^2_y \, \tilde \Phi_n (q,y)\!\! &+&\!\! \left( (L +1) + \frac{\la}{|\la|} \, y\right) \pa_y \tilde \Phi_n (q,y) +\\
 &&\frac{1}{4 |\la|}
\Big((4 - 2 J + 2 L) \la  + q^2)\Big) \Phi_n (q,y)=0\nn \enqa

which has indeed the form of Kummer's equation  \req{kummer}

The solutions of \req{kummer} which vanish for $z\to \infty$ are:
\beq w(y) = U(a,b,y). \lb {solk1} \enq $U(a,b,y)$ is the
hypergeometric function (HypergeometriU[a,b,y] in Mathematica).

For the case of a positive  sign in front the the term $y\,w'(y)$
that is for the equation
   \beq
y \, w''(y) +(b+y) w'(y) - a\, w =0 \enq
   the solution is
   \beq  w(y) = e^{-y}\,U(a+b,b,y) \lb {solk2} \enq
    The solution $\tilde \Phi (q,y)$ of \req{ELG} has thus the form:
\beq \lb{final3} \tilde \Phi(q,z)=
\rho(z)U\left(a_\la,L+1,|\la|z^2\right)\enq
     with $ \rho(z) = z^{L-J+2} e^{-(|\la|+\la) z^2/2}$.

The constant $a_\la$ depends on the sign of $\la$:

for $\la <0 $ we have \req{solk1} and from \req{EL-final} \beqa
\la <0 \quad  a_\la &=& -\frac{q^2}{4 |\la|} - \frac{\la}{4 |\la|}(4-2J+ 2L) \nn \\
 \lb{a-} &=& -\frac{q^2}{4 |\la|} + \frac{1}{4}(4-2J+ 2L)
\enqa for $\la >0 $ we have \req{solk2} and from \req{EL-final}
\beqa
 \la>0 \quad  a_\la &=& -\frac{q^2}{4 |\la|} -\frac{\la}{4 |\la|}(4-2J+ 2L)+(L+1) \nn\\
 &=& -\frac{q^2}{4 |\la|} + \frac{1}{4}(2J+ 2L)
  \lb{a+}
\enqa

\subsection{The propagator for the conserved current in the holographic soft wall model \lb{psw}}

We use the general result \req{final2} and insert \req{final3}
(omitting terms, which are 1 in the limit $\ep\to 0$. \beqa
\Si[q^2] &=& \left(\frac{R}{z}\right)^{3-2J}\pa_z[\log[\rh[z] U(a_\la ,L+1,\la z^2)]_{/z\to \ep}\\
&=&\left(\frac{R}{z}\right)^{3-2J}\pa_z[\log[U(a_\la ,L+1,\la z^2)]_{/z\to \ep}+\pa_z[\log[\rh[z] ]_{\big/ z= \ep \to 0}\\
&=& \left(\frac{R}{z}\right)^{3-2J}\frac{U'(a_\la,L+1,|\la| z^2)\,
2 |\la| z}{U(a_\la,L+1,|\la| z^2) }_{\big/ z= \ep \to 0}+
\makebox{non-dynamical terms} \enqa where non-dynamical terms are
(possibly infinite) terms which are independent of $q$.

We see that multiplying the solution \req{final3} with a function
$f(z)$ which is regular at $z=0$ has no influence on the two point
function.

In order to perform the limit $\ep \to 0$ we have to expand the
ratio $U'/U$ around $z=0$. It is most transparent to use the
identity (\cite{AS}, 13.4.21) \beq  U'(a,b,y)= - a
\,U(a+1,b+1,y)\enq from which we get the final
result\cite{Jugeau:2013zza,Brodsky:2014yha}: \beq \lb{finalfinal}
\Si[q]= \left(\frac{R}{z}\right)^{3-2J}{\frac{-a_\la
\,U(a_\la+1,L+2,|\la| z^2)\; 2 |\la| z}{U(a_\la,L+1,|\la|
z^2)}}_{\big/ z= \ep \to 0}\enq

First we consider the AdS conserved vector current, i.e $J=1,\;
\mu=0$;  from $L^2=(J-2)^2 + (\mu R)^2 $, see \req{LLmu},  follows
that $L=1$: Both for $\la >0$ and $\la <0$ we obtain: \beq a_\la =
-\frac{q^2}{4 |\la|} +1 \enq

We insert into mathematica, with $y=|\la| z^2$:
\begin{quote}
{\tt Simplify[
 Series[a  HypergeometricU[a + 1, 3, y]/HypergeometricU[a, 2, y],(y,
   0, 0)]]}
\end{quote}

   The result is

   \beq
   \frac{1}{y}-(a-1) (\psi ^{(0)}(a)+\log (y)+2 \gamma )+O\left(y^1\right)
   \enq

The function $\psi^{(0)}(x) = \pa_x \log( \Ga[x])$  is a
meromorpic function with poles at $ x=0, -1, -2 \dots$ It is
called psi-function, Digamma function or Polygamma function with
first argument 0, (Polygamma[0,x] in mathematica), we come later
to it.
% (AS 6.4).
%\includegraphics[width=20cm]{2-punkt-softwall.pdf}

The first term is diverging, but independent of $q^2$ and
therefore not dynamical, and we discard it. Putting pieces
together we obtain: \beqa
 \Si(q^2) &=&  - \frac{R}{z}\, ( 2 |\la| z) \frac{q^2}{4|\la|}\,\left(  \psi^{(0)}\big(-\frac{q^2}{4| \la|}+1\big)+
\log(|\la| z^2))\right)_{\big/ z=\ep} +  \makebox{nondynamical terms} \nn \\
&=&  - q^2 R\left(\half \psi^{(0)}\big(-\frac{q^2}{4 \la}+1\big) +  \log \ep \right)+  \makebox{nondynamical terms}
\enqa $q^2 \log \ep$ is an infinite substraction term

Normally the factor $q^2$ is absorbed  in the spin factor
\req{spf1} of the propagator which then becomes $q^2 \eta^{\mu\nu}
-q^\mu q^\nu $, we therefore obtain for the full vector current
propagator, see \req{final1}, discarding all nondynamical terms:
     \beq \lb{propfinal} \frac{2}{R} \Si_{\mu \nu}(q^2) = \frac{i}{ (2 \pi)^4}(
q^2 \eta^{\mu\nu} -q^\mu q^\nu) \, \psi\big(-\frac{q^2}{4
\la}+1\big) \enq

As mentioned above, the Digamma function  has no cuts, but only
poles, which correspond to the  intermediate states of the
two-point function.

\subsection{Physical relevance of conserved current \lb{prcc}}

The conserved current is a vector current ($J=1$) and therefore it
is natural to associate the poles in the Digamma function with the
vector particles ($\rho(770),\rho(1450)$ etc.). This has indeed be
assumed in the seminal paper of Karch, Katz, Son and
Stephanov~\cite{Karch:2006pv}.
 The scale $\la$ is then related to $M_\rho$, the mass of the $\rho(770)$ by
$
 M_\rh^2= 4 \la
$.
 On the other hand the value  of the slope $\frac{dM_\rh^2}{dJ}$ comes also out to be~\cite{Karch:2006pv} $4 \la$.
\begin{figure}
\includegraphics[width=5cm]{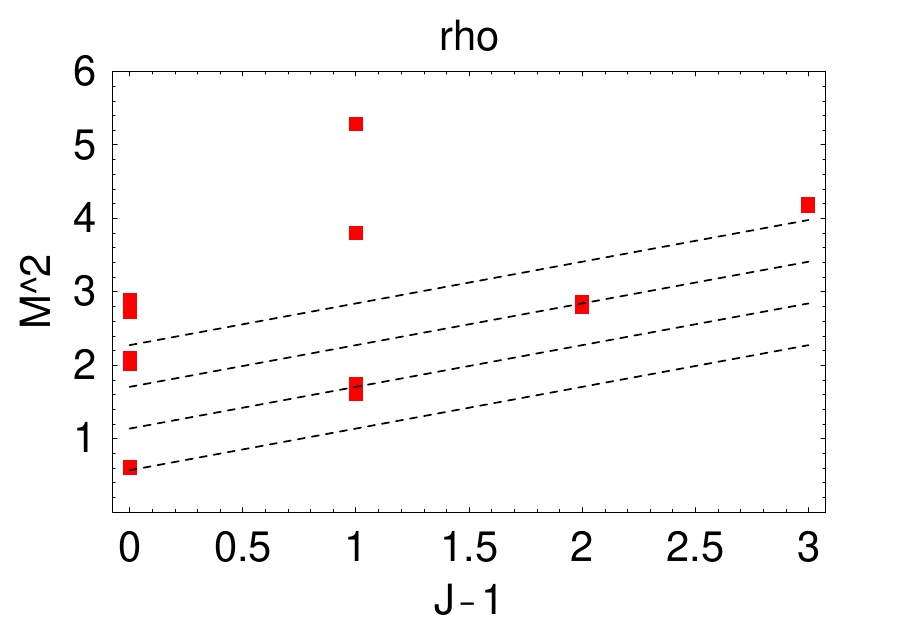} 
\includegraphics[width=5cm]{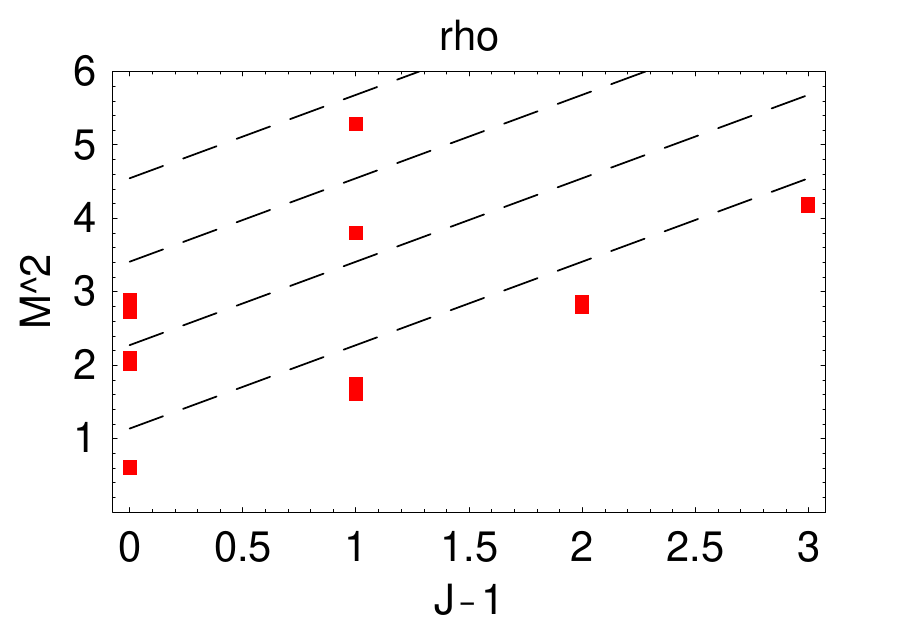} 
\includegraphics[width=5cm]{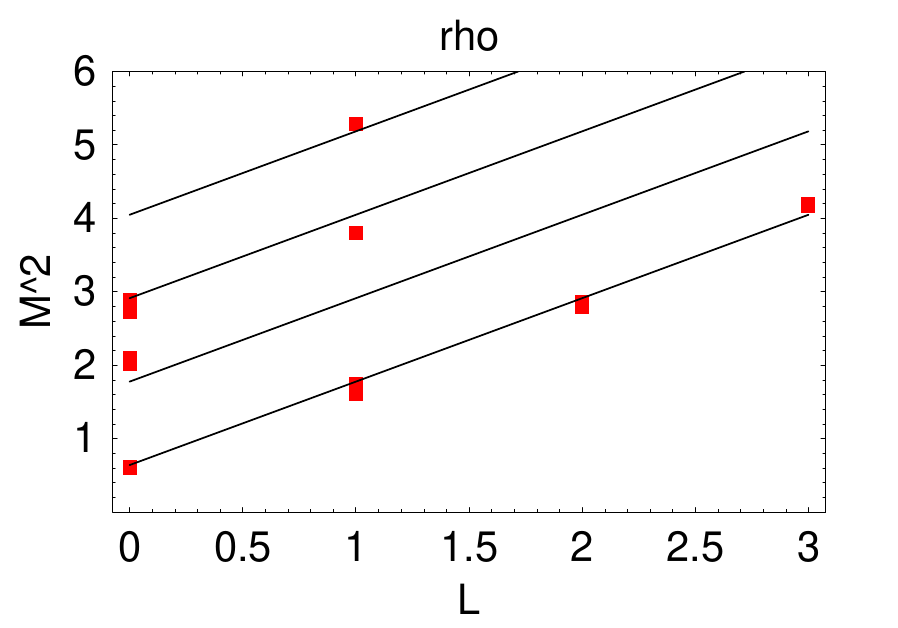}
\caption{\lb{rho-mod} a) and b) The rho-trajectory from the
conserved vector current ($J=L=1$) with correct rho mass a) and
with  with correct slope c) , Fig. c the trajectories from LFHQCD,
with $J=1,\,L=0$.  }
\end{figure}
From  Fig. \ref{rho-mod} one sees that both assignments are not
compatible with each other. Either the mass of the rho comes out
correctly and the slope is wrong,   (\ref{rho-mod}.a) or the slope
is correct, and the mass comes out to large,  (\ref{rho-mod}.b).
So this interpretation is not well compatible with the data.

In LFHQCD the $\rho$-field must have $J=1, \,L=0$, since it has
parity $1^-$. In that case the squared mass of the $\rho$ is $2
\la$ and the slope $4\la$, see \req{spff} or \req{mesbarfin}. As
can be seen from Fig. \ref{pirho}, repeated as (\ref{rho-mod}.c)
the agreement with the data is in that  case much better.

\subsection{\lb{exth} Propagators for other  currents in LFHQCD}

 For the rho current with $J=1,\,L=0$ we have from  \ref{a+}:
 \beq a_\la =\frac{Q^2}{4 |\la| } +\frac{1}{4}(2J+2L) = \frac{Q^2}{4 |\la| } + \half \enq and we obtain for the propagator function
 \req{finalfinal}
\beq \lb{LF-1} \Si[q]= \left(\frac{R}{z}\right)\,{\frac{a_\la\,
U(a_\la+1,2,|\la| z^2)\;2 |\la| z} {U(a_\la ,1,|\la| z^2)}}{\big/ z=
\ep \to 0}\enq

 We again expand around $z=0$ ( $y=|\la|z^2$ ) with Mathematica

{\tt  Simplify[
 Series[a HypergeometricU[a + 1, 2, y]/HypergeometricU[a, 1, y], {y,
   0, 0}]]; }

 The result:

  \beq
  -\frac{1}{y (\psi ^{(0)}(a)+\log (y)+2 \gamma )}+O[1]
  \enq
  Now the leading dynamical term is proportional to $1/y =1/(|\la|z^2)$:

Putting pieces together we obtain: \beq \Si(q^2)=
\left(\frac{R}{z}\right)  \left( \frac{ 2 |\la| z }{|\la|z^2
(\psi(a) + \log( |\la| z^2) + 2 \ga^E)} + O(z) \right)\enq For
$z\to 0 $ the log diverges, therefore  we expand the denominator
     \beqa
\frac{1}{|\la|z^2 (\psi(a) + \log( \la z^2) + 2 \ga^E)}&=&\\
&& \hspace{-4cm}  \frac{1}{|\la| z^2 \log( |\la| z^2)} -
\frac{\psi(a) +  2 \ga^E}{|\la| z^2 \log^2( |\la| z^2)} + O\left[
\frac{1}{|\la| z^2 \log^3( |\la| z^2)}\right]\nn \enqa
    Taking into account only the leading dynamical term we arrive at~\cite{Brodsky:2014yha}:
\beq \frac{1}{R} \Si[Q^2] =   \frac{2}{ z^2 \log^2( |\la| z^2)}
\psi \left(\frac{Q^2}{4 |\la|} +\half \right) \enq "Renormalizing"
$\Si(Q^2)$ by the factor $\frac{1}{|\la| z^2 \log^2( z^2)} $ we
obtain for the vector propagator the result: \beq \frac{1}{R}
\Si_{\mu\nu}^{ren}[Q^2] = (\et_{\mu\nu} -\frac{q_\mu\,q_\nu}{q^2})
\psi \left(\frac{Q^2}{4 |\la|} +\half \right) \enq

The pseudoscalar current can be obtained in the same way, it has
also $L=0$,but  $J=0$  therefore only $a_\la$ and the prefactor
$\left(\frac{R}{z}\right)^{3-2J}$ change. The result for the
renormalized propagator of the pseudoscalar current is: \beq
\frac{1}{R^3} \Si^{ren}[Q^2] =  \psi \left(\frac{Q^2}{4 |\la|}
\right) \enq

\subsection{Asymptotic expansion of the two-point function. Comparison with QCD sum rules. \lb{csr}}
We have now an explicit expression for the propagator. For
comparison with other methods, most notably the QCD sum rule
method~\cite{Shifman:1978bx}, it is useful to consider an
asymptotic expression for the exact result.

\begin{quote} An {\bf asymptotic expansion}( $\sim$ ) is defined as follows:
\end{quote}
$$f(x)  \sim \sum_{n=0}^N a_n x^n \quad \mbox{means} \quad
f(x)-\sum_{n=0}^N a_n x^n = O(x^N)$$
\begin{quote}
The asymptotic series is generally diverging. Many series in
physics are only asymptotic, most notably the perturbation series
in QED and also, most probably in QCD. Sometimes asymptotic series
can by resummation be converted into converging series. Generally
one can say, the higher the order N, the better the approximation
but also the smaller the range, in which the series gives good
results.  We shall see this in the following  example.
\end{quote}

The asymptotic expansion of the Digamma function is \beq
\lb{asymuster}
 \psi^{(0)}(x) \sim (\log x - \frac{1}{2x} ) -\sum_{n=1} \frac{B_n}{2 n x^{2n}}= (\log x - \frac{1}{2x} )-\frac{1}{12 x^2} + \frac{1}{120x^4} + \dots
\enq In Fig. \ref{asyfig} we display the exact result for $
\psi^{(0)}(x) $ and the series with various truncations. We see
that for $x>0.6$ the inclusion of the second order term $1/(120
x^4)$(green)  improves the first order result, but that for
$x<0.6$ the first order result (blue) is better than the second
order result.
\begin{figure}
\bec
\includegraphics[width=7cm]{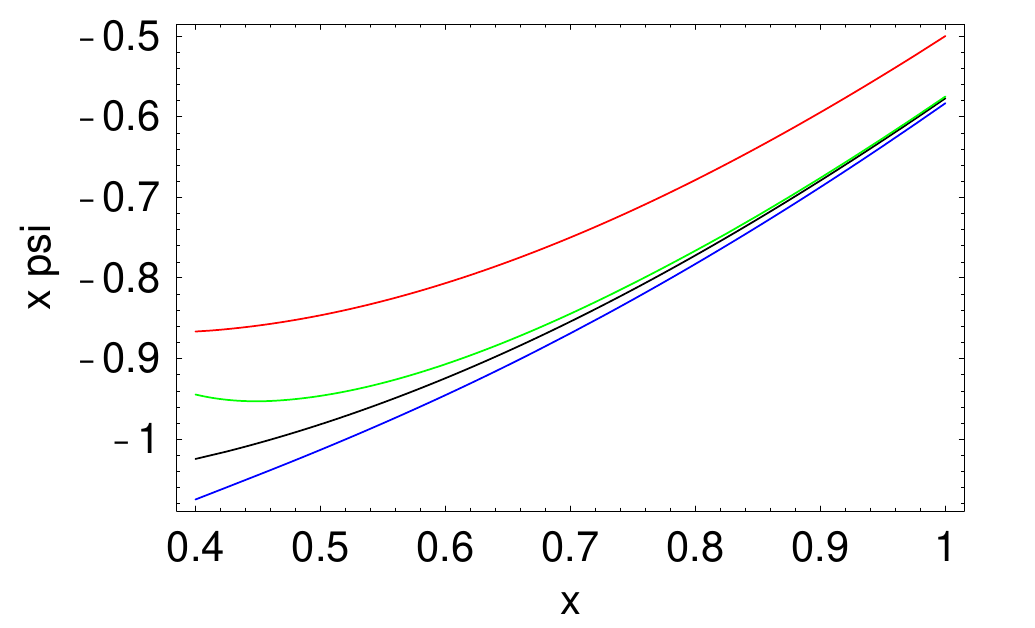}
\includegraphics[width=7cm]{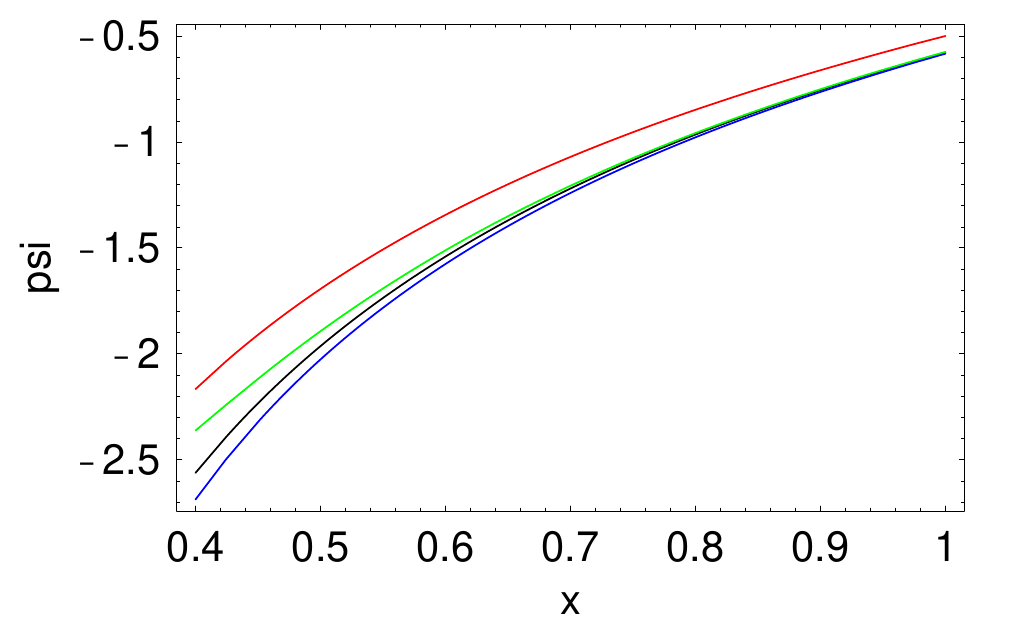}
\enc \caption{\lb{asyfig} Asymptotic expansion of $x \psi^{(0)}(x)
$(left) and $ \psi^{(0)}(x)  $(right) . Black: exact expression ,
red: zero order $\psi^{(0)} \approx \log x - \frac{1}{2 x}$ ,
blue: first term of AE included, green: first two terms included.}
\end{figure}

We now consider the scalar part of the asymptotic expansion of
several current propagators, introducing the abbreviation: \beq Y=
\frac{Q^2}{4 |\la|} =  -\frac{q^2}{4 |\la|} \enq

a) Conserved vector current ($L=J=1, \, \mu=0$): \beqa
\psi(Y+1) &=& \log(Y) +\frac{1}{2 Y}-\frac{1}{12 Y^2}+\frac{1}{120 Y^4}-\frac{1}{252 Y^6}+\frac{1}{240 Y^8} \\
&& -\frac{1}{132 Y^{10}}+\frac{691}{32760
Y^{12}}+O\left(\left(\frac{1}{Y}\right)^{13}\right) \enqa

b) $rho$ current in LFHQCD ($L=0,\,J=1$) \beq \psi \left(Y +\half
\right)=\log(Y) +\frac{1}{24 Y^2}-\frac{7}{960
   Y^4}+\frac{31}{8064 Y^6}+O\left(\left(\frac{1}{Y}\right)^8\right)
\enq

c) (Pseudo) scalar current in LFHQCD ($J=L=0$) \beq \psi \left(Y
\right)=\log(Y)-\frac{1}{2 Y}-\frac{1}{12
   Y^2}+\frac{1}{120 Y^4}-\frac{1}{252
   Y^6}+O\left(\left(\frac{1}{Y}\right)^8\right)
   \enq

d)  general asymptotic with open $L$ and $J$:
\beqa
\lb{holasy}
\psi \left(Y +\half(J+L)\right)&=&\\
&& \hspace{-3cm}\log(Y)+\frac{J+L-1}{2 Y}+\frac{-3 J^2-6 J L+6 J-3
L^2+6 L-2}{24
   Y^2}\nn \\
&& \hspace{-3cm}   +\frac{J^3+3 J^2 L-3 J^2+3 J L^2-6 J L+2
J+L^3-3 L^2+2 L}{24
   Y^3}
 +O\left(\left(\frac{1}{Y}\right)^4\right) \nn
\enqa

  %%%%%%%%%%%%%%%%%%%%%%%%%%%%%%%%%%

\subsubsection{The QCD sum rule method}
The sum rule method~\cite{Shifman:1978bx} is based on the operator
product expansion as an asymptotic expansion in the not so deep
Euclidean region, that is at $Q^2 \approx 1 GeV$. For very large
values of $Q^2$ perturbation theory (PT) is supposed to be
reliable because of asymptotic freedom. But for smaller values of
$Q^2$ nonperturbative terms are supposed to become relevant.
Therefore SVZ made for the propagator $\Si(Q^2)$ the following
theoretical ansatz:
\beq  \lb{SVZ}
 \Si_{SVZ}(Q^2) = \mbox{PT} +
\sum_{n=1} \frac{1}{Q^{2n}} F_n C_{2n}
 \enq
Here PT is the
perturbative expression, $C_{2n+2}$ are universal non-perturbative
QCD constants (condensates) and $F_n$ calculable expressions
depending on the quantum numbers of the propagating field.
Normally the series starts with the 4-dimensional gluon condensate
$C_4= \langle G^A_{\mu \nu} G^{A \mu \nu} \rangle$, but there are
also approaches which start with a two dimensional condensate, $
C_2$~\cite{Jugeau:2013zza}.
\begin{figure}
\bec
\includegraphics[scale=0.95]{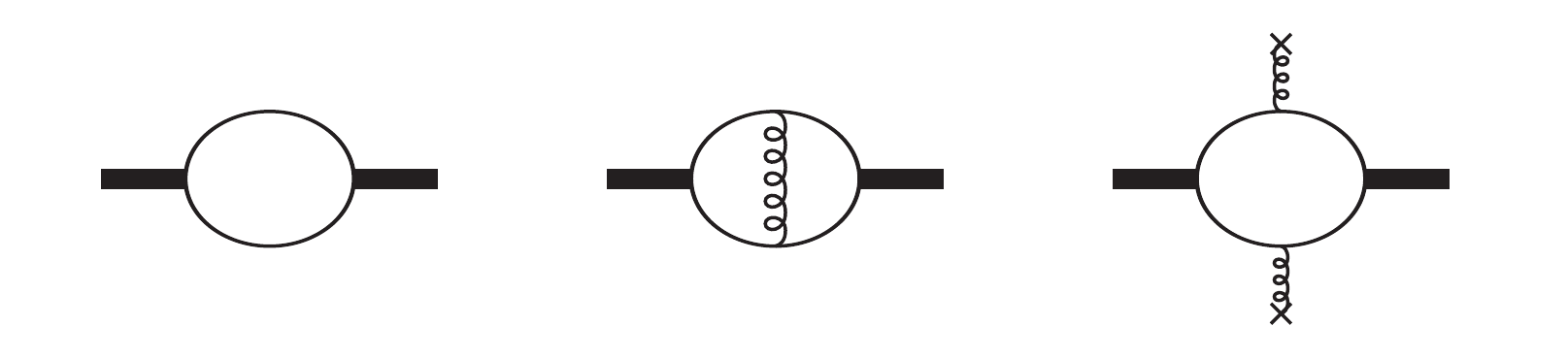}
\enc \vspace{0cm} \caption{Perturbative and non-perturbative terms
in a propagator. The thick lines represent a hadronic current, the
thin lines quarks, and the curled lines gluons. The first two
terms are from perturbation theory, the last term, where the gluon
lines end in the vacuum is an example of a non-perturbative
contribution \lb{svz}}
\end{figure}
The lowest order contribution to the expression from  perturbation
theory yields $PT \sim  \log Q^2$. Comparing \req{holasy}  and
\req{SVZ} we see the complete formal correspondence between the
asymptotic expansion of the holographic result and the sum rule
ansatz. This has been noticed and  partially exploited in HQCD
in~\cite{Colangelo:2008us,Jugeau:2013zza}. It should be noted,
that the analogy is formal. In the AdS/CFT case we have an exact
expression for which we construct an asymptotic expression. In the
case of QCD sum rules one  tries to approach the exact expression
(that is the observed two point function) by perturbation theory
plus a series of power corrections proportional to condensates.
These condensates are the difference of expectation values
obtained in the perturbative and the physical vacuum.

\subsection{The propagator in the hard wall model}
The propagator in the hard wall model can be calculated along the
same lines as that of the soft wall model, see sect. \ref{psw}.
The result for the conserved current
is~\cite{Erlich:2006hq}: \beq \frac{1}{R} \Si(q^2)
\sim  q^2\left( \log(q) -\frac{ \pi\,  Y_0(q z_0)}{2\,J_0(q z_0)}\right) +
\mbox{divergent non-dynamical terms} \enq It is a meromorphic
function with poles at the zeros of the Bessel function $J_0$. The
logarithmic cut of the Bessel function $Y_0$ cancels the
explicitly occurring logarithm. In this case the propagator has no
asymptotic expansion with a logarithm and additional  power
corrections as in the soft wall model \req{asymuster} and there is
no formal similarity with the expression used in QCD sum rules.
Indeed, as has been shown in \cite{Erlich:2006hq},
the hard wall model is rather the equivalent of a model developed
by Migdal~\cite{Migdal:1977nu}. In this model the perturbative
expression (the logarithm) is approximated by a (finite) sum of
pole terms.

\section{Summary}
The propagator of a mesons in the soft wall model is a digamma
function, which is meromorphic. The asymptotic expansion of the
digamma function for negative arguments shows a remarkable
structural similarity with  the expansion of the propagator used
for QCD sum rules.
%%%%%%%%%%%%

%%%%%%%%%%%%%%%%%%%%%%%%%%%

%\end{document}

\chapter{Form factors in AdS}
This chapter is mainly based on
\cite{deTeramond:2012rt,Sufian:2016hwn} and  chapt. 6 in
\cite{Brodsky:2014yha}, We refer to these sources for
additional  literature.
\section{Form factors}
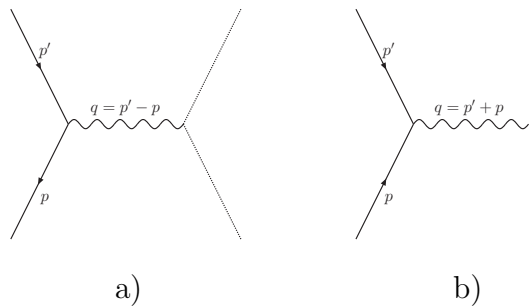
\begin{figure}
\bec
\includegraphics[width=8cm]{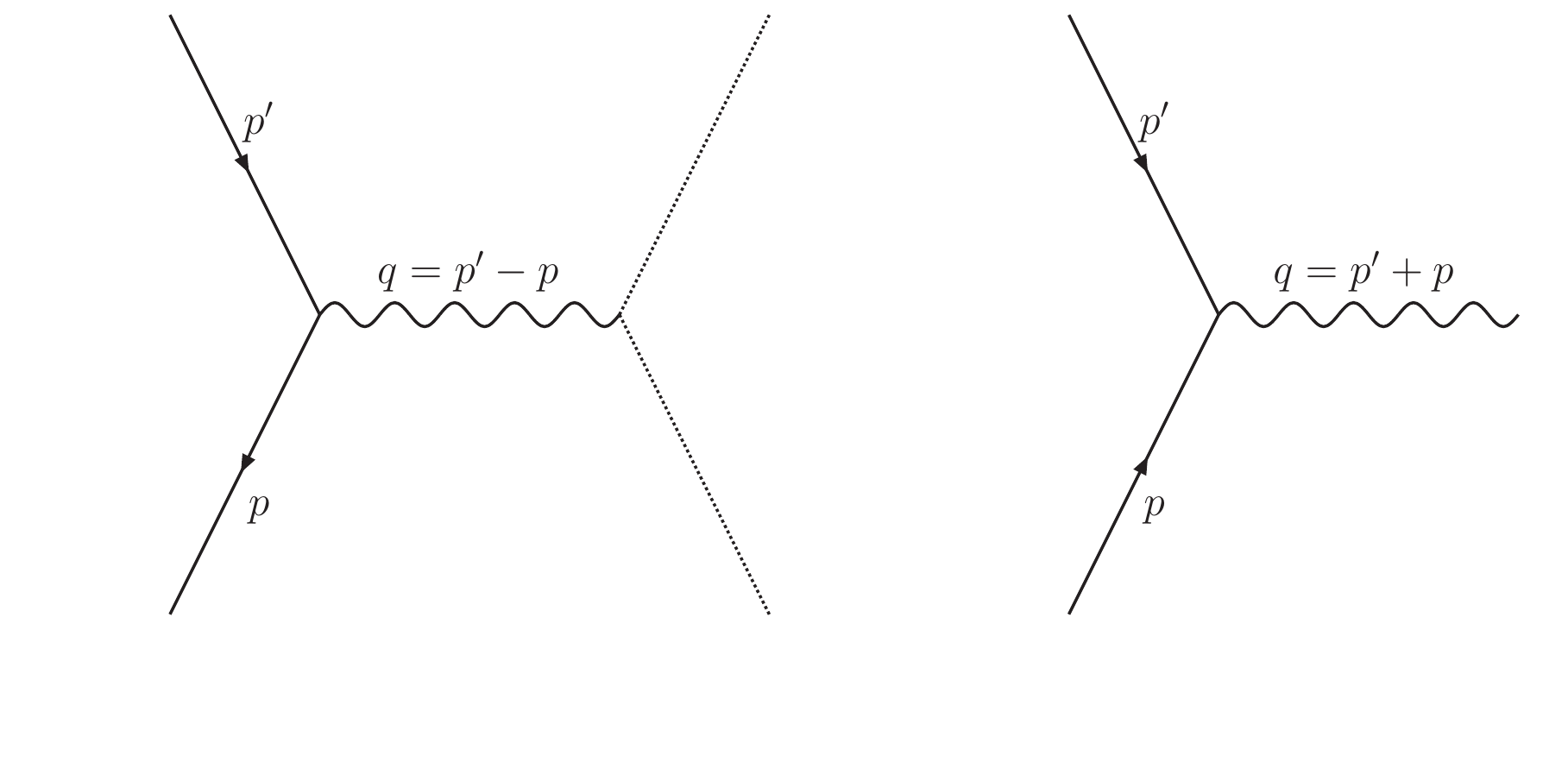}
\enc
\begin{picture}(0,0)
\setlength{\unitlength}{1mm} \put(65,10){a)} \put(110,10){b)}
\end{picture}
\caption{\lb{ FF-scheme} Measurement of the  form factor: a)  in the space-like region (scattering); b)
in the time-like region (annihilation or production).
% and the coupling of the virtual photon to a vector meson (same QN)
}
\end{figure}
The product of the electric charge {\tt e} and the form factor
$F(q^2) $ measures  the strength of the coupling of a virtual
photon with momentum $Q^2=-q^2$ to a hadron, the form factor  is
normalized $F(0)=1$. By scattering of leptons on hadrons one
probes the values for $q^2<0$ Fig. \ref{ FF-scheme}, a)  and by
annihilation the region $q^2>0$, Fig. \ref{ FF-scheme} b). The
form factor ( FF) gives information on the charge
distribution of the hadron, thus it is an important tool to
investigate the structure of hadrons. Indeed, three Nobel prizes
have been awarded for form-factor related investigations, namely
1961 to
 Hofstadter
  for his investigations of the neutron and proton form factor,
  and
  1990
  to  Friedman, Kendall and  Taylor  for investigations of the inelastic
   form factor, and 1976 to Richter and Ting  for investigations of the hadronic form factors
   in the time-like region.

Since the form factor  is classically the  Fourier transform of
the charge distribution, it shows  for  a homogeneous Gaussian
charge distribution of the form  $e^{-\la z^2}$  also a Gaussian
form. For a distribution of point-like charges, however, a power
behaviour $ \sim 1/(q^2)^n$ results. The results of  Friedman,
Kendall and  Taylor showed a power behaviour and were thus an
important corroboration of the parton model of hadrons.

\section{ Form factor  in HQCD and LFHQCD for a (pseudo-)scalar particle}
\subsection{The ``dressed" electromagnetic field in AdS/CFT} In order to calculate the  form factor in field theory one has to start from the interaction term of the electromagnetic with the hadron field in the action.
The interaction term in the modified  AdS action  is given by \beq
\lb{ form factor1} S_{int} = \int d^{d+1}x \, {\tt e}_5(z) \;
e^{\la z^2}  \; \sqrt{|g|}\,g^{NN'}\, i\, \big( (\pa_N \Phi(x))^*\,  \Phi(x) - \Phi^*(x) \, \pa_N \Phi(x)\big)\,
 A_{N'}(x), \enq {\tt
e}$(z)$ is the electric  charge in AdS$_5$ and $A_N$ the
electromagnetic current in AdS~\cite{Polchinski:2002jw}.
 We could in principle try to evaluate the generating functional for that three point function of two hadron fields and the electromagnetic one.  But we make a shortcut: We are interested in the form factors of on-shell particles, i.e. $p^2={p'}^2 = M^2$, and only  the momentum $q$ of the virtual photon is variable. Therefore we evaluate directly an expression corresponding to \req{ form factor1} where $\Phi(x)$ is not a general  hadron field, but the wave function of a specific hadron.

As before we Fourier transform the Minkowski variables, that is we
go from $\Phi(x,z)$ to $\tilde \Phi(p,z)$ and obtain
 \beq \lb{form factor2} S_{int} = \int \frac{d^4 p}{(2\pi)^4}   \frac{d^4 p'
}{(2\pi)^4} \,dz\, {\tt e}_5(z) \, \left(\frac{R}{z}\right)^3 \vep
\cdot (p-p') \; \tilde \Phi_\tau^*(p',z)  \, \tilde
\Phi_\tau(p,z) \tilde A(p-p',z), \enq here $ \vep $ is the
polarization vector of the Fourier transformed  field
  \beq
  \vep_\mu  \, \tilde A(q,z) = \int d^4x \,e^{- i q x}  \,A_\mu (x,z)
  \enq
and we  have introduced a modified wave function \beq \tilde
\Phi_\tau(q,z) = e^{\la z^2/2} \,\tilde \Phi(q,z)\enq in order to
compensate the dilaton factor $e^{\la z^2}$ in \req{ form
factor1} and we have also introduced a possibly $z$ dependent electric charge ${\tt
e}_5(z) $ in the bulk. We have used that in AdS$_5$ one has
$\sqrt{|g|}=\left(\frac{R}{z}\right)^5, \;
\,g^{LL'}=\left(\frac{z}{R}\right)^2 \,\eta^{L L'}$.

The expression for  the form factor of hadrons can be  read off
from\req{form factor2}:
 \beq \lb{ form factor2a} {\tt e} \,
F(Q^2)=  \int  dz\,  {\tt e}_5(z) \, \left(\frac{R}{z}\right)^3
\tilde \Phi_\tau^*(p',z))  \, \tilde \Phi_\tau(p,z) \tilde
A(p-p',z) \enq
where $Q^2 =-q^2= -(p'-p)^2$ and {\tt e} is the
total charge of the hadron.

In chapt. \ref{bsar}, \req{asso2} we have determined the bound
state wave functions  $\tilde \Phi(p,z) $ for $p^2 = M_{\it
hadron}^2 $. These wave function for the hadrons are \beq \lb{
form factor2b} \tilde \Phi_\tau (p,z)= \frac{1}{N}\,
\la^{(\tau-1)/2} z^\tau L_n^{(L)}(\la z^2) e^{-\la z^2/2} \quad
\mbox{ with }\; \tau = 2+L-J  \enq

The solution $\tilde A(q,z)$ for a conserved vector current, that
is with quantum numbers $J=1$ and  $L=1$, corresponding to $ \mu=0
$ is, see \req{final3}:
\beq \lb{ form factor3} \tilde A(q,z) =
z^{2} e^{-(|\la|+\la) z^2/2} U\left(\frac {-q^2}{4 |\la|
}+1,2,|\la| z^2\right)
\enq
 We use the relation~\cite{AS},
13.1.17, with $b=2$: $ U(a,2,x)= x^{-1}U(a-1, 0, x) $ and obtain
\beq \lb{ form factor4} \tilde A(Q^2,z) = \frac{1}{|\la|}
e^{-(\la+|\la|)z^2/2} U\left(\frac {Q^2}{4 |\la| },0,|\la|
z^2\right) \enq As normalization conditions we
impose~\cite{Brodsky:2014yha}: \beq \lb{bca} \frac{1}{\tt e}
\lim_{q^2\to 0} {\tt e_5}(z) \tilde A(Q^2,z) = \frac{1}{\tt e}
\lim_{z\to 0} {\tt e_5}(z) \tilde A(Q^2,z)=1 \enq Some useful
relations  for the function $U$ are, see e.g. \cite{AS}, 13:
\beqa U(\frac{Q^2}{4 |\la|},0, 0)&=& \frac{1}{\Gamma(1 + \frac{Q^2}{4 |\la|})}\\
U(0,0,z) &=& 1\\
\lim_{Q^2\to \infty} \Gamma( \frac{Q^2}{4 |\la|})U(\frac{Q^2}{4
|\la|},0, |\la| z^2) &=& Q z K_1(Q z)
 \label{asu form factor} \enqa

If we chose ${\tt e}_5(z) = e^{(|\la|+\la) z^2/2 }\, {\tt e}$ the
conditions \req{bca} are fulfilled for the solution $J(Q^2,z)$
with
 \beq \tilde J(Q^2,z)  = \Gamma(1 + \frac{Q^2}{4
|\la|})U(\frac{Q^2}{4 |\la|},0, |\la| z^2) .\lb{ form factorsw}
\enq

This function $ \tilde J(Q^2,z)  $ \req{ form factorsw} is a
Tricomi Hypergeometric function and  has  an integral
representation see \cite{AS}, (13.2), which is very useful to
perform the $z$ integration in \req{ form factor2a} \beq \tilde
J(Q^2,z) =|\la| z^2 \int_0^1 \frac{dx}{(1-x)^2 }x^{Q^2/(4 |\la|)} \,
e^{-|\la| z^2 x/(1-x)}. \enq

We confine ourself to orbital and radial ground states ($L=0,\,
n=0$) and have for the hadron wave functions: \beq \label{tau-f}
\Phi_\tau(z) = \sqrt{\frac{2}{\Ga(\ta-1) \la}} \;  (\la
z^2)^{\ta/2} e^{- \la z^2/2} \; \mbox{  with }\; \int_0^\infty
\frac{dz}{z^3} \Phi_\tau(z)^2 =1 \enq

We omit in the following the curvature radius in AdS$_5$, it can
be absorbed in the normalization of the wave functions.

 The form factor  \req{ form factor2a} is then
\beqa
F^\ta(Q^2)&=&\int \frac{dz}{z^3} \,  \Phi_\tau(z)^2 J(Q^2,z) dz \lb{ form factorxx}\\
&=& \int \frac{dz}{z^3}  \frac{2z^2}{\Ga(\ta-1)}\,(\la z^2)^\ta
e^{-\la z^2}\, \int dx\, \frac{ e^{-|\la| z^2 x/(1-x)}}{(1-x)^2
}x^{Q^2/(4 |\la|)}\lb{ form factorxxb} \enqa The $z$-integration
over the Gaussians can be performed analytically and one obtains:
\beq F^\ta(Q^2) = (\ta-1) \int_0^1  dx(1-x)^{\ta-2} x^{Q^2/(4
\la)} \enq

The integral on the right hand side of this equation has the form
of an integral representation of the Euler B function: (\cite{AS},
6.2)
\beq B[u,v]= \frac{\Ga[u]\, \Ga[v]}{\Ga[u+v]} \qquad
B[u,v]= \int_0^1 dx \, x^{u-1}\, (1-x)^{v-1} \enq

We thus can express the form factor analytically in terms of the
Beta function or in terms of Gamma functions. \beqa \lb{finb}
F^\tau[Q^2] &=&  (\tau-1) B[\ta-1,1+  Q^2/(4 \la)]\\
&=& (\ta-1) \frac{\Ga[\ta-1] \Ga[1+  Q^2/(4 \la)]}{\Ga[\ta +
Q^2/(4 \la)]} \enqa Since $\ta$ is an integer, the expression can
be transformed using the recursion relation $ u \Gamma(u) =
\Ga(u+1)$.

Consider e.g. particle  a (pseudo)scalar particle (pion): here we
have $\tau=2+L-J=2$. By making use of the recursion relation for
the denominator we obtain \beq \lb{fpi} F^\tau_\pi(Q^2) = 1
\frac{[\Ga[1+  Q^2/(4 \la)]}{\Ga[2+ Q^2/(4 \la)]}= 1 \frac{[\Ga[1+
Q^2/(4 \la)]} {(1+ Q^2/(4 \la) )\Ga[1+  Q^2/(4 \la)]}= \frac{1}{1+
Q^2/(4 \la) } \enq

\subsection{The scaling  twist}
There is a lot of discussion about the twist. The canonical  twist
of a field is defined as the dimension of the field minus spin
plus angular momentum. The dimension of a field can be determined
by the requirement that the action has the dimension of $\hbar$,
that is 0 in natural units.  From that follows that a scalar field
has  mass dimension 1 and a fermion field has  mass dimension 3/2. The
dimension of a quark (or antiquark) is $\frac{3}{2}$, therefore
the twist of a quark is one.

According to this counting the twist of a hadron is the number $N$
of the quarks it contains plus the angular momentum. We shall call
this quantity, $N+L$  the scaling twist, in following we shall
always use the scaling twist $\ta= N+L$.

For a scalar field with two particles in the ground state we have
$\tau=2$.

For the Nucleon or Delta the positive chirality component ($L=0$)
has twist $\tau=3$,  the negative chirality component with $L=1$
has  $\tau=4$. We come back to this point later in the discussion
of the nucleon  form factor.

\subsection{General results}
The procedure which led to \req{fpi} can be extended to any twist
by repeated application of the recursion formula and one obtains
\beq \lb{as1} F^\ta(Q^2) = \frac{(4
|\la|)^{\tau-1}(\tau-1)!}{(Q^2+4 |\la|) \cdots (Q^2 + (\ta-1)(4
|\la|))}.\enq For very high $Q^2$ one obtains: \beq \lb{cr1}
\lim_{Q^2\to \infty} F^\ta(Q^2) \sim \left(\frac{4 \la}{
Q^2}\right)^{\tau-1} \enq
 Holographic QCD therefore implies~\cite{Polchinski:2001tt} that the  form factor decreases with $(1/Q^2)^{N-1}$ where $ N$  is the number of constituents. This is in accordance with the famous quark counting  rule of Brodsky and Farrar~\cite{Brodsky:1973kr,Brodsky:1974vy}. This  rules can can easily be understood qualitatively. In  a hard elastic lepton hadron scattering the large  transferred momentum $q$ has to be distributed over all constituents. The photon interacts directly with only one  constituent. If the hadron stays intact, as is the case for the elastic form factor,   the momentum has to be transferred to the other constituents by gluons of momentum $\sim q$. This process creates a gluon propagator $\sim 1/Q^2$. For $N$ constituents the momentum has to be transferred from the active quark to the remaining $N-1$ passive constituents, that is we have a factor $1/(Q^2)^{N-1}$.

It is very remarkable that this lowest order result from
perturbative QCD  is also incorporated in the inherently
non-perturbative  HQCD and  LFHQCD~\cite{Brodsky:2003px}! Higher correction in
perturbation theory lead to logarithmic corrections, which,
however,  are not included in LFHQCD.

\subsection{ Final assumptions and results for the form factor \lb{farff}}

The most  delicate point is the following: The poles in the
time-like region of the form factor are  associated with hadrons
with the same quantum numbers as the photon $J^P=1^-$, that is
with vector mesons:  the $\rho$ and its radial excitations.  On
the other hand we have seen in the last chapter, sect. \ref{prcc},
especially Fig. \ref{rho-mod},  that the conserved current does
not give give an adequate description of the $\rho$ and its
trajectory, whereas the current with $J=1,L=0$ in LFHQCD does so.
The easiest solution of the problem is to replace the argument
$Q^2+1$ in the Beta function \req{finb} by
$Q^2+\half$~\cite{Brodsky:2014yha}; this preserves all the
positive results but shifts the $\rh$ poles to the right position.

In this way we obtain as final general expression for the form
factor  of a hadron with twist $\tau$ rthe very simple result:
     \beq \lb{finalfinalB}
F^\ta[Q^2] =  \frac{\tau-1}{\cN} B[\ta-1,\half+  Q^2/(4 \la)]
   \enq
where the normalization constant $\cN$ is the rational number:
\beq \cN= \frac{\Ga(\ta)\,\Ga(\half)}{\Ga(\ta+\half)} \enq This
shift of the pole positions  amounts to a purely numerical
modification of \req{as1} and \req{cr1} to \beq \lb{as1n}
F^\ta(Q^2) = \frac{1}{(1+ Q^2/M_0^2) \cdots ) (1+
Q^2/M_{\tau-2}^2)} \enq where $M_n^2$ are the theoretical values
for the squared masses of the rho and its radial excitations,
$M_n^2=(4n +2)|\la|$.

Equation \req{cr1} reads now: \beq \lb{cr1n} \lim_{Q^2\to \infty}
F^\ta(Q^2) = M_0^2 \cdots M_{\ta-2}^2 \left(\frac{1}{
Q^2}\right)^{\ta -1} \enq

Another step to a realistic theory is to replace the theoretical
masses $M_i$ by the observed ones. For the vector current: \beqa
M_0= \sqrt{2 \la} \approx 0.76 \;  &\Rightarrow& \;   M_\rho = 0.77 \;{\rm GeV}^2\\
M_1= \sqrt{6 \la} \approx 1.32 \;  &\Rightarrow& \;   M_\rho = 1.45 \;{\rm GeV}^2\\
M_2= \sqrt{10 \la} \approx 1.70 \;  &\Rightarrow& \;   M_\rho = 1.70 \;{\rm GeV}^2\\
M_3= \sqrt{14 \la} \approx 2.02 \;  &\Rightarrow& \;   M_\rho =
2.15 \;{\rm GeV}^2 \enqa

This shift to the observed masses has little influence on the form
factors in the space-like region, but it is important in the  timelike
region. There the inclusion of the observed
widths is  ven more important. LFHQCD as a zero width ($N_c\to \infty$) theory
predicts form factors  with poles on the real axis, but the
physical form factor has cuts and poles in the complex plane. This
again has no great influence  on the space-like behaviour but is
crucial for the time-like region. Therefore for a realistic
treatment of the time-like region one must take into account the
experimentally observed finite widths of the rho and its radial
excitations. This will be illustrated in the next subsection.

The valence quark (leading twist) approximation might be not so
appropriate for the treatment of form factors, especially again in
the time-like region. If one takes an additional  meson cloud into
account, each additional meson in the Fock state increases the
number of constituents and hence also the twist by two. Taking
into account only one additional meson in the cloud, one obtains:
\beq \lb{fs} F(Q^2) = (1-P) F^{\tau}(Q^2) + P\, F^{\tau +2}(Q^2) \enq
where $P$ is the probability of the higher twist
contribution and $\ta$ the leading twist. Since the higher twist
does not contribute asymptotically compared to the leading twist,
the asymptotic formula \req{cr1n} is reduced  by the factor
$(1-P)$.

\subsection{Comparison of $\pi$  form factor with experiment}
The results obtained by (modified) LFHQCD for the pion form factor
in the space and time-like region is displayed in Fig. \ref{pi-
form factor}.  The occurrence of a (weak) contribution of the
$\rh(1450)$ in the pion pair production makes a  small  higher
twist contribution  necessary, with $P=0.12$, see \req{fs}, For
further details we refer to \cite{Brodsky:2007hb,Brodsky:2014yha}.
The importance of inclusion of the finite observed  width becomes
evident by comparing Fig. \ref{pi- form factor}b) and d).

\begin{figure}
\bec
\includegraphics[width=6.5cm]{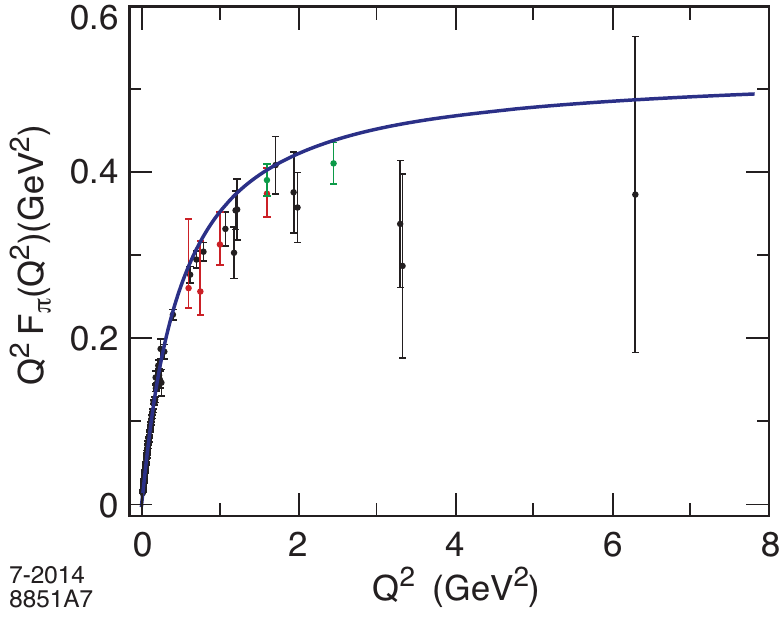}
\includegraphics[width=8cm]{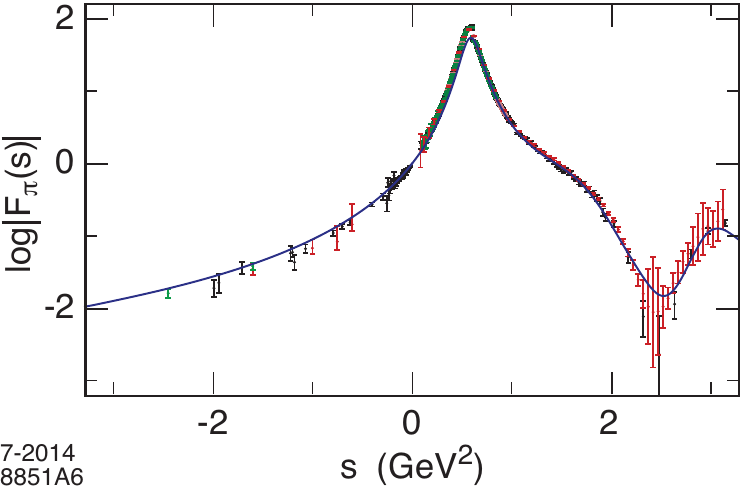}

\includegraphics[width=6.5cm]{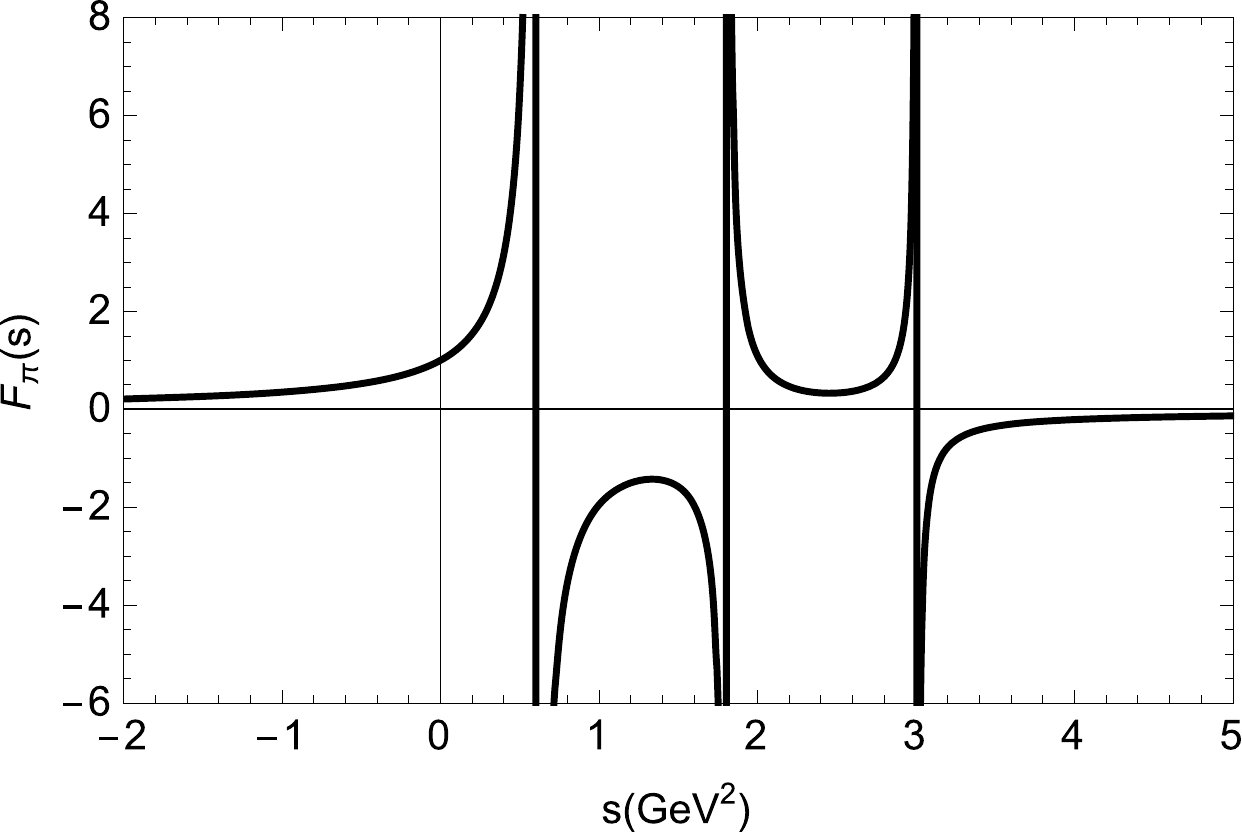}
\includegraphics[width=6.5cm]{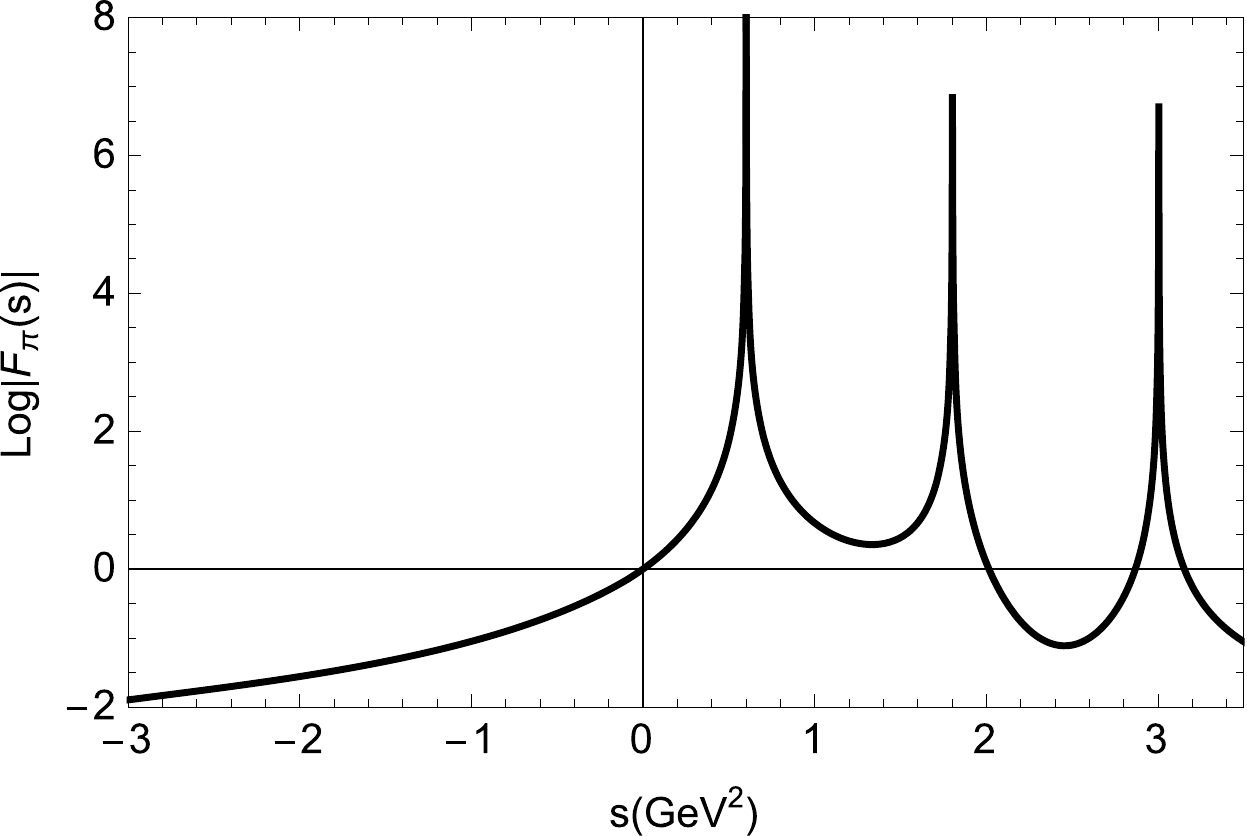}
\enc
\begin{picture}(0,0)
\setlength{\unitlength}{1mm} \put(30,99){a)} \put(95,99){b)}
\put(30,46){c)} \put(95,46){d)}
%\put(133,18){\line(0,1){40}}
\end{picture}
\caption{\lb{pi- form factor}The form factor of the pion in the
spacelike region (a) and in the timelike region (b), where
information about width from experiment and some threshold effects
are incorporated. c) and d) The formfactor from LFHQCD without any
corrections in the timelike region.  a) and b) come from
\cite{Brodsky:2014yha}.}
\end{figure}
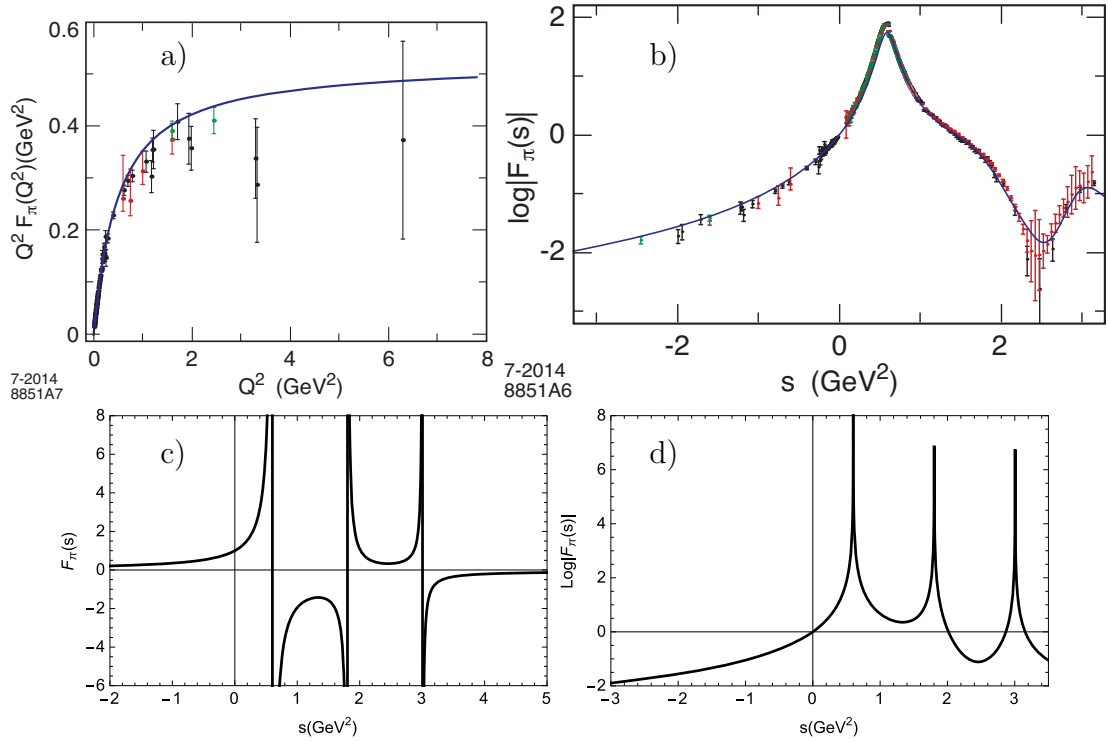
%%%%%%%%%%%%%%%%%%%%%%%%%%%%%%%%%%

\subsection{The form factor in the parton model, effective  wave functions \lb{pmod}}
The form factor in the parton model has been studied by Drell and
Yan~\cite{Drell:1969km} and West~\cite{West:1970av}. The Drell-Yan
West (DYW) expression for the form factor calculated from light
front wave functions is:
 \beq F(Q^2) = \int dx \, b db d\theta e^{i x |\vec q_\perp| b \cos
\theta}|\phi^{LF}(x,b)|^2= 2 \pi \int  \int dx \, b db J_0(x Q b)
|\phi^{LF}(x,b)|^2. \lb{ form factordyw} \enq

In Fig. \ref{form4} we show the result for the form factor of the
pion in the space-like region as  obtained by LFHQCD (solid line)
and the result of the DYW expression \req{ form factordyw} where
the LF wave functions obtained from AdS wave functions, see
\req{slf}, have been used.

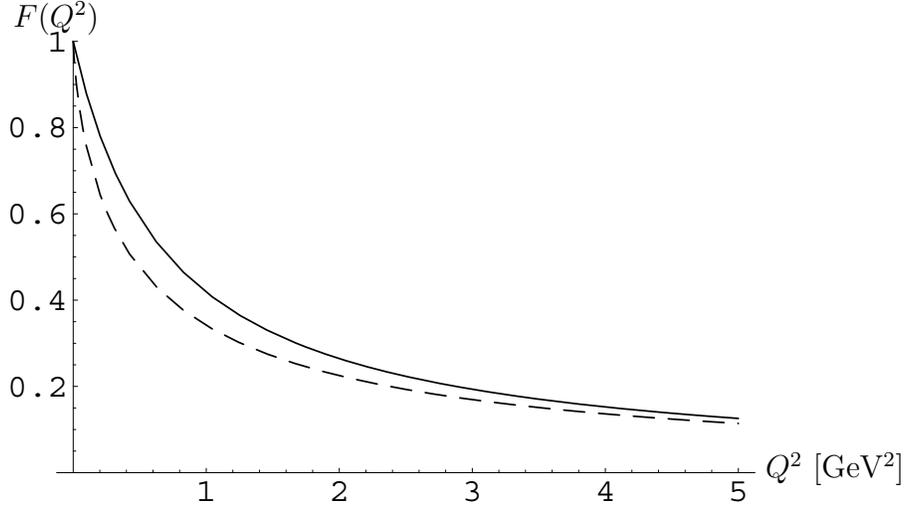
\begin{figure}[t]
\setlength{\unitlength}{1cm}
\begin{center}
\includegraphics*[width=10cm]{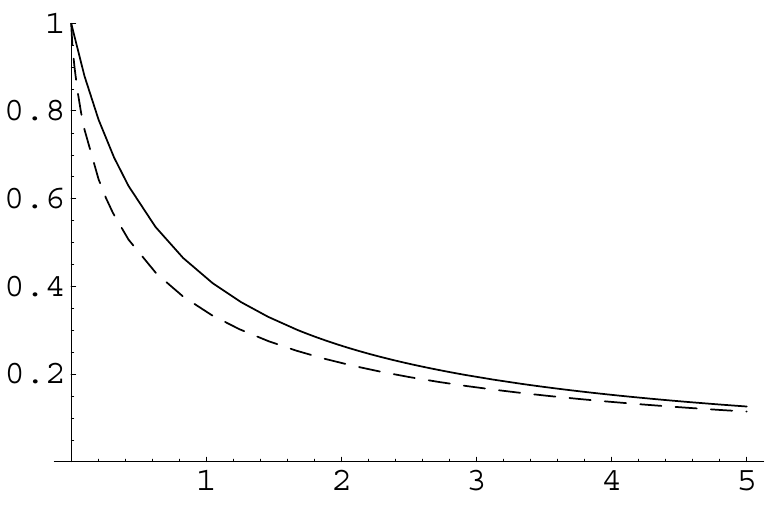}
\begin{picture}(0,0)(0,0)
 \put(0,0.5){$Q^2$ [GeV$^2$]}
 \put(-10,6.5){$F(Q^2)$}
\end{picture}
\end{center}
\vspace{-6mm} \caption {Solid line: The  form factor in the soft
wall model for the ground state , obtained directly from AdS/CFT.
Dashed line:  the form factor obtained from the soft-wall LF
function with the Drell-Yan West expression (\ref{ form
factordyw}) \lb{form4}.}
\end{figure}

 The DYW expression evaluated with wave functions as obtained in AdS %LFHQCD ??
 fall off to fast at small values of $Q^2$, the resulting  charge radius is  infinite. For large $Q^2$, however, the two expressions agree asymptotically. The failure of the
wave functions obtained from the bound state wave equations of AdS
shows that they  may describe very well global features, as the
masses of the hadrons,  but are not adequate to describe reliably
the inner structure precisely. Therefore it is useful to introduce
{ effective wave functions} $\phi^{\it eff}(\ze)$, which inserted
into the DYW expression \req{ form factordyw} reproduce the result
obtained above, \req{finalfinalB}, with the dressed
electromagnetic current~\cite{Brodsky:2014yha}. They are defined
by the condition: \beq F_{AdS}(Q^2) =\frac{ (\tau-1)}{\cN}
B\left[\tau-1,\frac{Q^2}{4\la}+\half\right] = 2 \pi \int  \int dx
\, b db J_0(x Q b) \, |\ph^{\it eff}_\tau(x,b)|^2. \lb{e form
factor} \enq .

\section{Nucleon Form Factors }
\subsection{Form factors for spin $\half$ fields in AdS/CFT}
The coupling of the electromagnetic field to fermions is in AdS
given analogously to \req{ form factor1} by: \beq \lb{ form
factorD}
 \int d^4x \, dz \,  \sqrt{g}   \,  \bar\Psi_{P'}(x,z)
 \,  e^M_A  \, \Gamma^A \, A_M(x,z) \Psi_{P}(x,z)
\enq

The notation is the same as in sect. \ref{bsebas}, below \req{af}.
It leads in the 4-dim space to
 the Dirac form factor $F_1$.
\beq
 (2 \pi)^4 \delta^4 \left( P'  \! - P - q\right) \epsilon_\mu \, \bar u(P') \gamma^\mu F_1(q^2) u({P}),
\enq

In the parton model it  is the spin-conserving matrix element
of the quark current $J^\mu = \sum_q e_q \bar q \gamma^\mu q$.

In  AdS$_5$ nucleons have  positive and negative chirality
components, $\Psi^+$ and $\Psi^-$, as described in sect.
\ref{bsebas}. The spin non-flip nucleon elastic form factor $F_1$
(Dirac form factor) is diagonal in chirality and follows from
(\ref{ form factorD}): \beq \lb{ form factorDpm} F_1^N(Q^2) =
\sum_\pm g^N_{\pm}  \int  \frac{dz}{z^4} \, J(Q^2, z)  \,
\Psi_\pm^2(z). \enq The current is given by  \req{ form
factorsw},but the same pole shift as in \req{finalfinalB} will be
performed. Notice that there is an additional scaling power  in
(\ref{ form factorDpm}), as compared with \req{ form factorxx}).
%%%%%%%%%%%%%%%%%%%%%%%%%%%%%%

 The effective  charges $g_\pm^N$ have to be determined by the specific spin-flavor
  structure which is not contained in the holographic  principle. For example, in the SU(6)
   symmetry approximation the effective charges are computed by the sum of the charges of the
   struck quark convoluted by the corresponding probability for the $L=0$ and $L=1$
    components $\Psi_+$ and $\Psi_-$ respectively. The result is~\cite{Brodsky:2014yha}
    \beq \lb{Neff}
g_+^p = 1,   ~~~~  g_-^p = 0,    ~~~~ g_+^n = - \frac{1}{3}, ~~~~
g_-^n = \frac{1}{3}.
    \enq
    Since the structure of  (\ref{ form factorD}) can only account for $F_1$, one should
     include an effective  gauge-invariant interaction in the five-dimensional gravity
     action to describe the spin-flip amplitude~\cite{Abidin:2009hr}.
\beq \lb{ form factorsf}
 \int d^4x \, dz \,  \sqrt{g}   \,  \bar\Psi_{P'}(x,z)
 \,  e^M_A   \,  e^N_B \, \left[\Gamma^A,  \Gamma^B \right]\, F_{M N}(x,z) \, \Psi_{P}(x,z)
\enq where the resulting expression in 4 dimensions is the Pauli
form factor $F_2$. \beq
 (2 \pi)^4 \delta^4 \left( P'  \! - P - q\right) \epsilon_\mu \bar u(P') \, \frac{\sigma^{\mu \nu} q_\nu}{2 M_N} \, F_2(q^2) u({P})
\lb{form factorP} \enq
 It corresponds to the spin-flip matrix element. Since (\ref{ form factorsf}) represents
 an effective interaction, its overall strength has to be fitted to the observed static
 values of the anomalous magnetic moments $\chi_p$ and $\chi_n$~\cite{Brodsky:2014yha, Abidin:2009hr}.

Extracting the factor $(2 \pi)^4 \delta^4 \left( P'  \! - P -
q\right)$ from momentum conservation in (\ref{form factorP}) we
find~\cite{Abidin:2009hr} \beqa \lb{ form factorPpm}
 F_2^N(Q^2) = \chi_N  \int  \frac{dz}{z^3} \, \Psi_+(z) \,J(Q^2, z)  \, \Psi_-(z),
\enqa where $N = p,n$.

Since $\Psi_+(z) \sim z^{\tau +\half}$ and $\Psi_-(z) \sim z^{\tau
+1+ \half}$ the total power of $z$ in the Pauli form factor \req{
form factorPpm} is $z^{2(\tau +1)-3}$, that is $F_2$ is a form
factor with twist $\tau +1$.

\subsection{A Simple Light-Front Holographic Model for Nucleon Form Factors} \lb{NH form factormodel}

From \req{nwf3} follows for the behaviour of the positive and
negative component of the nucleon  field in the ground state, note
that $T=J-\half=0$ for the nucleon: \beq \tilde \Psi^+(P,z)  \sim
z^{2+\half} ; \quad \tilde \Psi^-(P,z)  \sim z^{3+\half} \enq The
additional factor $z^\half$ compensates the additional power of
$z$ in \req{ form factorDpm} and we obtain the same expression as
for a meson. But we expect for the leading twist, $L=0$, the value
$\tau =3$ since  the baryon has 3 constituents. The discrepancy
has the following reason: The wave functions derived in sect. 4.4
are wave function of a two body system consisting of a quark and a
two-quark cluster. Since the electromagnetic field interacts with
both components of the cluster separately, the latter has for the
form factor to be resolved and this gives the additional factor
$z$ in the wave function. This fact corroborates our statement
made in sect \ref{pmod} that the bound state wave functions of AdS
describe only the collective properties. We use the realistic
twist for the nucleon wave functions and therefore put: \beqa
\tilde \Psi^+ &=& z^\half\, \Phi_3(z) = z^\half\sqrt{\frac{2}{\la}} \;  (\la z^2)^{3/2}\\
\tilde \Psi^- &=& z^\half\, \Phi_4(z) = z^\half\sqrt{\frac{2}{2
\la}} \;  (\la z^2)^{2} \enqa

%%%%%%%%%%%%%%%%%%%%%%
Analogous to the  pion  we will consider the case   where we only
include the valence contribution (with probability $1-P$) and
possibly a contribution with two more constituents (pion cloud),
probability $P$, that is in leading twist $\tau=3, \, 5$ for
$\Psi^+$. For  $\Psi^-$ no higher twist is asked for by the data.

We obtain for the proton
     \beqa \lb{protonF1} F_1^p(Q^2) &=&
(1-P_p^1)F^{(\ta=3)}(Q^2)+P^1_pF^{(\ta=5)}(Q^2)  \\  \lb{protonF2}
   F_2^p(Q^2) &=& \chi_p [(1-P^2_p)F^{(\ta=4)}(Q^2) +P^2_p
F^{(\ta=6)}(Q^2)] \enqa
 where $\chi_p=\mu_p-1=1.793$ is the proton anomalous moment.

The valence contribution alone gives a good description of the
proton Dirac form factor, see Fig. \ref{fig:q4f1}, blue line, that
is $P_p^1 \approx 0$, but the fit for the Pauli form factor can be
improved by choosing a rather large higher-Fock-state probability,
$P^2_p \approx  0.27$, see Fig. \ref{fig:q6f2}, blue line. It
is of course not satisfactory to have different cloud
contributions for the same particle, but it  necessary to have
good agreement with the data.

 It turns out that for the neutron higher twist contributions cannot ameliorate
  the agreement with the experiment and  we use
\beqa \lb{neutronF1}
F_1^n(Q^2) &=&  -\frac{1}{3}(1-P_n^1)\left[F^{(\ta=3)}(Q^2) - F^{(\ta=4)}(Q^2) \right]\\
&& \qquad -\frac{1}{3}\,P_n^1 \left[F^{(\ta=5)}(Q^2) -
F^{(\ta=6)}(Q^2) \right]
   \\ \lb{neutronF2}
F_2^n(Q^2) &=& \ch_n  \left[(1- \gamma_n)F^{(\ta=4)}(Q^2) +
\gamma_n F^{(\ta=6)}(Q^2) \right] \enqa with the experimental
value $\chi_n=\mu_n=-1.913$.

The result for the neutron Dirac form factor comes out by a
constant factor 2.1 to small,see Fig. \ref{fig:q4f1}, yellow line.
One  expects indeed  that the theory is less reliable for the
neutron than for the proton, since the result for the neutron form
factor is the difference of two two theoretical curves which
compensate exactly at $Q^2=0$. Therefore the neutron is much more
sensitive to uncertainties  of the theory, notably to the
determination of the effective charges from $SU(6)$ symmetry.

The result for the neutron Pauli form factor is not a difference
of two theoretical curves and indeed quite satisfactory, see Fig.
\ref{fig:q6f2}, green line,  though also here a rather large
higher Fock-state probability has to be assumed, $P^2_n=0.38$.

\begin{figure}
\begin{center}
\includegraphics[height=10cm,width=15cm]{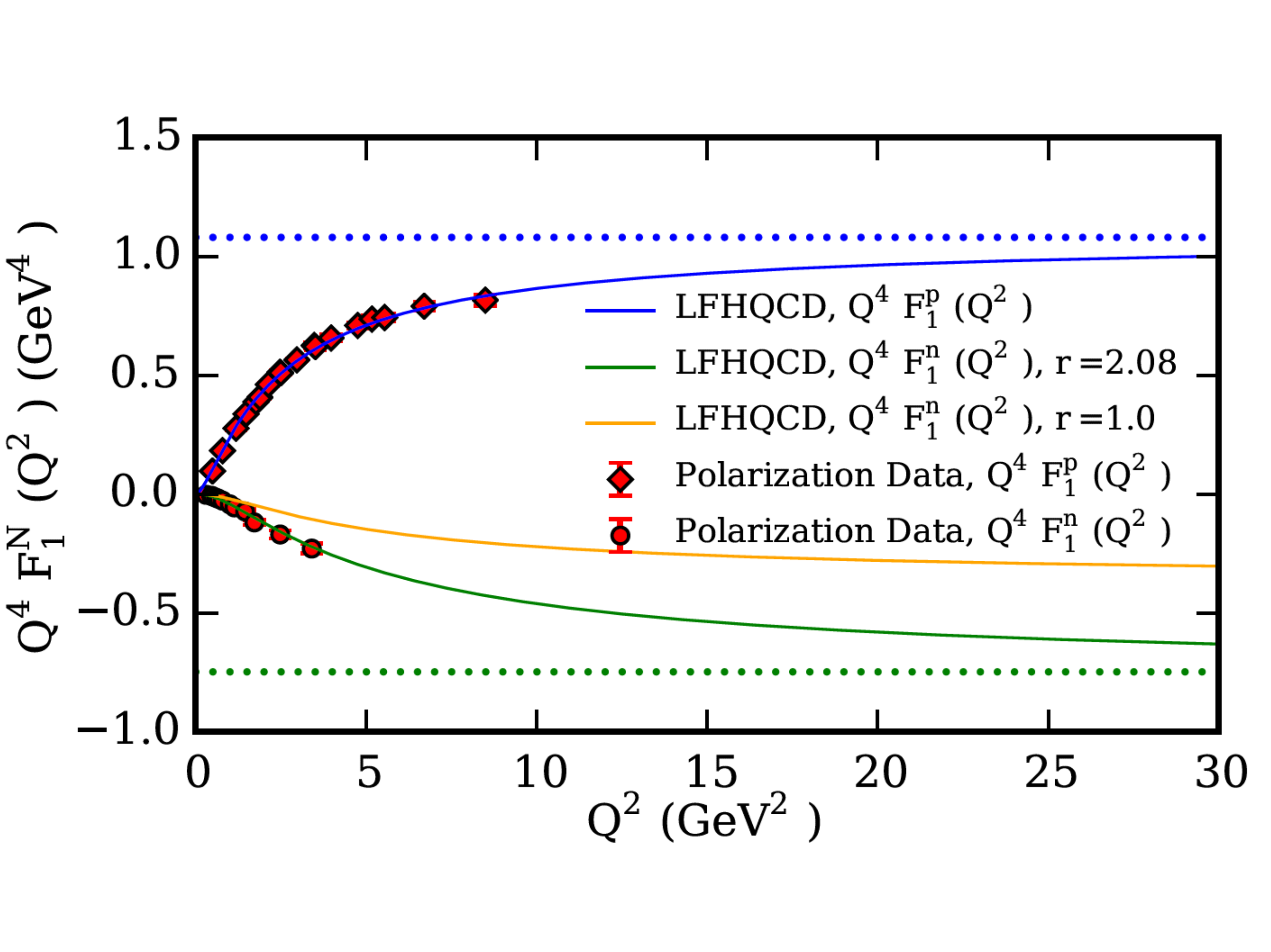}
\end{center}
\caption{ \lb{fig:q4f1} Polarization measurements and predictions
for the proton and neutron Dirac form factors. The blue line is
the prediction of the proton Dirac  form factor from LFHQCD,
\req{protonF1} with $P^1_p=0$, multiplied by $Q^4$. The orange
line is the predictions for the neutron Dirac  form factor,
$Q^4F_1^n(Q^2)$ with $P_n^1=0$, from \req{neutronF1}. The green
line is the prediction multiplied by a factor 2.1.  The dotted
lines are the asymptotic values, from \cite{Sufian:2016hwn}.}
\end{figure}

\begin{figure}
\begin{center}
\includegraphics[height=10cm,width=15cm]{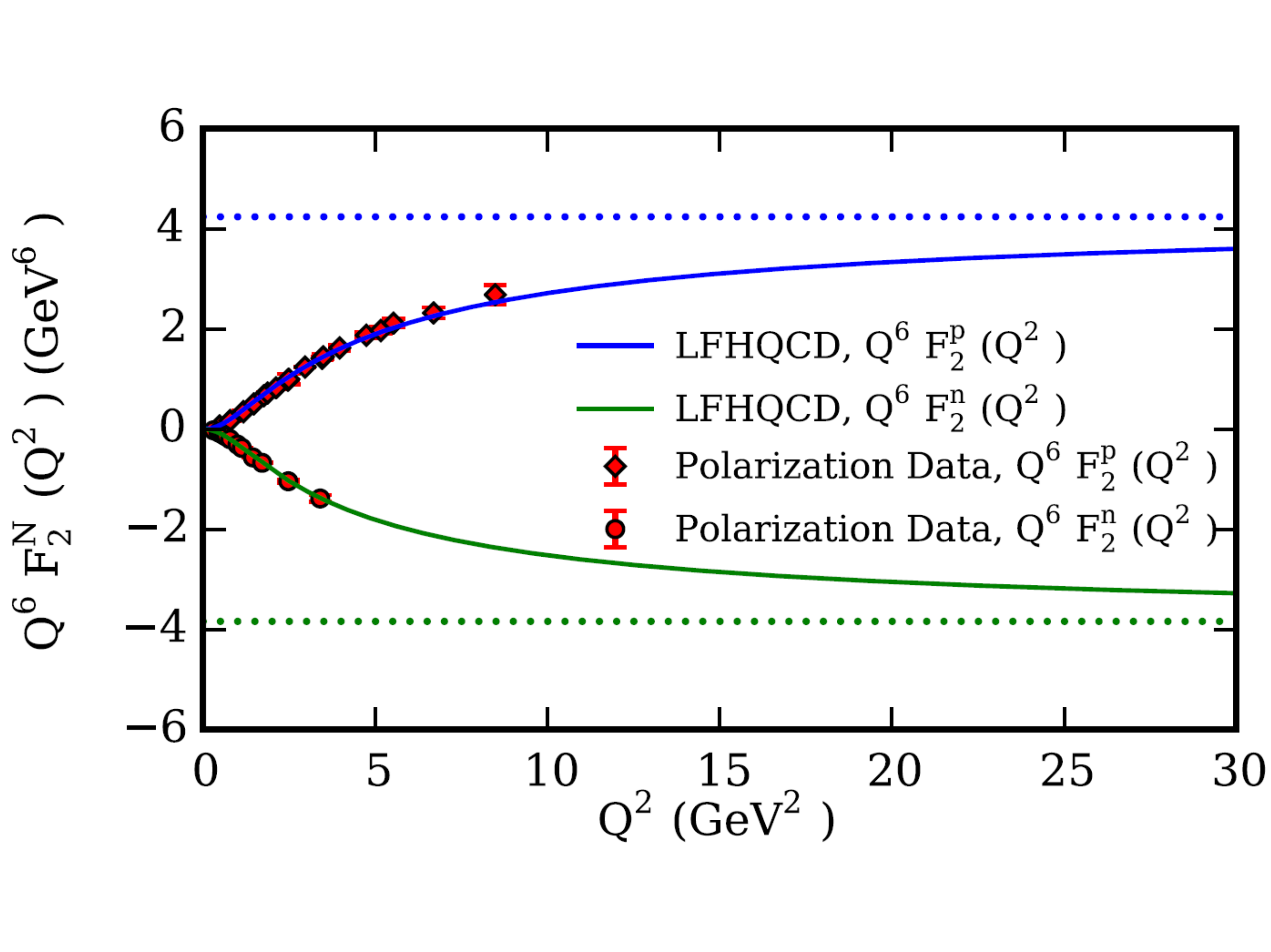}
\end{center}
\setlength\abovecaptionskip{-9pt}
\setlength\belowcaptionskip{-5pt} \caption{
\lb{fig:q6f2}Polarization measurements and predictions for the
proton and neutron Pauli form factors. The blue line is the proton
Pauli  form factor, $Q^6F_2^p(Q^2)$ prediction, with
$\gamma_p=0.27$ in Eq.~(\ref{protonF2}). The green line is the
prediction for the neutron Pauli  form factor, $Q^6F_2^n(Q^2)$, in
Eq.~(\ref{neutronF2}) from LFHQCD  with a higher Fock-state
probability $P_n^2 \approx 0.38$. The dotted lines are the
asymptotic predictions, from \cite{Sufian:2016hwn}.}
\end{figure}
\clearpage
\section{Summary} LFHQCD yields a very simple and elegant analytical formula for
form-factors of light hadrons in the space-like region as well as in the time-like region.
 The Form factor is a Euler  function which is determined by the  twist,
 see \req{finalfinalB}. To arrive at this formula from the originally derived one, \req{finb}
 a  shift from the argument
 $Q^2+1$ in \req{finb} to $Q^2+\half$  in \req{finalfinalB} has to be performed.

The structure of hadron form factors is very well described by
this analytical formula. In order to have quantitative agreement
with experiments additional parameters have to be introduced, as
the probability of a higher Fock state, containing an additional
meson. For the neutron Dirac form factor a strong deviation from
$SU(6)$ wave functions has to be parameterized by a multiplicative
factor, since LFHQCD cannot make predictions for the spin-isospin
structure of the the nucleon.

\chapter*{Acknowledgement}

One of the authors (HGD) wants to thank the Institute of Modern Physics and the Lanzhou
University for the warm hospitality and support, he also is greatly indebted to Guy de Teramond
and Stan Brodsky for countless instructive discussions. Our special thanks are also due to Prof.
Zhang Pengming, who initiated this lecture, for many fruitful discussions and suggestions.
\appendix
%\input{vorspann}
%\begin{document}
\chapter{Collection of wave functions}

We use different forms of wave functions, depending on the treated problem.
\section{Mesons}
For hadron wave functions we regard only such solutions which are normalizable and regular at $z=0$.
This implies that the spectrum has  discrete values given by \req{assp}.
The solutions $\tilde \Phi$ of the Euler-Lagrange equation  \req{HEL} derived directly from the action \req{action2} are
\beq
 \lb{asso2a}
\tilde \Ph_{nL_{AdS}}(z) =   z^{2 +L_{AdS}-J} L_n^{(L_{AdS})}(|\la| z^2)
e^{-(|\la|+\la) z^2/2} \enq
they are normalized as:
\beq  \lb{normHELa}
\int_0^\infty dz\, e^{\la z^2} z^{2 J-3} \Ph_{nL_{AdS}}(z)^2 =1
\enq
$L_n^{\ell}(x)$  are the associated Laguerre polynomials.

By the rescaling \req{resc1} we obtain solutions $\phi$ of a Schr\"odinger-like equqtion \req{HSX} which are:
\beq \lb{asso1a} \ph_{nL_{AdS}}(z) = 1/N
\,z^{L_{AdS}+1/2} L_n^{L_{AdS}}(|\la| z^2) e^{-|\la| z^2/2} \mbox{ with }\, N=\sqrt{\frac{(n+L) !}{2 n!}}\;|\la|^{-(L+1)/2}  \enq
They are normalized to  $\int_0^\infty dz \, (\ph_{nL_{AdS}}(z) )^2=1$

The light front wave functions $\phi^{LF}$ are solutions of the two dimensional LF Hamiltonian \req{LFH} for massless constituents. They are related to the Schr\"odinger like wave functions by \req{slf}:
\beq \label{slfa} \phi^{LF}(x, b_\perp)
=\sqrt{\frac{x(1-x)}{2 \pi \ze}}\, \ph(\ze). \enq
and are normalized to
\beq \int_0^\infty  dx \, \int d^2 b_\perp |\phi^{LF}[x,b_\perp]|^2 =1 \enq
where $\ze = \sqrt{x(1-x)}\, b_\perp$.

For the form factor it is convenient to introduce the twist wave functions $\Phi_\tau$ by
\beq \Phi_\tau(z) = e^{-\la z^2/2} \tilde \Phi(z) \enq
in order to compensate the factor $e^{\la z^2} $ in the interaction action \req{ form
factor1}.

\section{Baryons}
For baryons the Euler Lagrange equation \req{DE2} can be brought into the form \req{asso1} with the positive and negative chirality solutions,  which are regular and normalizable:
    \beq \lb{solbarb} \ps^+_{nL}(q,z)
= \ph_{nL}(z), \quad  \ps^-_{nL}(q,z) = \ph_{nL+1}(z) \enq
    The  solutions of \req{DE2} are
\req{rsbar})
      \beqa \lb{anwf3}
\Psi^+(z) &=& z^{2+L+1/2-T} L_n^{(L)}(|\la| z^2) e^{-|\la| z^2/2}\\
\Psi^-(z) &=& z^{3+L+1/2-T} L_n^{(L+1)}(|\la| z^2) e^{-|\la|
z^2/2} \nn \enqa where $T=J-\half$.

\section{Currents}
Here we look for solutions opf the Euler Lagrange equations \req{ELG}  which are defined for any value of $q^2$ but vanish for $z\to \infty$. Such a solution is:
\beq \lb{afinal3} \tilde \Phi(q,z)=
\rho(z)U\left(a_\la,L+1,|\la|z^2\right)\enq
     with $ \rho(z) = z^{L-J+2} e^{-(|\la|+\la) z^2/2}$.

The constant $a_\la$ depends on the sign of $\la$:
\beqa
\la <0 \quad  a_\la &=& -\frac{q^2}{4 |\la|} - \frac{\la}{4 |\la|}(4-2J+ 2L) \nn \\
 \lb{aa-} &=& -\frac{q^2}{4 |\la|} + \frac{1}{4}(4-2J+ 2L)
\enqa
\beqa
 \la>0 \quad  a_\la &=& -\frac{q^2}{4 |\la|} -\frac{\la}{4 |\la|}(4-2J+ 2L)+(L+1) \nn\\
 &=& -\frac{q^2}{4 |\la|} + \frac{1}{4}(2J+ 2L)  \lb{aa+}
\enqa
$U(a,b,x)$ is Kummers hypergeometric function.
%\end{document}

\end{document}